\documentclass[article]{aa}     

\usepackage{graphicx}
\usepackage{txfonts}
\usepackage{lipsum}
\usepackage{booktabs}
\usepackage{longtable}
\usepackage{caption}
\usepackage{subcaption}         
\usepackage{lscape}             
\usepackage{placeins}           
\usepackage{xcolor}

\usepackage{hyperref}

\begin{document}
	
	\makeatletter
	\let\do@linenumbers\relax
	\let\linenumbers\relax
	\makeatother
        
        \title{Signatures of planets and Galactic subpopulations in solar analogs}
        
        \subtitle{Precise chemical abundances with neural networks}
        
        \author{Giulia Martos,\inst{1, 2,3}\thanks{\email{gimartos@mpia.de}} Jorge Meléndez,\inst{3} Lorenzo Spina\inst{4,5} and Sara Lucatello\inst{4}
        }
        
        \institute{Max-Planck-Institut für Astronomie, Königstuhl 17, 69117 Heidelberg, Germany
                \and Fakultät für Physik und Astronomie, Universität Heidelberg, Im Neuenheimer Feld 226, 69120 Heidelberg, Germany  
                \and Department of Astronomy, Universidade de São Paulo, Rua do Matão 1226, 05508-090 São Paulo, Brazil   
                \and INAF-Osservatorio Astronomico di Padova, Vicolo dell'Osservatorio 5, 35122, Padova, Italy
                \and INAF-Osservatorio Astrofisico di Arcetri, Largo E. Fermi 5, 50125, Firenze, Italy\\}
        
        \date{Received March 21, 2025; accepted May 23, 2025}

        \abstract
        {} 
        {The aim of this work is to obtain precise atmospheric parameters and chemical abundances automatically for solar twins and solar analogs to find signatures of exoplanets, as well as to assess how peculiar the Sun is compared to these stars and to analyze any possible fine structures in the Galactic thin disk.} 
        {We developed a neural network (NN) algorithm using \texttt{Python} to derive atmospheric parameters and chemical abundances for a sample of 99 solar twins and solar analogs previously studied in the literature from normalized high-quality spectra from HARPS, with a resolving power of R $\sim$ 115000 and a signal-to-noise ratio of S/N > 400.}
        {We obtained precise atmospheric parameters and abundance ratios [X/Fe] of 20 chemical elements (Li, C, O, Na, Mg, Al, Si, S, Ca, Sc, Ti, V, Cr, Mn, Co, Ni, Cu, Zn, Y, and Ba). The results we obtained are in line with the literature, with average differences and standard deviations of $(2 \pm 27)$ K for T$_{\rm eff}$, $(0.00 \pm 0.06)$ dex for log g, $(0.00 \pm 0.02)$ dex for [Fe/H], $(-0.01 \pm 0.05)$ km s$^{-1}$ for microturbulence velocity (v$_t$),  $(0.02 \pm 0.08)$ km s$^{-1}$ for the macro turbulence velocity (vmacro), and $(-0.12 \pm 0.26)$ km s$^{-1}$ for the projected rotational velocity (vsin$i$). Regarding the chemical abundances, most of the elements agree with the literature within 0.01 - 0.02 dex. The abundances were corrected from the effects of the Galactic chemical evolution through a fitting versus the age of the stars and analyzed with the condensation temperature (T$_{\rm cond}$) to verify whether the stars presented depletion of refractories compared to volatiles.}
        {We found that the Sun is more depleted in refractory elements compared to volatiles than 89\% of the studied solar analogs, with a significance of 9.5$\sigma$ when compared to the stars without detected exoplanets. We also found the possible presence of three subpopulations in the solar analogs: one Cu-rich, one Cu-poor, and the last one being slightly older and poor in Na.}
        
        
        \titlerunning{Signatures of planets and Galactic subpopulations in solar analogs}
        \authorrunning{G. Martos et al.}
        \maketitle

        \section{Introduction}
        
        Precise chemical abundances are key for the characterization of planet hosts, since the formation and the presence of planets around the stars can alter their chemical composition, leaving chemical fingerprints that can be revealed via detailed analyses of the spectrum. A widely accepted hypothesis in the field of star formation is that stars and planets are formed at approximately the same epoch, from the gravitational collapse of an unstable molecular cloud. In this way, the chemical elements used to form the planets are sequestered in smaller amounts by the star in the last stages of stellar formation, resulting in a slight deficit of these elements in the stellar convective zone.
        
        The works of \citet{Melendez_2009}, \citet{Ramirez_2009}, \citet{Gonzalez_2010}, \citet{Nissen_2015}, \citet{Bedell_2018}, and more recently \citet{Rampalli_2024} showed that the Sun presents a depletion of $\sim$ 0.05 - 0.08 dex (12 - 20$\%$) of refractory elements in relation to volatile elements when compared to solar twins or solar analogs. According to \citet{Melendez_2009}, this difference in abundances is of the same order as the mass of refractory elements present in the rocky planets of the Solar System. They also found a strong correlation between this depletion and the T$_{\rm cond}$ of the elements in the protosolar nebula \citep{Lodders_2003}, with a higher depletion for the refractories, and a break in the trend in T$_{\rm cond} \sim 1200$ K (Fig. 1 of \citet{Melendez_2009}). This temperature is only found in the interior part of protoplanetary disks, where the rocky planets are located. \citet{Chambers_2010} showed that the peculiar solar abundance pattern relative to the average of solar twins could be erased by adding about four Earth masses of rocky material into the Sun's convection zone.
        
        The most plausible explanation for the chemical anomalies of the Sun is the presence of planets around it in a relatively stable configuration, preserving part of the deficiency of refractories imprinted from the formation of the planets, while other planetary systems could have had important engulfment events, enriching the star with refractory elements, resulting in no deficit. Among the alternative explanations for the depletion are the effects of the Galactic Chemical Evolution - GCE (for example, the contamination by supernovae), and the vanishing of the dust from the primordial cloud where the Sun was formed by the radiative pressure of luminous nearby stars \citep{Gustafsson_2018}. However, it is unknown why these events occurred with the Sun in particular.         
        
        Trends of the depletion of refractories with the condensation temperature were found by \citet{Ramirez_2015}, \citet{Teske_2016}, \citet{Maia_2019}, \citet{Galarza_2021}, \citet{Jofre_2021}, and \citet{Miquelarena_2024}, among others, in binary systems of twins or similar stars. As these systems are formed from the same molecular cloud approximately at the same time, it is expected that both stars present similar chemical composition during the main sequence (except for lithium and beryllium, which are the lighter elements that are destroyed in the stellar interior), and should not be affected by GCE. Therefore, the chemical inhomogeneity between the stars of the system suggests the occurrence of physical processes not necessarily related to the stellar evolution, such as the formation of planets or the engulfment of planets by one of the components \citep{Saffe_2017, Oh_2018, Liu_2020, Flores_2024}, with less and more refractory material, respectively.         
        
        The chemical signatures of planets are on the order of few 0.01 dex \citep{Maia_2019}. Thus, they can be revealed only from a thorough analysis of stellar spectra. The current precision of 0.01 - 0.02 dex is achieved employing the method of differential line-by-line spectroscopy \citep{Bedell_2014}, where the equivalent width (EW) of a spectral line is measured in strictly the same manner for the star under study and the comparison star, by comparing each line in the spectrum of the star and in the reference spectrum. The differential approach diminishes the statistical (observational) errors and the uncertainties associated with the \textit{log (gf)} of each line are totally erased, while the use of similar stars minimizes the systematic errors associated with stellar models, as they are mostly canceled in the comparison. Yet, it is necessary to employ spectra of high quality, with high resolution and high signal-to-noise ratio (S/N) and with attenuated contamination of telluric lines. The importance of revisiting high precision abundances is shown by the recent results of \citet{Cowley_2022}, who suggested that the Sun is not depleted in refractories. However, they analyze abundances from different works altogether, that introduces more errors on the analysis. As the line-by-line method is affected by subjective decisions, such as the definition of the continuum, as well as how the EW is measured, combining measurements made by different authors can increase the errors and suppress the subtle planetary effects. Also, the authors force the GCE trend to produce zero abundances for stars of solar age; hence, all stars will have similar abundance when compared to the Sun.
        
        Although the line-by-line differential method provides precise chemical abundances, it is unfeasible to apply it to large samples, such as those from large-scale surveys with thousands of stars and many elements, since manual measurements are time consuming. Some codes, such as ARES \citep{ARES}, can be used to measure EWs automatically. However, the results obtained using them lead to only moderate precision. This has a great impact on the study of very weak signatures, such as the planet's imprint on the photosphere of the star. In the literature, some automatic codes such as the Cannon \citep{Casey_2016} and the Payne \citep{Ting_2019} are being used to improve the precision of abundances of catalogs with hundreds of thousands of stars, such as APOGEE, using machine learning and spectra with medium to moderately high resolution and low-to-medium S/N. They were able to refine the abundances calculated with dedicated pipelines for the surveys and even reproduce results from the literature for chemical evolution and globular clusters with higher significance. The typical precision achieved for individual abundances was of 0.03 to 0.04 dex, and up to 0.1 dex for some elements. Also, in a more recent paper, \citet{angelo_2024} used the Cannon to obtain the atmospheric parameters of main sequence stars from Gaia DR3, with precisions of 72 K in T$_{\rm eff}$, 0.09 dex in log g, 0.06 dex in [Fe/H] and 1.9 km s$^{-1}$ in broadening velocity. However, they did not report chemical abundances. The performance of these codes should improve by employing data with higher resolution and S/N, resulting in more precise stellar parameters and abundances.
        
        In this work, we developed a neural network algorithm to obtain chemical abundances automatically at the same level of precision (0.01 - 0.02 dex) as the manual differential method, allowing a refined analysis of the chemical composition of solar twins and analogs. We used the derived abundances to compare the Sun to the solar twins and analogs and also study possible subpopulations present in these stars.

        \section{Sample and data}
        
        The sample is composed of 99 solar twins and close analogs, with temperatures within  T$_{\rm eff, \odot} \pm 150$ K and metallicities [Fe/H]$_{\odot} \pm 0.15$ dex  \citep{Strobel_1996,Ramirez_2009}. Of these, 79 were previously studied by \cite{Bedell_2018} and \cite{Spina_2018}, 6 by \citet{Martos_2023}, and 14 by \citet{Rathsam_2023}. The atmospheric parameters are around solar values, with the range of ages spanning the main sequence to permit the study of the GCE: 5600 $\leq$ T$_{\rm eff}$ (K) $\leq$ 5900, 4.10 $\leq$ logg (dex) $\leq$ 4.60, -0.15 $\leq$ [Fe/H] (dex) $\leq$ 0.15, 0.9 $\leq$ v$_t$ (km s$^{-1}$) $\leq$ 1.2, 0.95 $\leq$ Mass (M$_{\odot}$) $\leq$ 1.08, and 0.45 $\leq$ Age (Gyr) $\leq$ 9.80. This limitation in stellar parameters is important in that it allows us to avoid  possible anomalies found in the abundances that stem from systematic errors (and not due to the presence of planets). Finally, 11 stars of the sample have confirmed exoplanets.
        
        We employed high-resolution (R $\sim$ 115000) and high-S/N (> 400) spectra obtained with the  High Accuracy Radial velocity
        Planet Searcher (HARPS) spectrograph, downloaded from the ESO Science Archive Facility \footnote{\url{http://archive.eso.org/wdb/wdb/adp/phase3\_main/form}}. Around 20-30 individual spectra of minimum S/N > 40 obtained in different dates were combined in order to achieve the highest S/N possible and mitigate the contamination of telluric lines. The spectra were corrected by the radial velocity of the star and normalized using IRAF \citep{iraf}. We distinguished between spectra taken before and after the HARPS upgrade occurred in June 3rd 2015, when the optical fibers of the instrument were changed, as there are differences in the continuum. The spectra cover the region from 378 nm to 691 nm, where the spectral lines of many elements are observable.    
        
        We adopted the reflected spectrum from the asteroid Vesta as the reference solar spectrum. Also, the spectrum of the Moon and Ganymede were employed to obtain the uncertainties of the solar abundances, as explained in Section \ref{sec:solar_abundances}.

        \section{Method}
        
        \subsection{The algorithm}
        
        The algorithm developed to obtain atmospheric parameters and chemical abundances automatically was divided in two parts. The first was a neural network (NN), employed as a fast interpolator between the stellar labels (atmospheric parameters and abundance ratios [X/Fe]) and the flux of the star. Instead of using the normalized flux directly, we used the differential flux of the star in relation to the Sun, given in Equation \ref{eq:F_diff}, where $F_s$ and $F_{\odot}$ are the flux of the star and the Sun at a certain pixel, respectively. In this way, we mimic the way that the abundances were obtained in the differential method, where the differential abundance is given approximately by Equation \ref{eq:diff_abund} and replacing the definition of equivalent width ($EW = \int^\infty_0 \left( 1 - \frac{F_\lambda}{F_c}\right) d \lambda$), we obtain Equation \ref{eq:F_diff}. These equations are expressed as
        \begin{equation}
                \delta A_X \sim log_{10}\frac{EW_{X, \lambda}^{star}}{EW_{X, \lambda}^{\odot}}
                \label{eq:diff_abund}
        ,\end{equation}
        \begin{equation}
                F_{\rm diff} = log_{10} \frac{1-F_{s}}{1-F_{\odot}}
                \label{eq:F_diff}
        .\end{equation}
        
        The NN was built using the \texttt{Python} libraries \texttt{scikitlearn} and \texttt{tensorflow}, and it was composed of three hidden layers with 300 neurons each. We used the LeakyReLu activation function for each layer, except for the output, where we chose a linear activation function. For the loss or cost function, we used mean absolute errors. The NN was trained using ten thousand synthetic spectra generated with iSpec \citep{iSpec1, iSpec2}, separated as 80\% for the training set and 20\% for the test set. We adopted MARCS model atmospheres \citep{MARCS_used} and TurboSpectrum \citep{turbospectrum} as the synthetic spectra generator. The line list we adopted is version 6 of the Gaia-ESO Survey (GES) \citep{linelist_GES_v6}, with hyperfine splitting and isotopic and molecular data included. The labels used as input parameters were randomly sampled from a linear distribution around solar values to reflect the parameters of solar-type stars, as follows: T$_{\rm eff}$ = T$_{\odot}\ \pm$ 300 K; log g = log g$_{\odot}\ \pm$ 0.3 dex; $[$Fe/H$]$ = [Fe/H]$_{\odot}\ \pm$ 0.3 dex;  v$_t$ = v$_{t, \odot}\ \pm$ 0.5 km s$^{-1}$; vsin$i$ = 0.1 - 10 km s$^{-1}$ (the solar value being vsin$i_{\odot}$ = 1.9 km s$^{-1}$); and $[$X/Fe$]$ = [X/Fe]$_{\odot}\ \pm$ 0.3 dex. The macro turbulence velocity (vmac), was calculated using the relation obtained by \citet{dosSantos_2016} for solar analogs: $\rm vmac = \rm vmac_{\odot} - (0.00707\ \rm T_{eff}) + (9.2422 \times 10^{-7}\ \rm T_{eff}^2) + 10 +$$ (k1(logg - 4.44)) + k2$, where $\rm vmac_{\odot} = 3.1$, $k1 = -1.81$ and $k2 = 0$. For the determination of the atmospheric parameters, we adopted model atmospheres with solar-scaled ratios.
        
        The spectra used in the training of the NN were generated in a region of 500 pixels (5 \AA) around the center of each spectral line. For the determination of the atmospheric parameters, 87 Fe I and 17 Fe II were employed, as these lines are sensitive indicators of effective temperature, gravity, and microturbulence \citep{Gray_2008}. To  determine the abundance ratios, we computed the spectra around the lines of each element, based on the line list of \citet{Melendez_2014}. The 20 elements considered were the ones with atomic number Z $\leq$ 30, to account for both volatile and refractory elements, and with lines present in the range of HARPS spectra: Li, C, O, Na, Mg, Al, Si, S, Ca, Sc, Ti, V, Cr, Mn, Co, Ni, Cu, Zn, and two heavier elements, Y and Ba. 
        
        The second part of the algorithm was a fitter, responsible for finding the labels that best reproduce the observed $F_{\rm diff}$ of the star. For this, models resulting from the training of the NN were fitted to the observed data using the library \texttt{lmfit}. We used the Least Squares Method to minimize a residual function, defined as the difference between the model and the observed data, weighted by the errors in the data. The uncertainties in $F_{\rm diff}$ are given by Eq. \ref{eq:uncertainty}, obtained by propagating the errors of $F_{\rm diff}$. $\sigma_{F_{star}}$ and $\sigma_{F_{\odot}}$ are the errors of the fluxes of the star and the Sun. The adopted value for both $\sigma$s was 0.002, corresponding to a S/N = 500.
        
        \begin{equation}
                \sigma_{F} = \frac{1}{|(1-F_{star})(1-F_{\odot})ln10|} \sqrt{\left( \frac{\sigma_{F_{star}}}{1-F_{star}}\right)^2 \left( \frac{\sigma_{F_{\odot}}}{1-F_{\odot}}\right)^2}
                \label{eq:uncertainty}
        .\end{equation}\\
        
        A region of 30 pixels around the center of the line, corresponding to 0.3 \AA, was considered in the fit. This was found to be the optimal range to work with, so that features surrounding the line that do not correspond to the chemical element are disregarded and yet a small portion of the continuum is considered. Also, the fluxes were interpolated in this range because a certain pixel may not correspond to the exact same wavelength in the spectrum of the star and the Sun, causing small shifts between both fluxes, which are increased when calculating the log of differential fluxes. In addition, line blends were masked automatically by assigning an infinite error to the flux to mitigate the contamination of other elements in the line of interest. A blend was identified as an extra region of absorption next to the central line, where the flux starts to decrease again after it was increasing towards the continuum.The blend masking was applied only for the abundances, as the Fe lines used to obtain the atmospheric parameters are well isolated.
        
        The final labels were obtained in two steps. First, the data was fitted without any processing besides the interpolation, to have a first guess of the result. Next, the values were refined. The first guess of the parameters was used to generate a synthetic spectrum around the lines and the observed fluxes were re-normalized, considering the continuum of the synthetic spectrum as the reference. To do this, the value of the 95\% percentile of a region of 50 pixels around the line, corresponding to 0.5 \AA, was used as the continuum value for both the observed and synthetic spectrum. The ratio between these two values was the normalization factor used to multiply the observed fluxes to do the re-normalization.         
        
        The atmospheric parameters and abundances for each element were obtained separately, both generally and on a line-by-line basis, respectively. For the first, the $F_{\rm diff}$ values for all Fe lines were concatenated and fitted as a single data array. This forces the fitting algorithm to obtain a set of atmospheric parameters that is appropriate for all Fe lines simultaneously. Indeed, when using spectroscopic equilibrium, it needs to be achieved while considering all the Fe lines. The values adopted as the parameter uncertainties were taken as the uncertainties of the fit. 
        
        To calculate the abundance ratios, the atmospheric parameters were fixed to the values fitted before and each line of the element was fitted individually. The final abundance adopted for the element was the average of all abundances obtained individually with each line, calculated in linear space. The uncertainties  were calculated as the quadratic sum of the error associated with the fit and the systematic errors due to the uncertainties of the atmospheric parameters. The first one was calculated as the standard deviation of the abundances obtained with each line, divided by $\sqrt{N}$, where $N$ is the number of lines. For the second, the abundances were calculated again using the value of each parameter plus its error, for each parameter individually. 
        
        To test the method, we fit 100 synthetic spectra from the test set to assess how well the NN can recover parameters from simulated data. The results are in excellent agreement with the original values. The average residuals and standard deviations are $(-0.2 \pm 3.9)$ K for T$_{\rm eff}$, $(-0.001 \pm 0.010)$ dex for log g, $(0.000 \pm 0.004)$ dex for [Fe/H], $(0.000 \pm 0.007)$ km s$^{-1}$ for v$_t$, $(0.000 \pm 0.015)$ km s$^{-1}$ for vmacro, and $(0.011 \pm 0.064)$ km s$^{-1}$ for vsin$i$.
        
        \subsection{Obtaining the stellar labels}
        
        The fit of the concatenated 104 Fe lines for the first guess of the atmospheric parameters takes $\sim 30$ seconds for each star and the refined fit takes $\sim 80$ seconds. Thus, all the atmospheric parameters for one single star are obtained in $\sim 2$ minutes. If each line in the manual method is measured in 1 minute, our method is more than 50 times faster just considering the measurements, as the parameters would need to be further calculated using spectroscopic equilibrium.
        
        The results for the atmospheric parameters (Table \ref{tab:params_NN}) were compared with the 79 stars in common with \citet{Spina_2018} (Figure \ref{fig:result_params}). The average residuals and standard deviations are $(2.0 \pm 27.1)$ K for T$_{\rm eff}$, $(0.00 \pm 0.06)$ dex for log g, $(0.00 \pm 0.02)$ dex for [Fe/H], $(-0.01 \pm 0.05)$ km s$^{-1}$ for v$_t$, $(0.02 \pm 0.08)$ km s$^{-1}$ for vmacro, and $(-0.12 \pm 0.26)$ km s$^{-1}$ for vsin$i$. The dispersion (i.e., the standard deviation of the parameters calculated with NN in relation to the literature values) is similar and even smaller than typical dispersions reported in the literature. There are some outliers, for example, the star HIP 102040, for which the values of log g and v$_t$ found (4.18 dex and 0.716 km s$^{-1}$) are much smaller than in the literature (4.48 dex and 1.05 km s$^{-1}$). Overall, the global results follow the 1:1 expected behavior. In general, the NN is able to identify solar twins because the star-to-star scatter (e.g., 27 K in T$_{\rm eff}$) is smaller than the typical definition of solar twins (within 100 K of the Sun), and the case is similar for the other parameters.
        
        Regarding the abundance ratios, the generation of the synthetic spectrum around a spectral line used to do the re-normalization takes $\sim 3-4$ seconds and the fit of the line is done in $\sim 0.3$ seconds. An individual line is fitted in less than 5 seconds. Considering that we can take 1 minute to measure a line manually in the differential line-by-line method, our method is more than ten times faster. The abundance ratios obtained (Table \ref{tab:abundances}) were compared to the results of \citet{Bedell_2018} and \citet{nissen_2020}, with 79 and 27 stars in common, respectively. For approximately half of the elements (Na, Mg, Al, Si, Ca, Ti, Cr, Co, Ni, and Cu), we reached a dispersion on the order of 0.01 dex and 0.02 dex for the majority of the remaining elements. Figure \ref{fig:comp_abunds} shows the comparison of the abundances with the literature.
        
        Higher dispersions, for Zn (0.028 dex) and S (0.031), for example, can be due to normalization issues and the small number of spectral lines available (three lines for Zn and two lines for S, while other elements  usually have more than five lines available). In the case of O, the higher dispersion (0.058 dex) is due to the fact that the only line adopted was the forbidden line in 6300 \AA, which is very weak and has a blend with Ni, making both the manual and automatic measurements difficult. The discrepancies in the comparison can also be due to the observational differences with other authors, as \citet{Bedell_2018} used the O triplet in 7774 \AA\, observed at a lower spectral resolution, R $\sim$ 60 k.
        
        The abundances of the heavier elements Y and Ba were compared with \citet{Spina_2018}. A larger dispersion of around 0.03 dex for these elements can be explained by the fact that they are affected by the hyperfine splitting, which could not be well described in the line list used to generate the synthetic spectra for the training. The Li abundance was compared with \citet{Martos_2023}. This is the element with the higher dispersion, ten times the dispersion of other elements ($\sim$ 0.1 dex). The Li line in 6707.8 \AA\ is extremely shallow and the line region is affected by species of other elements, such as CN and C$_2$, as shown in Figure 2 of \citet{Carlos_2016}, which could have influenced the results. The larger differences are also for the lower Li abundances, which have larger uncertainties, as the determination for the vanishing absorption line is more uncertain and sometimes it is only an upper limit. This difficulties in the calculation of low Li abundances is also reflected by the larger uncertainties in the results reported in \citet{Martos_2023} for the stars with lower Li content. 
        
        The differences between our results and the literature may also be explained by the assumptions that were made by the respective authors. For example, they assume that the shape of the lines are Gaussian functions. However, the lines are closer to Voigt profiles or in some cases, neither of these functions are able to reproduce the observed profile, for example, due to blends. In this sense, the NN method is more robust, because no assumptions are made about the shape of the line; the pixels are fitted independent of any defined function, just using the weights provided by the training of the NN. It is also important to stress that the NN method is fully reproducible, while past works using the differential line-by-line method were subject to human assumptions and decisions on how to do the manual measurements; for example, the placement of the continuum. Nevertheless, the errors are minimized in the                 line-by-line technique, as the human assumptions made are the same for the star and the Sun.  For example, as can be seen in Figure 1 from \citet{Carvalho-Silva_2025}, the human differential line-by-line results for the solar twin 18 Sco show excellent agreement among different authors using different spectra, agreeing in T$_{\rm eff}$ within 6 K (1-sigma). Our NN result for 18 Sco also agrees within 6 K with that average value.

\begin{figure}[htbp]
                \center
                \begin{tabular}{cc} 
                        \includegraphics[width=0.85\linewidth]{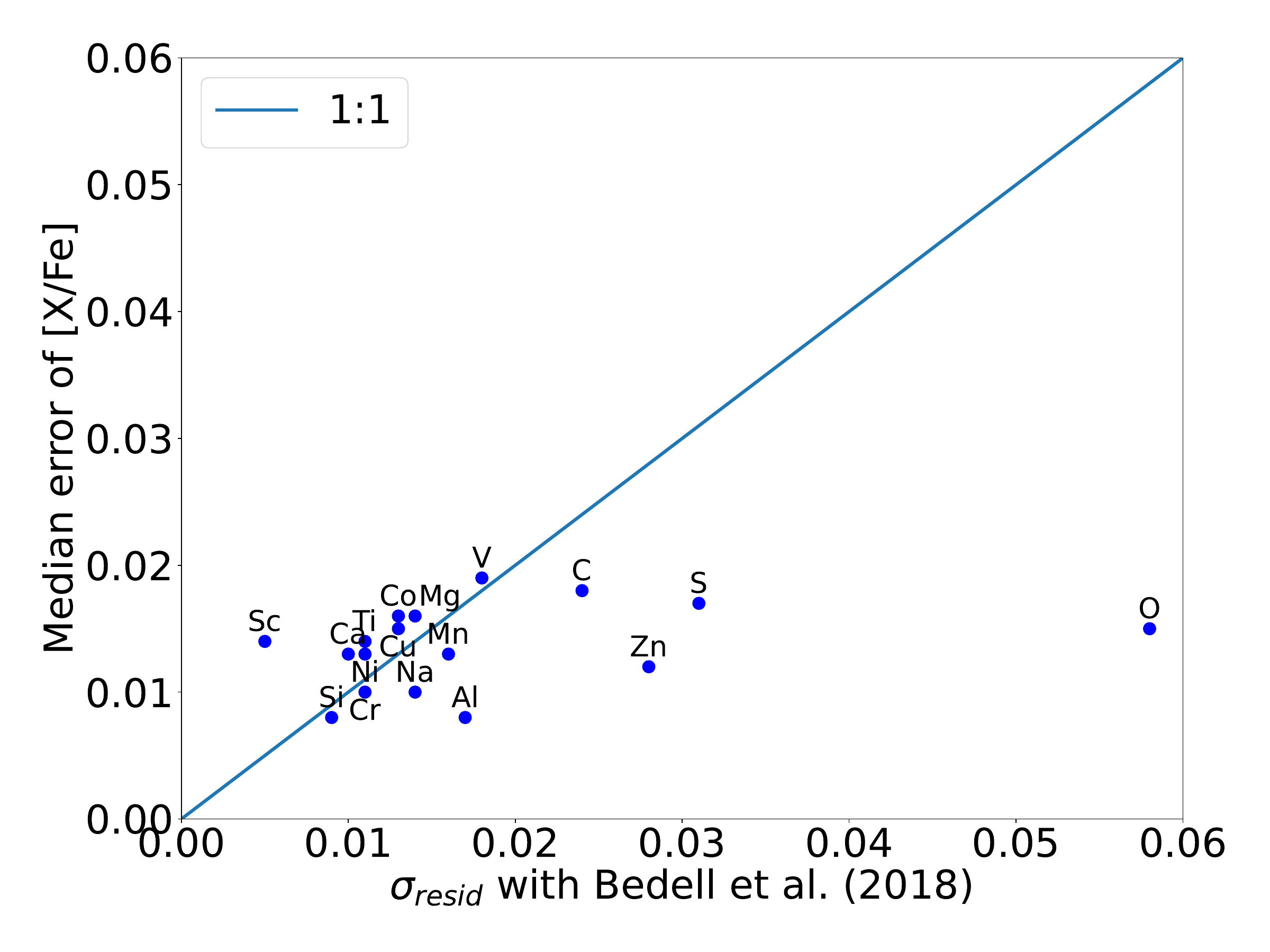}\\
                \end{tabular}
                \caption{Median error of the abundance ratios [X/Fe] of the elements versus the standard deviation of the residuals when comparing our automatic abundances with \citet{Bedell_2018}}
                \label{fig:err_elem_sigma_bedell}
        \end{figure}

        Figure \ref{fig:err_elem_sigma_bedell} shows the median error of each element against the standard deviation of the residuals of the comparison with \citet{Bedell_2018}. For the elements located on the left of the blue line, such as Sc, our method may be overestimating the error bars. For the elements around the blue line, the error bars are comparable with $\sigma$, indicating that they are realistic, as the errors from the authors also contribute to the dispersion. On the other hand, for the elements more complicated to be measured, such as C, O, S, and Zn, our error bars could be underestimated.
        
        \subsection{Limitations of the method}
        
        Our results are affected by the synthetic gap, which is the difference that exists between real and synthetic data and that cannot be surpassed. It is introduced when we use synthetic spectra generated with models that are not the ideal representation of the atmospheres to train the NN instead of real data. In particular, the model used is a 1D LTE model. Thus, the 3D hydrodynamical structure of the atmosphere and non-LTE effects are disregarded, but corrections could be performed afterwards when necessary. 
        
        Besides that, the line list used to generate the spectra depends on atomic data that sometimes can be poorly estimated, as some regions were not identical to the observed spectrum. This particular issue did not impact  our results in a significant way as the work was done in a differential way and the stars have comparable stellar parameters, so this type of difference was roughly canceled out in the calculations. In a few cases, the line list either introduced additional spectral lines that do not exist on real spectra (which had to be masked) or specific features of the region were not accounted for, such as the calcium autoionization feature around the Zn line in 6362.5 \AA. These differences in the observed and synthetic spectrum affect the determination of the abundances because the NN learns what is present in the synthetic spectrum and when it fails to find a certain pattern or behavior in the observed spectrum, it provides an abundance value that may not be correct.
        
        Possible improvements for this method are the use of more robust line lists, with more complete atomic and hyperfine splitting data for the spectral lines, along with with a detailed revision around each important region to remove spurious features. Also, the use of other codes or model atmospheres to generate the synthetic spectra (including 3D, magnetic, and NLTE effects when possible) to work as close as possible to a realistic scenario and to diminish the influence of the synthetic gap.

        \section{Results and analysis}
        
        \subsection{Solar abundances}
        \label{sec:solar_abundances}
        
        The reflected spectrum of Vesta was employed in this work as the reference solar spectrum. To assess the effect of using different solar spectra and the error due to the choice of Vesta as the reference, the solar abundances of the 20 elements were also obtained using the reflected spectrum of the Moon and Ganymede, both in relation to Vesta. The three solar spectra employed were obtained with the HARPS spectrograph and post-processed in the same manner as the stellar spectra. The results are shown in Table \ref{tab:solar_abundances}. The last column shows the error in the solar abundances, calculated as the average of the absolute value of the abundance obtained from the Moon and Ganymede spectra. If the abundances were exactly zero, the error was assumed as the standard deviation of all the abundances obtained using the Vesta spectrum. The latter can also be considered as the error of the method itself. 
        
        All the abundances are compatible with the zero abundance expected for the Sun within the 3$\sigma$ range defined by the uncertainties listed in the last column of Table \ref{tab:solar_abundances}. The elements with higher standard deviations are C, S, Ca, Sc, Y, and Ba. The larger uncertainty is that of Li, being $>$ 0.10 dex. Apart from that case, most of the elements have errors of around 0.01 dex.

        \subsection{Galactic chemical evolution and the possible presence of Galactic thin-disk subpopulations}
        
        The abundance ratios obtained are in large part due to the evolution of the Milky Way, as they should reflect the chemical composition of the interstellar medium (ISM) when the star was born; in other words, the abundances depend on the age of the star. We refer to this effect as the Galactic chemical evolution (GCE). The effects of metallicity for GCE in this work are small, as the [Fe/H] of the stars of the sample span a limited range around solar values. Also, chemical evolution depends on the galactic location, with a faster enrichment occurring in the inner regions of the Galaxy. Our sample is in the solar neighborhood,  thereby avoiding large variations due to the Galactic chemo-dynamics, but some of our sample stars may have suffered radial migration, meaning that they could have originated elsewhere and migrated towards near the solar vicinity \citep{Prantzos_2023, Plotnikova_2024}.
        
\begin{figure}[htbp]
                \includegraphics[width=0.85\linewidth]{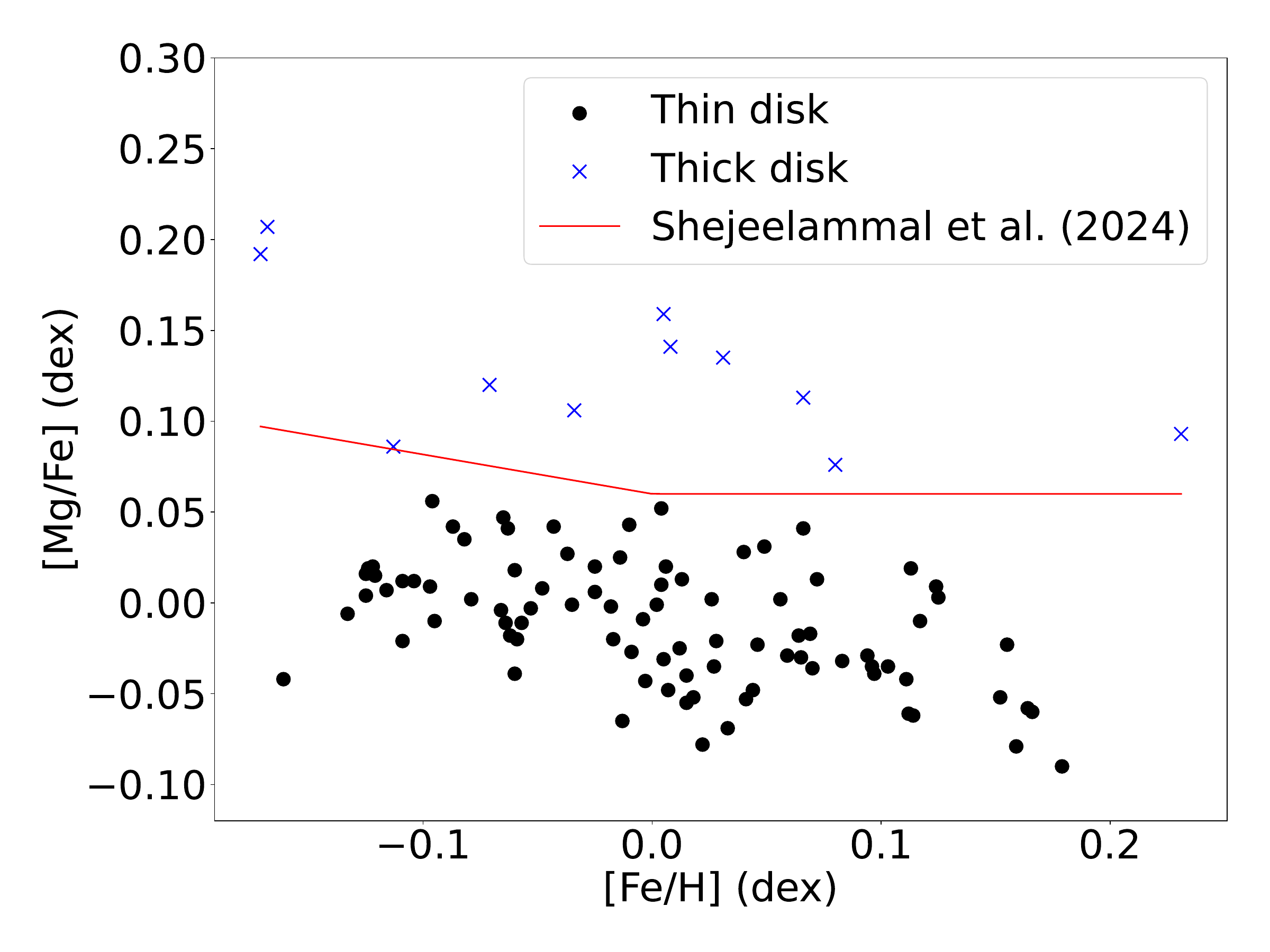}\\
                \caption{Separation of the stars of the sample between thin and thick disk according to their [Mg/Fe] ratio in relation to [Fe/H]. The stars above the red line and represented by blue crosses belong to the thick disk.}
                \label{fig:sep_disks}
        \end{figure}

        The abundance ratios were correlated with the age of the stars to remove the influence of the GCE and minimize its influence on the detection of planetary signatures. First, the stars were separated between thin and thick disk populations using a chemical criterion according to their [Mg/Fe] abundance ratios in relation to [Fe/H], using the proposed equation from \cite{Adibekyan_2011} as adapted by \citet{Shejeela_2024}. The eleven stars identified as part of the thick disk are shown as blue crosses in Figure \ref{fig:sep_disks} (HIP 14501, HIP 28066, HIP 30476, HIP 33094, HIP 65708, HIP 73241, HIP 74432, HIP99115, HIP 108158, HIP 109821, and HIP 115577). These stars had already been singled out by \citet{Bedell_2018} as coming from the thick-disk population  as part of their empirical selection of stars older than 8 Gyr and visual enhancement in $\alpha$ elements. 
        
        Only the stars of the thin disk were considered in the correlation with age. A linear fit ($\rm [X/Fe] = A \times Age + B$) was performed using the Python library \texttt{scipy} with the orthogonal distance method, which takes into account the error bars of both variables. The fits for all elements are plotted in Figure \ref{fig:linear_fit_GCE} and the coefficients are shown in Table \ref{tab:coefs_GCE_linear}), where the last column contains the standard deviation of the residuals. The GCE correction is valid for the entire range of metallicity of our sample (-0.15 $\leq$ [Fe/H] $\leq$ 0.15 dex). We also tried a quadratic fit. However, since both linear and quadratic fits presented similar $\chi^2$ values and a very similar behavior for some elements as well (based on a visual inspection), the linear fit was chosen for simplicity. The correction was done by subtracting the linear function from the abundances replacing the age of each star.
        
\begin{figure}[htbp]
                \center
                \includegraphics[width=0.85\linewidth]{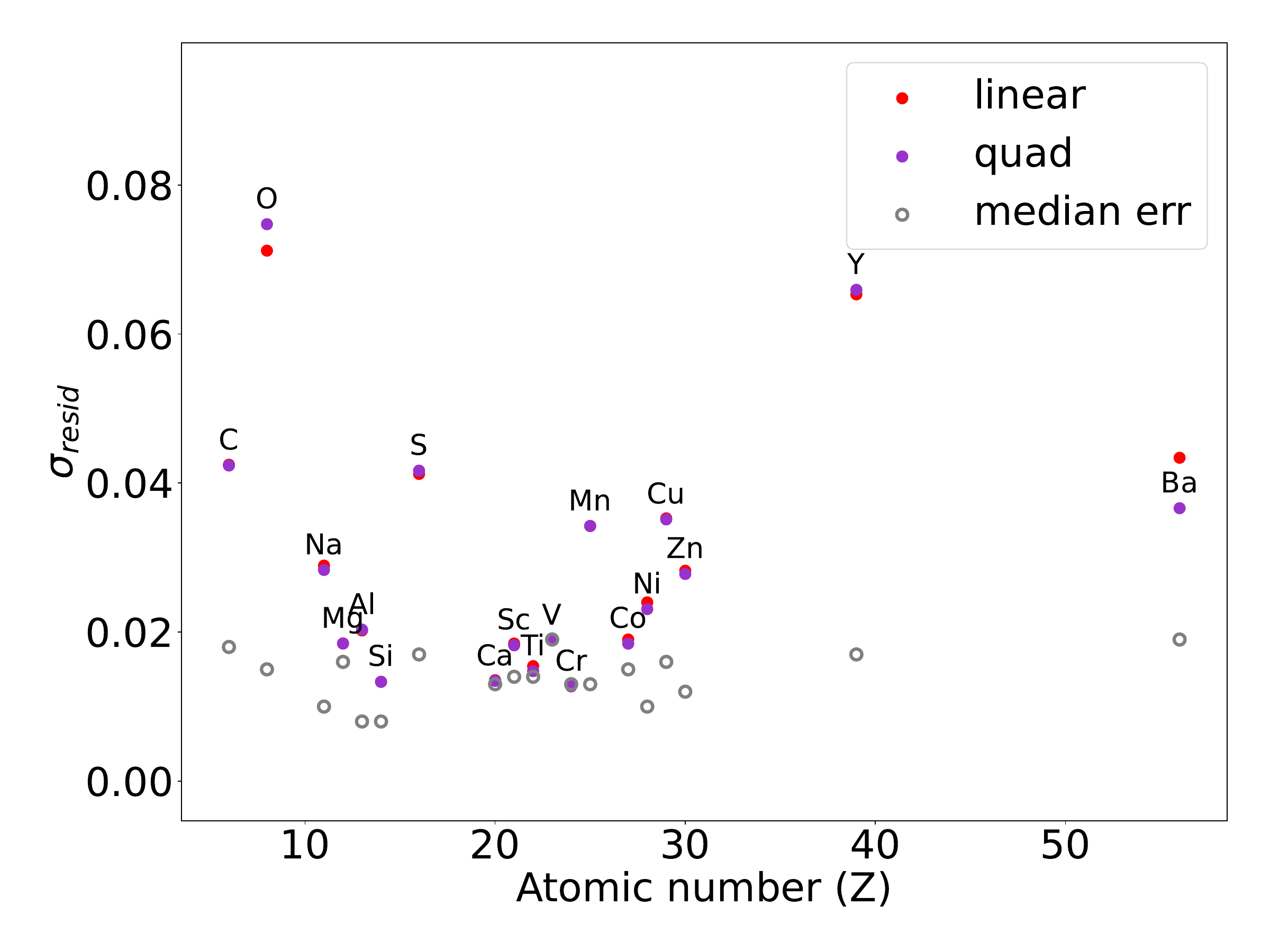}\\
                \caption{Standard deviation of the residuals of the fit of [X/Fe] versus age and median error according to the atomic number (Z) of the elements.}
                \label{fig:std_Z}
        \end{figure}

        Figure \ref{fig:std_Z} shows the standard deviation of the residuals of the fit for each element, for both linear and quadratic fits. Due to our high precision, we can discern better cosmic scatter from measurement errors. For Ca, Ti, Cr, and V, the errors of the method are about the same as the dispersion of the abundances, being thus harder to detect abundance peculiarities. On the other hand, other elements have a dispersion larger than the errors, singled out based on the evidence  of scatter of an astrophysical nature -- except in the cases where our measurement errors could be underestimated, such as in the case of oxygen. Considering our error bars, the somewhat larger scatter in some plots of Figure \ref{fig:linear_fit_GCE} may be of astrophysical nature, rather than due to errors in our analysis.
        
\begin{figure*}
                \center
                \begin{tabular}{cc} 
                        \includegraphics[width=0.4\linewidth]{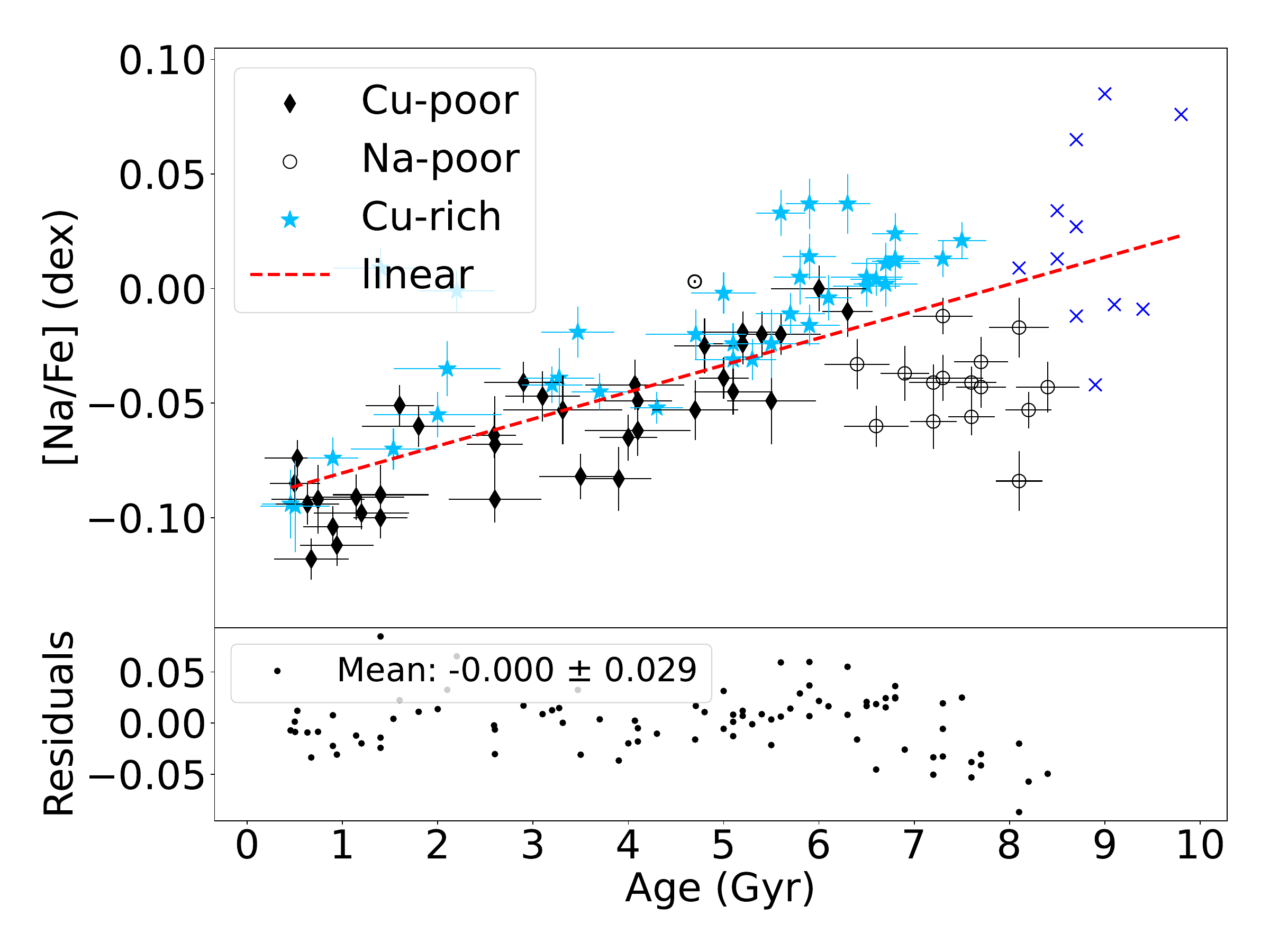} & \includegraphics[width=0.4\linewidth]{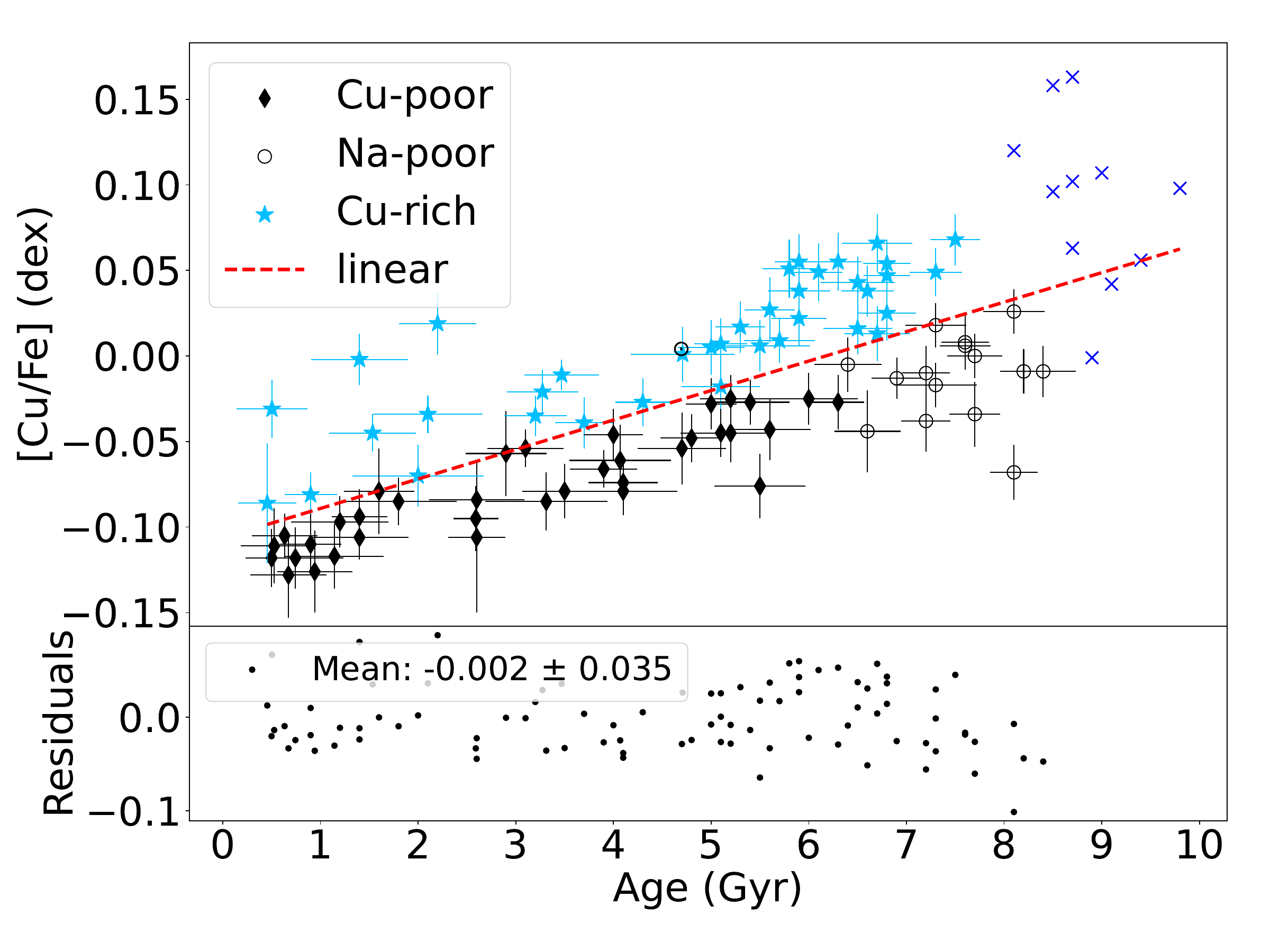}\\
                \end{tabular}
                \caption{Fit of [Na/Fe] and [Cu/Fe] versus age. The open circles indicate a possible older subpopulation poor in Na, and the blue stars and black diamonds represent possible subpopulations enhanced and reduced in Cu, respectively. The blue crosses correspond to thick disk stars that were not considered in the fit. The bottom panel shows the residuals, as well as their average and standard deviation.}
                \label{fig:subpops}
        \end{figure*}

\begin{figure*}
                \center
                \begin{tabular}{cc} 
                        \includegraphics[width=0.4\linewidth]{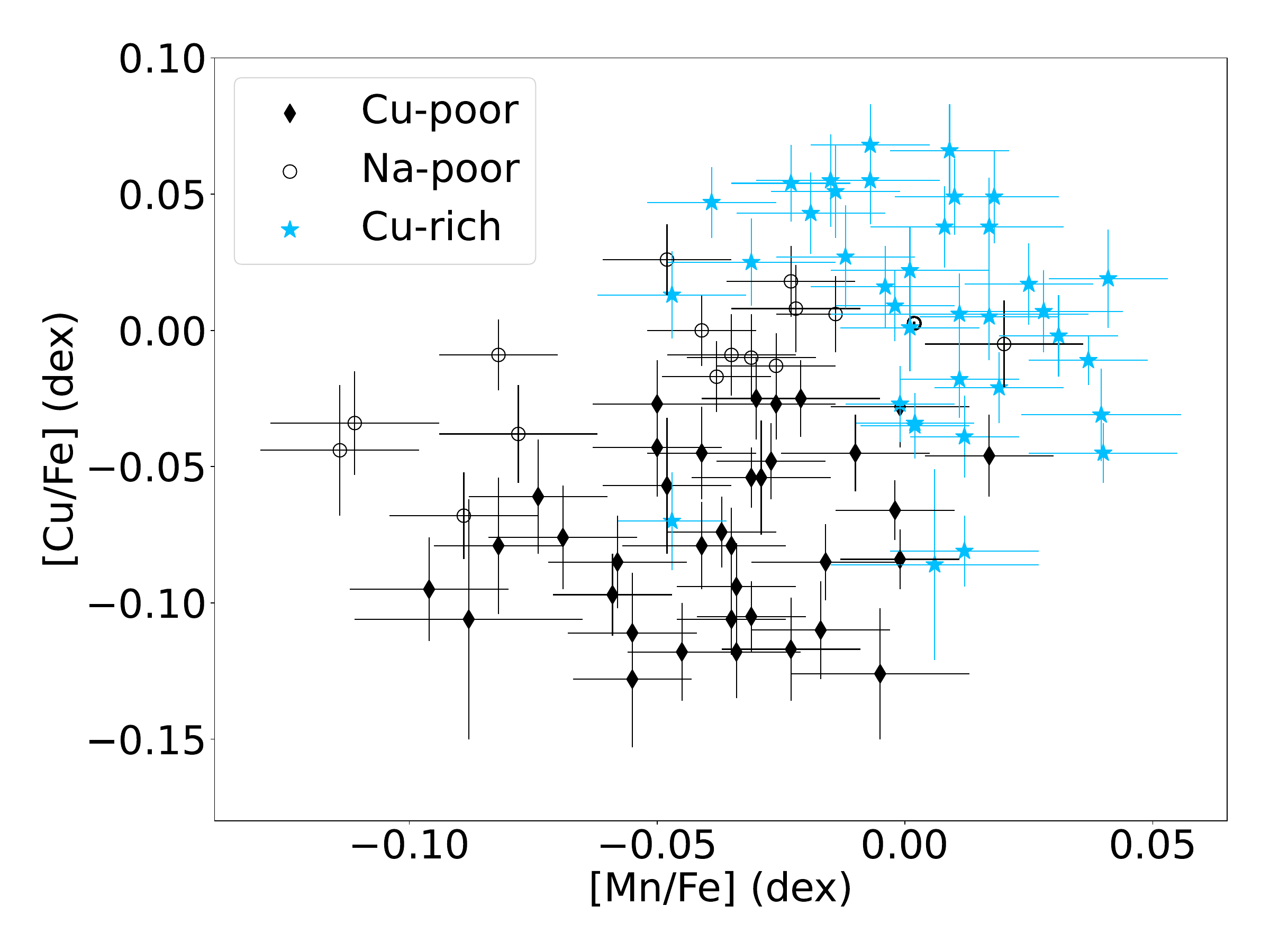} & \includegraphics[width=0.4\linewidth]{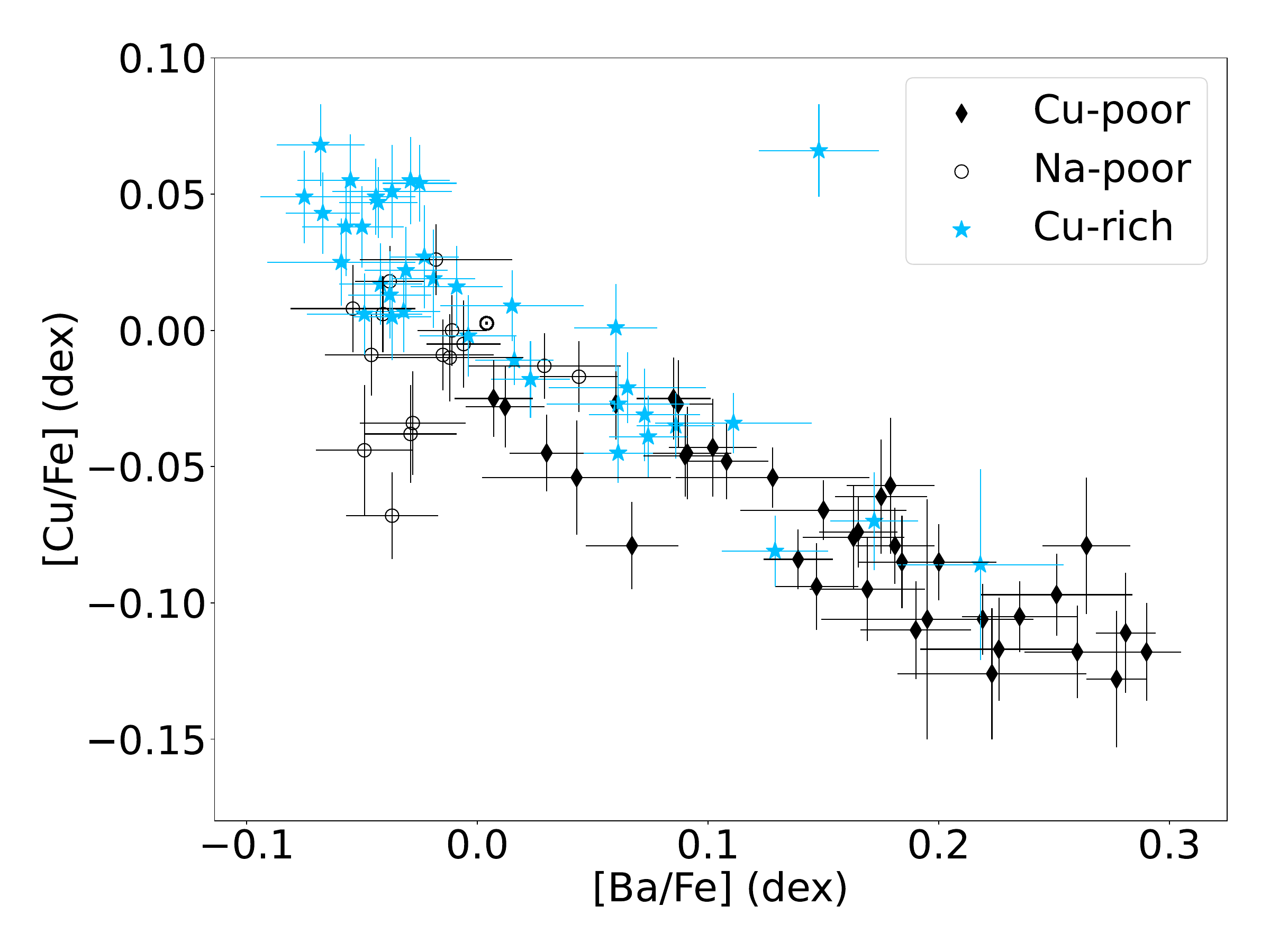}\\
                \end{tabular}
                \caption{Relation of [Cu/Fe] abundances with [Mn/Fe] and [Ba/Fe], where it is also possible to note the separation of the stars into the subgroups.}
                \label{fig:Cu_Mn_Ba}
        \end{figure*}

        From Figure \ref{fig:subpops}, we noted the possible presence of three Galactic subpopulations in our sample of solar analogs. In the [Na/Fe] plot, we identified a group of older stars (age $>$ 6 Gyr) and lower abundances, represented as open circles. Other possible subpopulations were identified, as the richer stars above the linear fit in the [Cu/Fe] plot, represented by blue star markers, and the poor stars below the fit, shown as black diamonds. These groups can also be identified in equivalent regions in the plots of other elements, such as Al, Si, Mn, Co, Ni, and Zn (Figure \ref{fig:linear_fit_GCE}). It is interesting that for [Ba/Fe] the behavior is reversed. So far, these subpopulations have not been reported in the literature. From Figure \ref{fig:Cu_Mn_Ba}, it is also possible to note the separation of the subgroups, with Cu and Mn varying in the same direction and Cu and Ba exhibiting an opposite behavior with respect to each other. 
        
\begin{figure}[htbp]
                \center
                \includegraphics[width=0.85\linewidth]{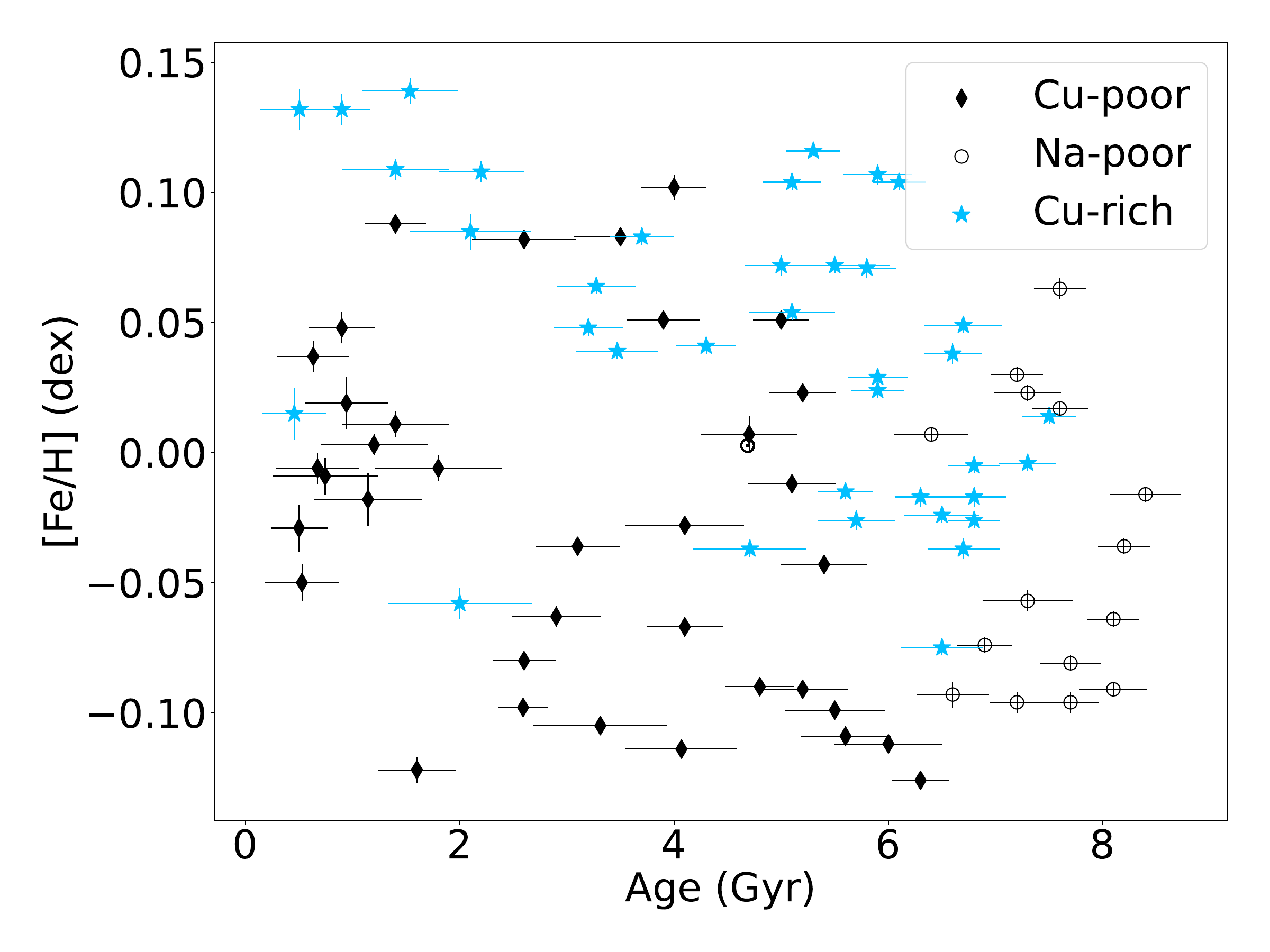} 
                \caption{Distribution of [Fe/H] versus age for the stars of our sample, distinguished by the possible subgroups.}
                \label{fig:feh_age}
        \end{figure}

        In Figure 4 of \citet{nissen_2020}, there is a region with smaller [Na/Fe] ratios for older stars. However, the authors distinguish only two populations, based on a separation  identified in a [Fe/H] versus age plot (their Figure 3). We did not find this separation in our sample, as shown in Figure \ref{fig:feh_age}. Also, in their subgroups, there is a overlap of the stars that would be part of our Na-poor group with some Cu-rich stars. 
        
\begin{figure}[htbp]
                \center
                \includegraphics[width=0.85\linewidth]{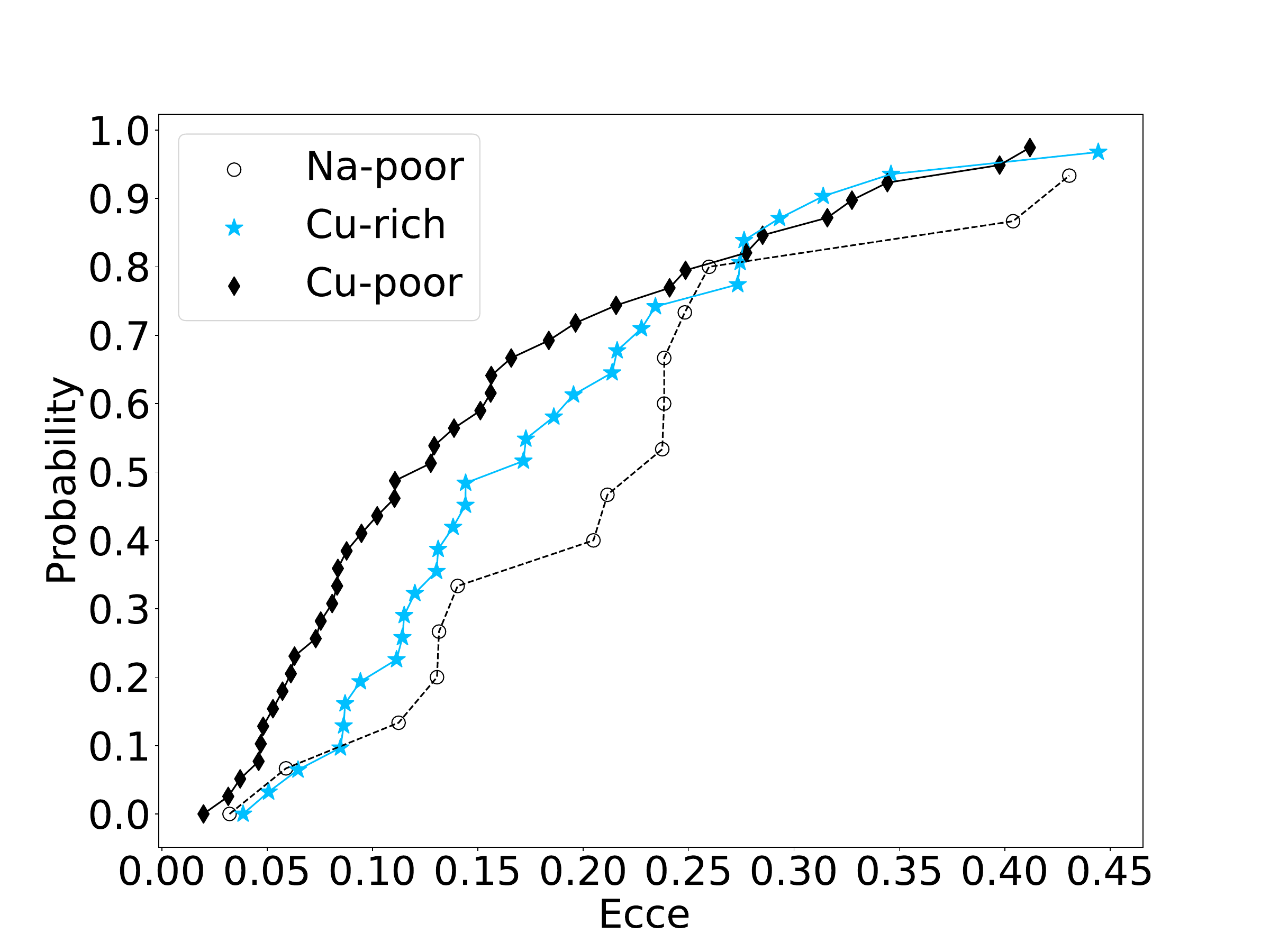} 
                \caption{Cumulative distribution of the eccentricity of the stars that belong to the Na-poor (open circles), Cu-rich (blue stars), and Cu-poor (black diamonds) samples.}
                \label{fig:ecce}
        \end{figure}

\begin{figure}[htbp]
                \center
                \includegraphics[width=0.85\linewidth]{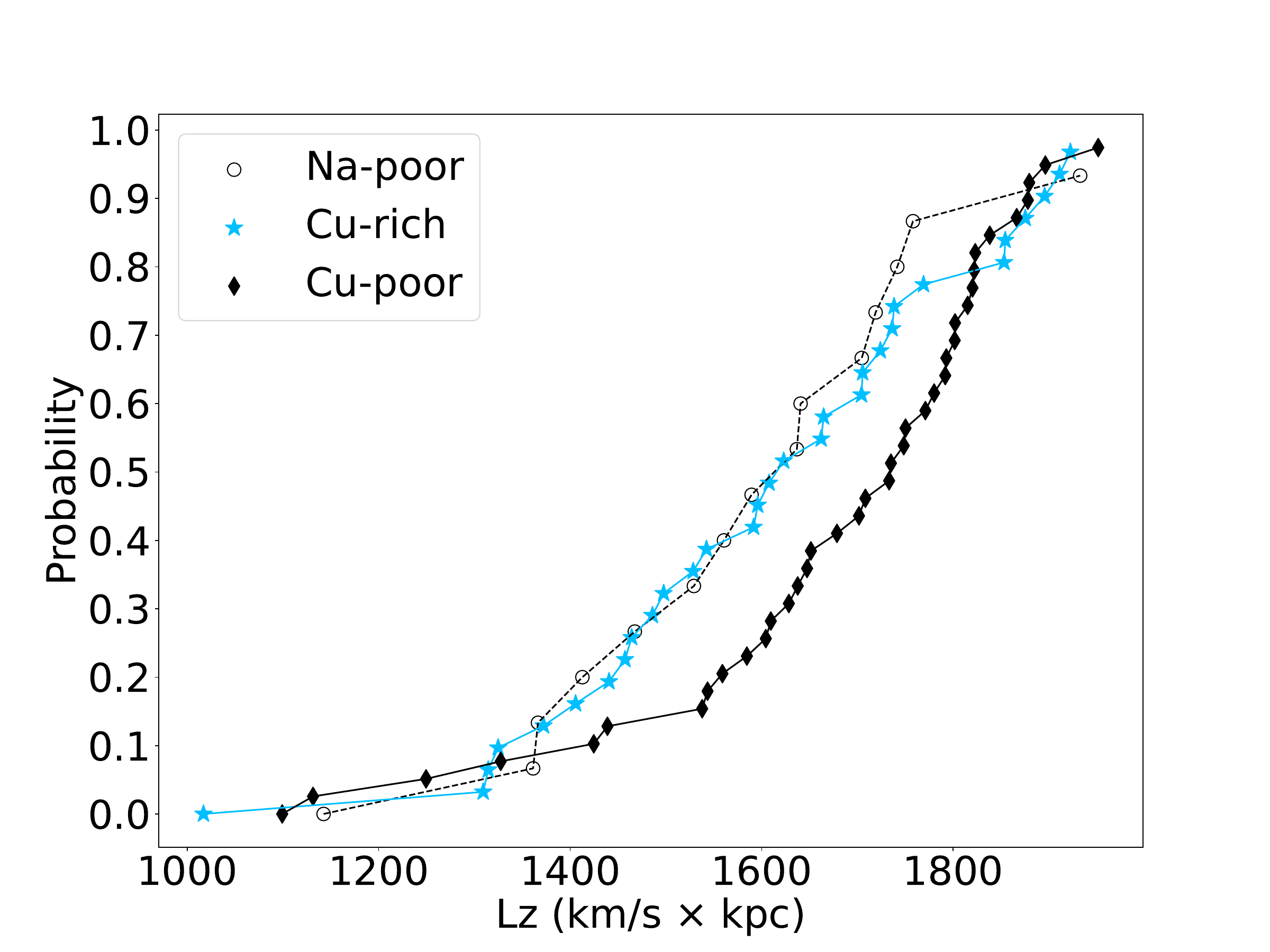} 
                \caption{Cumulative distribution of the z direction of the angular momentum (L$_z$) of the stars that belong to the Na-poor (open circles), Cu-rich (blue stars), and Cu-poor (black diamonds) samples.}
                \label{fig:Lz}
        \end{figure}

        To investigate the origin of these groups, we generated the galactic parameters' eccentricity (Ecce), energy, the z component of the angular momentum (Lz), galactocentric distance and maximum height above the Galactic plane using the Python code \texttt{galpy} \citep{galpy}, following the same procedure as \citet{Shejeela_galpy}. Figures \ref{fig:ecce} and \ref{fig:Lz} show the cumulative distributions of Ecce and L$_z$ of the stars. It is possible to note that the Na-poor and Cu-rich groups present higher Ecce and smaller L$_z$ than the Cu-poor group, indicating that perhaps the stars are originated from the inner regions of the galactic disk and (due to the eccentric orbits together with migration) they were observed in the solar vicinity. Thus, their chemical composition could reflect the chemistry of thin-disk stars in the inner regions.
        
        The Na-poor subpopulation seems to form a plateau along with older stars of the thick disk. Migration of the stars from the inner part of the disk, which is richer in metals and with a distinct abundance pattern due to a different supernova enrichment history, could also have contributed. The Sun itself may have been formed in the inner disk, around 5 - 6 kpc, and migrated to its current position \citep{Nieva_Przybilla_2012, Tsujimoto_Baba_2020, Baba_2023,Prantzos_2023, Lu_2024}.

        \subsection{Comparison of the Sun with the solar twins and analogs and the signature of exoplanets}
        
        Refractory elements are the building blocks of rocky planets, whose signatures can be revealed by evaluating how the abundance of such species behave according to T$_{\rm cond}$. If there is a decrease in the abundances of more refractory species, this may indicate that they were used to form planets and that the star accreted the refractory-depleted material from the protonebula. 
        
        In an attempt to find these signatures, the abundance ratios were analyzed with the $50\%$ condensation temperature from \citet{Lodders_2003}. We performed a linear fit using the Python library \texttt{lmfit} with the least squares method for each of the 88 stars of the thin disk. As mentioned above and in other papers (e.g., \citet{Bedell_2018}), the highly volatile elements have larger uncertainties. Therefore, we followed \citet{Bedell_2018} and considered in the fit only the elements with T$_{\rm cond} >$ 900 K (\citet{Ramirez_2009} also took a similar approach). Thus, we started the fit with the semi-volatile element sodium (T$_{\rm cond}$ = 958 K) and included all elements of higher condensation temperature until the highly refractory elements (T$_{\rm cond} \sim$ 1650 K; Sc, Al). Since most elements included in the fit are refractories (T$_{\rm cond} \geq$ 1300 K; \citet{Lodders_2003}), we  refer to both the semi-volatiles and refractories elements considered in the fit as refractories henceforth.
        
        Before performing the fit, a correction was made to the abundances, so that they could reflect the abundance that the star would have if it had the solar age (4.6 Gyr), using the same linear fit of the GCE correction. The volatiles were excluded from the fit because the determination of their abundances is more uncertain: there are few spectral lines available and the line regions are more complicated to do the measurements at high precision (0.01 - 0.02 dex level). Also, Li, Ba, and Y were not considered in the fit. Overall, Li is a very sensitive element, varying significantly with stellar parameters such as age, [Fe/H] \citep{Martos_2023}, T$_{\rm eff}$, and mass \citep{Rathsam_2023} and, thus, it is not reliable to study the global influence of exoplanets in the abundances. Regarding Y and Ba, at solar metallicity, they are s-process elements generated mainly in AGB stars, while the other elements are mostly produced through nuclear reactions and released by supernovae. The retrieval of the elements from AGBs is slower, as it is a less energetic event, and it could create inhomogeneities that are not specifically associated with any planets.
        
\begin{figure}[htbp]
                \center
                \begin{tabular}{cc} 
                        \includegraphics[width=0.9\linewidth]{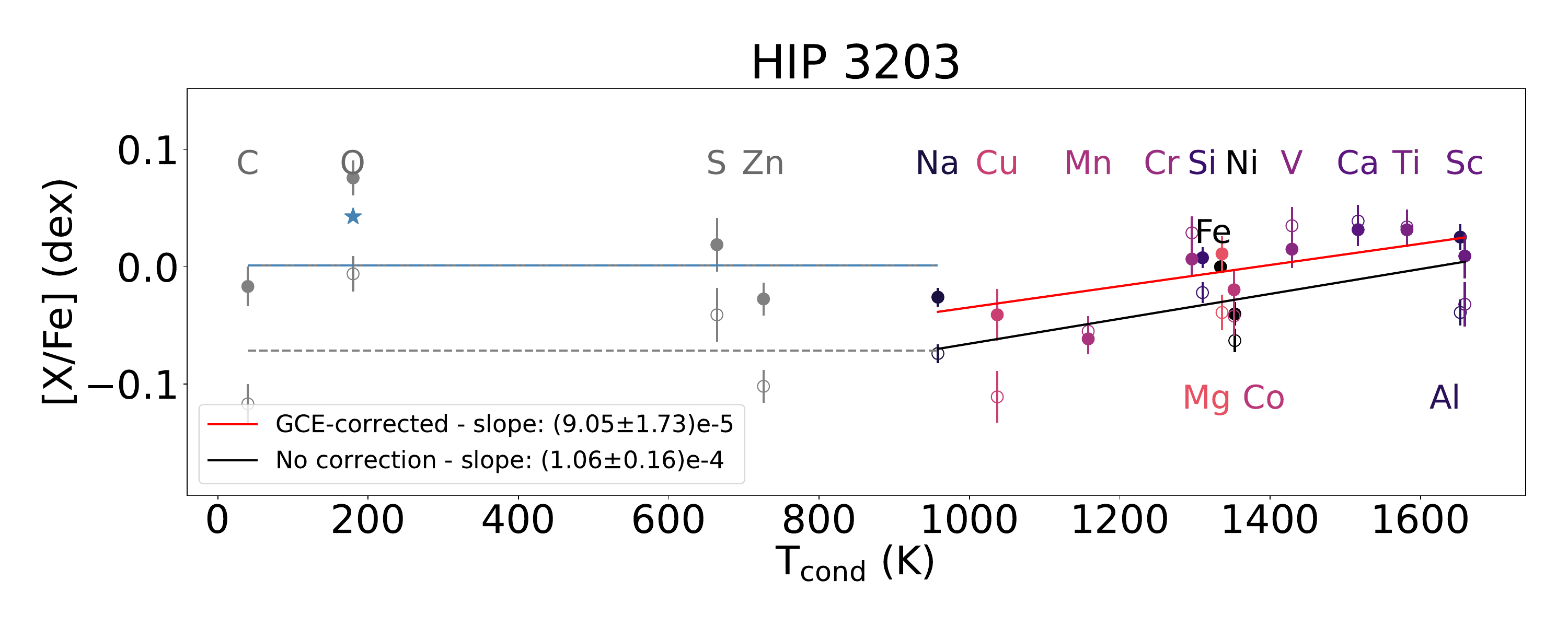} \\
                \end{tabular}
                \caption{Example fit of the refractory elements of the star HIP 3203 versus T$_{\rm cond}$.}
                \label{fig:tcond_HIP3203}
        \end{figure}

        Figure \ref{fig:tcond_HIP3203} is an example plot for the star HIP 3203. For the volatiles (C, O, S, and Zn), the median value of the solar-age abundances is represented by the solid grey line and the median of the non-corrected abundances is shown by the dashed grey line. The blue star represents the oxygen abundance corrected by the offset of 0.033 dex between this work and \citet{nissen_2020}. The blue dotted-dashed line is the median considering this value for the abundance of oxygen. This was done as an example to show the sensitivity of the volatiles, which have a significant impact on the median value and support the argument why the volatiles were not included in the fit.
        
\begin{figure}[htbp]
                \center
                \includegraphics[width=0.85\linewidth]{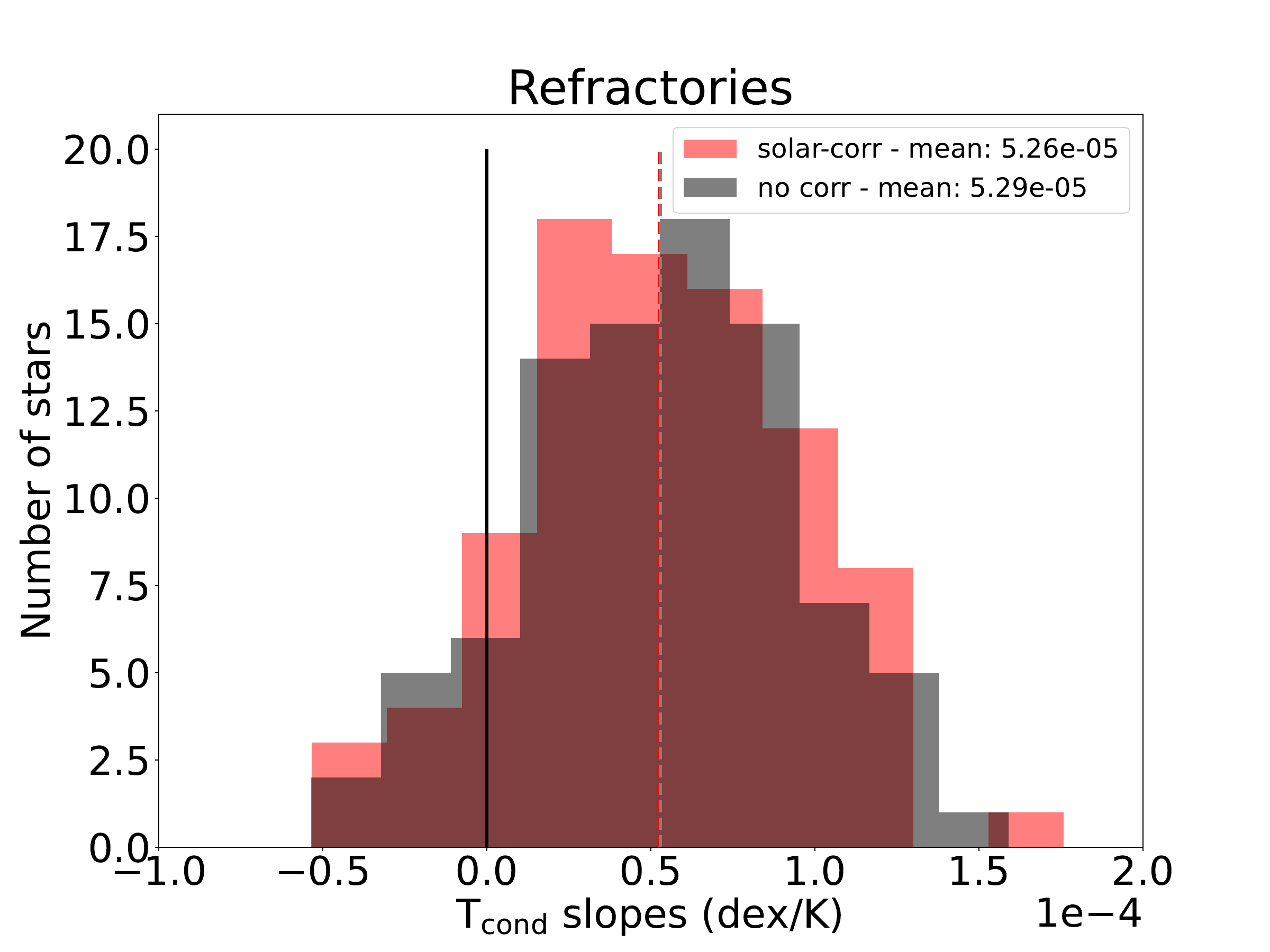} 
                \caption{Slopes of the fit of [X/Fe] versus T$_{\rm cond}$ for all stars. Here, ``solar-corr'' refers to the solar-age abundances and ``no-corr'' refers to the abundances without any GCE or age correction. The vertical line represents the Sun.}
                \label{fig:slopes_tcond}
        \end{figure}

        The distribution of slopes resulting from the fit of the refractories versus T$_{\rm cond}$ is shown in Figure \ref{fig:slopes_tcond}. The slope of the Sun, which is zero by definition, is smaller than 89\% of the stars, meaning that the Sun is more depleted in refractories compared to volatiles than 89\% of the studied stars. A negative slope, meaning a slope smaller than solar, indicates that a smaller amount of refractory material has been accreted compared to the mass of refractories in the planets of the Solar System. 
        
        In Figure 8 of \citet{Bedell_2018}, the authors provide a wider range of slopes, from $-1 \times 10^{-4}$ to $3.5 \times 10^{-4}$, approximately, whereas in this work the slopes range from around $-0.5 \times 10^{-4}$ to $1.75 \times 10^{-4}$, perhaps due to our consistent low uncertainties on the abundances. Also, the mean value of their slopes for the corrected (solar-age) and non-corrected abundances present a higher difference than this work. This happens because the sample of \citet{Bedell_2018} is more populated with stars older than the Sun and the ages of our sample are more homogeneous, resulting in similar average values for the slopes for the cases with and without correction for the abundances. In their work, \citet{Bedell_2018} found that the Sun is more depleted than 93\% of the solar twins and close analogs, whereas we found 89\%. These percentages are compatible, considering the uncertainties in the abundances in both works. Our sample has four additional planet hosts (HIP 669, HIP 70965, HIP 77358, and HIP 99115) and it is interesting to note that the number of stars more depleted than the Sun increased compared to the sample with fewer planets hosts from \citet{Bedell_2018}. However, we found no correlation between the slope and the presence of planets.
        
        To assess how peculiar the Sun is compared to the solar twins and analogs, we analyzed the difference between the solar abundances and the abundance of an ``average solar analog,'' calculated as the average of the abundances of the stars in the linear space of number of atoms, given in Equation \ref{eq:average_twin}. The error adopted was the quadratic sum of the standard deviation of the average ($\sigma(10^{[X/Fe]})/\sqrt{N}$), where N is the number of stars whose abundances were considered in the sum, with the errors of the solar abundances, given in the last column of Table \ref{tab:solar_abundances}. The equation is expressed as
        
        \begin{equation}
                \bigg\langle\left[\frac{X}{Fe}\right]\bigg\rangle = log_{10} \left(\frac{1}{N} \sum_{n=0}^{N}10^{\left[\frac{X}{Fe}\right]_n}\right)
                \label{eq:average_twin}
        .\end{equation}
        
\begin{figure}[htbp]
                \begin{tabular}{c}                                      \includegraphics[width=0.9\linewidth]{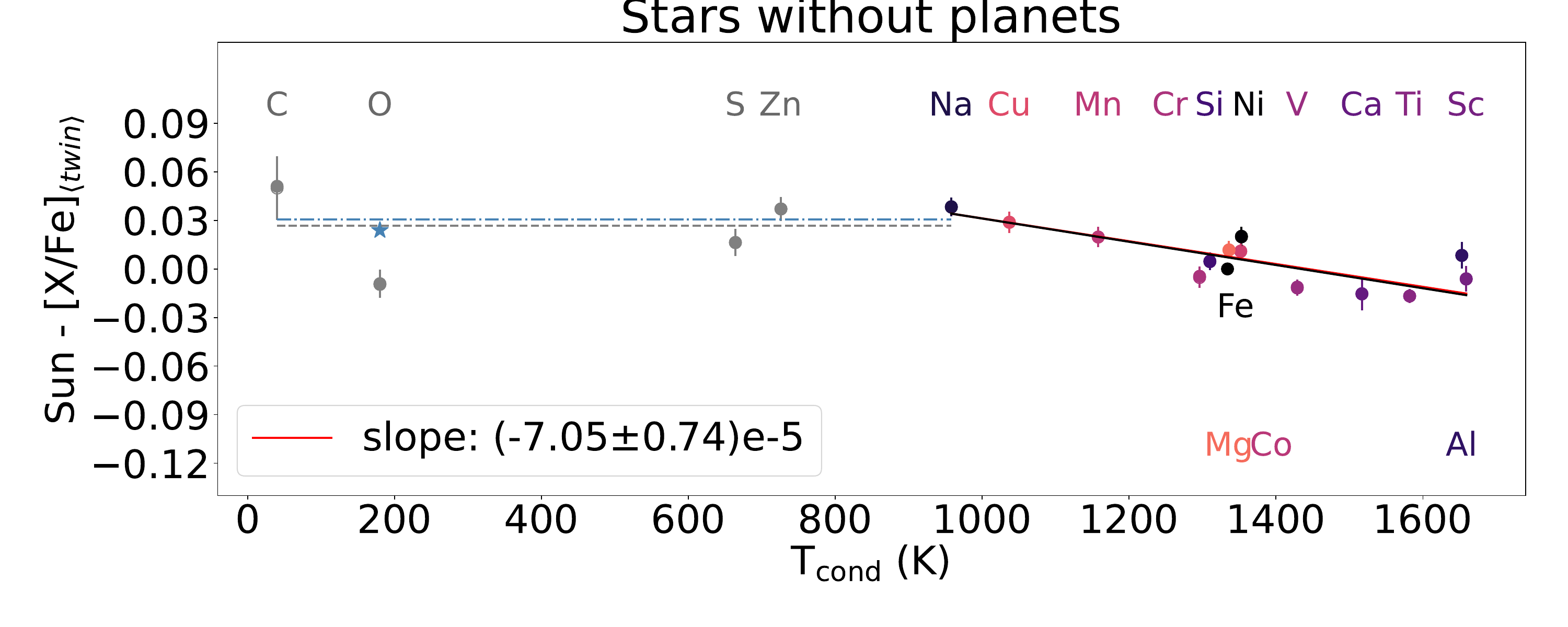}\\
                        \includegraphics[width=0.9\linewidth]{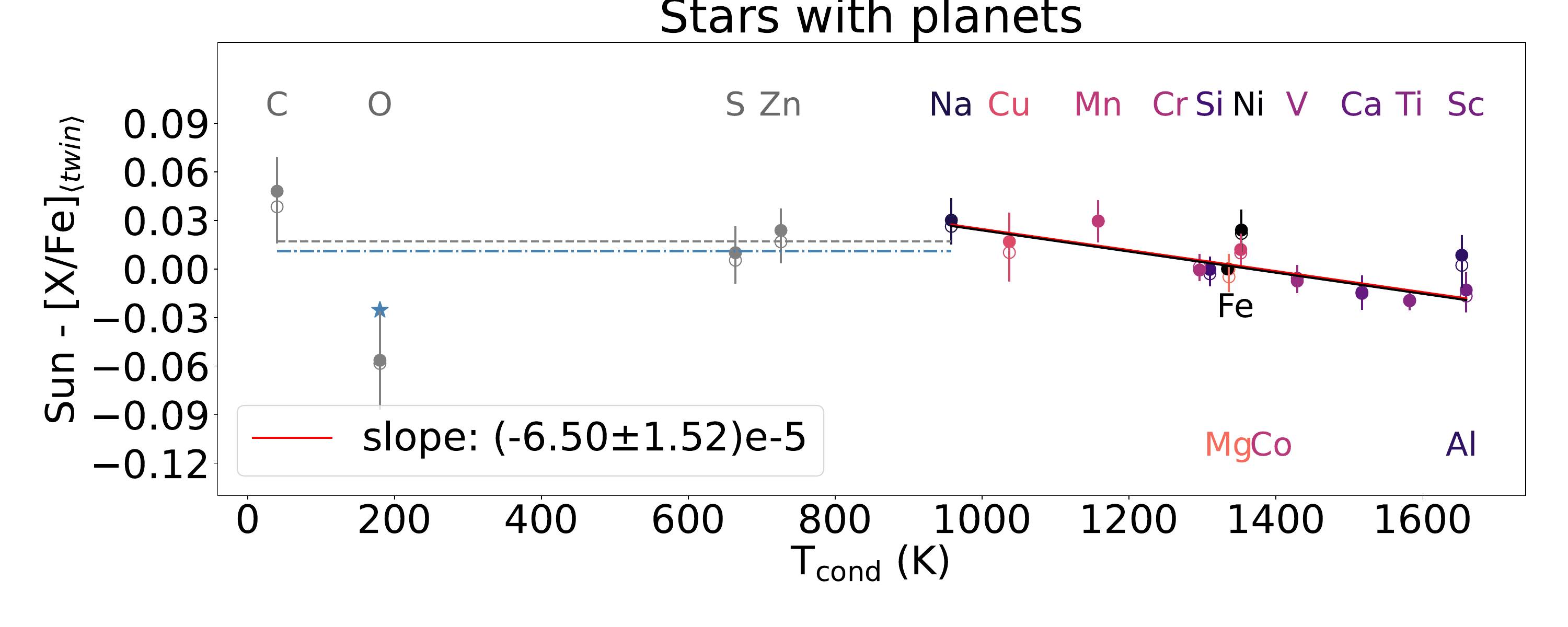}\\
                \end{tabular}
                \caption{Comparison of the abundances of the Sun with the average solar twin, considering the stars without detected exoplanets (upper panel) and the stars with planets (bottom panel). The refractory elements are shown as colored markers, and the volatiles, which were not fitted, are shown in grey. The solid circles are the solar-age abundances and the empty circles are the abundances without any correction. The red lines are linear fits to the solar-age abundances, and the black lines are the linear fits to the non-corrected abundances. In both cases, only the refractories were fitted. For the volatiles (C, O, S, and Zn), the median value of the solar-age abundances is represented by the solid grey line and the median of the non-corrected abundances is shown by the dashed grey line. The blue star represents the oxygen abundance corrected by the offset of 0.033 dex between this work and \citet{nissen_2020}. The blue dotted-dashed line is the median when considering this value for the abundance of oxygen.}
                \label{fig:fit_tcond_av_twin}
        \end{figure}

        The analysis was done in three cases: considering all the 88 stars of the thin disk, only the stars without planets, and only the stars with planets. For each case, the difference in the refractory abundances between the Sun (zero abundances, by definition) and the average solar twin was fitted against T$_{\rm cond}$. Ten stars have confirmed exoplanets in orbit: HIP 669 (1), HIP 5301 (1), HIP 11915 (1), HIP 15527 (3), HIP 68468(2), HIP 70695 (1), HIP 77358 (2), HIP 96160(1), HIP 99115 (1), and HIP 116906(1). Figure \ref{fig:fit_tcond_av_twin} shows the case considering the stars without detected exoplanets the case for planet hosts. In the first, the significance of the correlation is of 9.5$\sigma$. Regarding the stars with planets, the depletion is less significant (4.6$\sigma$). This means that the composition of the Sun is more similar to the stars with detected planets, although they are not as depleted as the Sun.        
        
        \section{Conclusions}
        
        We used neural networks (NNs) to obtain precise atmospheric parameters and abundance ratios [X/Fe] of 20 elements for a sample of 99 solar twins and analogs, using high-quality spectra from HARPS. The results obtained are in line with the literature, with average residuals and standard deviations of $(2.0 \pm 27.1)$ K for T$_{\rm eff}$, $(0.00 \pm 0.06)$ dex for log g, $(0.00 \pm 0.02)$ dex for [Fe/H], $(-0.01 \pm 0.05)$ km s$^{-1}$ for v$_t$, $(0.02 \pm 0.08)$ km s$^{-1}$ for vmacro, and $(-0.12 \pm 0.26)$ km s$^{-1}$ for vsin$i$. It was possible to achieve the desired precision of 0.01 dex for approximately half of the elements (Na, Mg, Al, Si, Ca, Ti, Cr, Co, Ni, and Cu) and about 0.02 dex for the rest. 
        
        We identified the possible presence of three Galactic thin disk stellar subpopulations in our sample from the plots of [Na/Fe] and [Cu/Fe] versus age. The first is slightly older (age $>$ 6 Gyr) and poor in Na, while the others are more and less enriched in Cu. These groups are also present in the plots of Al, Si, Mn, Co, Ni, and Zn. One possible explanation is that these stars belong to inner regions of the thin disk and, due to their eccentric orbits (associated with migration), they were observed in the solar vicinity. Due to radial migration, the stars actually reflect the local supernova+AGBs enrichment of their birthplace, resulting thus in a distinct chemistry.
        
        Finally, the solar-age abundances of refractories were correlated with the condensation temperature of the elements to compare the Sun with the solar twins and close analogs. We found that the Sun is more depleted in refractories in relation to volatiles than 89\% of the studied stars, with a significance of 9.5 $\sigma$ when comparing to the stars without detected exoplanets. When comparing the Sun with the stars that are planet hosts, the significance is 4.3$\sigma$. This means that the Sun’s composition is more similar to that of stars that have exoplanets, although they are not as depleted as the Sun. With the detection of new exoplanets in the following years, the results can be refined, clarifying the relationship between chemical abundances and exoplanets and potentially offering valuable insights into the uniqueness of the Solar System.

        \begin{acknowledgements}
                We thank Shejeelammal Jamela for sharing the Galactic parameters computed with \texttt{galpy} and Chiaki Kobayashi for discussions on the observed abundance patterns. G.M. and J.M. acknowledges FAPESP (São Paulo State Research Foundation) for the MSc grants (processes 2022/05833-0, 2023/15044-5) and Temático research grant (2018/04055-8), respectively. This work made use of data collected at the European Southern Observatory under ESO programs: 072.C-0488, 183.C-0972, 192.C-0852, 188.C-0265, 188.C-0265, 188.C-0265, 183.D-0729, 188.C-0265, 188.C-0265, 188.C-0265, 092.C-0721, 188.C-0265, 093.C-0409, 077.C-0364, 188.C-0265, 188.C-0265, 188.C-0265, 188.C-0265, 188.C-0265, 188.C-0265, 188.C-0265, 188.C-0265, 188.C-0265, 0103.D-0445, 0103.D-0445, 0100.D-0444, 076.C-0155, 1102.C-0923, 188.C-0265, 074.C-0364, 188.C-0265, 183.D-0729, 089.C-0732, 091.C-0034, 087.C-0831, 090.C-0421, 111.24ZQ.001, 198.C-0836, 075.C-0202, 0102.C-0584, 0103.C-0206, 292.C-5004, 075.C-0332, 099.C-0491, 196.C-1006, 106.21TJ.001, 60.A-9036, 60.A-9709, 192.C-0852, 0104.C-0090, 106.215E.004, 098.C-0739, 106.215E.002, 105.20AK.002, 0104.C-0090, 192.C-0224, 0101.C-0275, 085.C-0019, 098.C-0366, 0103.C-0432, 0100.C-0097, 0102.C-0558, 106.21R4.001, 0101.C-0379, 095.C-0551, 099.C-0458, 108.222V.001, 096.C-0460, 091.C-0936, 097.C-0571, 0100.C-0474, 0101.C-0275.
                
        \end{acknowledgements}

        \bibliographystyle{aa} 
        \bibliography{ref.bib} 

\begin{thebibliography}{51}
\expandafter\ifx\csname natexlab\endcsname\relax\def\natexlab#1{#1}\fi

\bibitem[{{Adibekyan} {et~al.}(2011){Adibekyan}, {Santos}, {Sousa}, \&
  {Israelian}}]{Adibekyan_2011}
{Adibekyan}, V.~Z., {Santos}, N.~C., {Sousa}, S.~G., \& {Israelian}, G. 2011,
  \aap, 535, L11

\bibitem[{{Angelo} {et~al.}(2024){Angelo}, {Bedell}, {Petigura}, \&
  {Ness}}]{angelo_2024}
{Angelo}, I., {Bedell}, M., {Petigura}, E., \& {Ness}, M. 2024, \apj, 974, 43

\bibitem[{{Baba} {et~al.}(2023){Baba}, {Saitoh}, \& {Tsujimoto}}]{Baba_2023}
{Baba}, J., {Saitoh}, T.~R., \& {Tsujimoto}, T. 2023, \mnras, 526, 6088

\bibitem[{Bedell {et~al.}(2018)Bedell, Bean, Mel{\'{e}}ndez, Spina,
  Ram{\'{\i}}rez, Asplund, Alves-Brito, dos Santos, Dreizler, Yong, Monroe, \&
  Casagrande}]{Bedell_2018}
Bedell, M., Bean, J.~L., Mel{\'{e}}ndez, J., {et~al.} 2018, \apj, 865, 68

\bibitem[{{Bedell} {et~al.}(2014){Bedell}, {Mel{\'e}ndez}, {Bean},
  {Ram{\'\i}rez}, {Leite}, \& {Asplund}}]{Bedell_2014}
{Bedell}, M., {Mel{\'e}ndez}, J., {Bean}, J.~L., {et~al.} 2014, \apj, 795, 23

\bibitem[{{Blanco-Cuaresma}(2019)}]{iSpec2}
{Blanco-Cuaresma}, S. 2019, \mnras, 486, 2075

\bibitem[{{Blanco-Cuaresma} {et~al.}(2014){Blanco-Cuaresma}, {Soubiran},
  {Heiter}, \& {Jofr{\'e}}}]{iSpec1}
{Blanco-Cuaresma}, S., {Soubiran}, C., {Heiter}, U., \& {Jofr{\'e}}, P. 2014,
  \aap, 569, A111

\bibitem[{{Bovy}(2015)}]{galpy}
{Bovy}, J. 2015, \apjs, 216, 29

\bibitem[{{Carlos} {et~al.}(2016){Carlos}, {Nissen}, \&
  {Mel{\'e}ndez}}]{Carlos_2016}
{Carlos}, M., {Nissen}, P.~E., \& {Mel{\'e}ndez}, J. 2016, \aap, 587, A100

\bibitem[{Carvalho-Silva {et~al.}(2025)Carvalho-Silva, Meléndez, Rathsam,
  Shejeelammal, Martos, Lorenzo-Oliveira, Spina, \&
  Ribeiro~Alves}]{Carvalho-Silva_2025}
Carvalho-Silva, G., Meléndez, J., Rathsam, A., {et~al.} 2025, \apj Letters,
  983, L31

\bibitem[{{Casey} {et~al.}(2016){Casey}, {Hogg}, {Ness}, {Rix}, {Ho}, \&
  {Gilmore}}]{Casey_2016}
{Casey}, A.~R., {Hogg}, D.~W., {Ness}, M., {et~al.} 2016, arXiv e-prints,
  submitted to AAS (ApJ), arXiv:1603.03040

\bibitem[{{Cayrel de Strobel}(1996)}]{Strobel_1996}
{Cayrel de Strobel}, G. 1996, \aapr, 7, 243

\bibitem[{{Chambers}(2010)}]{Chambers_2010}
{Chambers}, J.~E. 2010, \apj, 724, 92

\bibitem[{Cowley \& Yüce(2022)}]{Cowley_2022}
Cowley, C.~R. \& Yüce, K. 2022, \mnras, 512, 3684

\bibitem[{{dos Santos} {et~al.}(2016){dos Santos}, {Mel{\'e}ndez}, {do
  Nascimento}, {Bedell}, {Ram{\'\i}rez}, {Bean}, {Asplund}, {Spina},
  {Dreizler}, {Alves-Brito}, \& {Casagrande}}]{dosSantos_2016}
{dos Santos}, L.~A., {Mel{\'e}ndez}, J., {do Nascimento}, J.-D., {et~al.} 2016,
  \aap, 592, A156

\bibitem[{{Flores} {et~al.}(2024){Flores}, {Yana Galarza}, {Miquelarena},
  {Saffe}, {Jaque Arancibia}, {Iba{\~n}ez Bustos}, {Jofr{\'e}}, {Alacoria}, \&
  {Gunella}}]{Flores_2024}
{Flores}, M., {Yana Galarza}, J., {Miquelarena}, P., {et~al.} 2024, \mnras,
  527, 10016

\bibitem[{{Gonzalez} {et~al.}(2010){Gonzalez}, {Carlson}, \&
  {Tobin}}]{Gonzalez_2010}
{Gonzalez}, G., {Carlson}, M.~K., \& {Tobin}, R.~W. 2010, \mnras, 407, 314

\bibitem[{{Gray}(2008)}]{Gray_2008}
{Gray}, D.~F. 2008, {The Observation and Analysis of Stellar Photospheres}
  (Cambridge University Press)

\bibitem[{{Gustafsson}(2018)}]{Gustafsson_2018}
{Gustafsson}, B. 2018, \aap, 616, A91

\bibitem[{{Gustafsson} {et~al.}(2008){Gustafsson}, {Edvardsson}, {Eriksson},
  {J{\o}rgensen}, {Nordlund}, \& {Plez}}]{MARCS_used}
{Gustafsson}, B., {Edvardsson}, B., {Eriksson}, K., {et~al.} 2008, \aap, 486,
  951

\bibitem[{{Heiter} {et~al.}(2021){Heiter}, {Lind}, {Bergemann}, {Asplund},
  {Mikolaitis}, {Barklem}, {Masseron}, {de Laverny}, {Magrini}, {Edvardsson},
  {J{\"o}nsson}, {Pickering}, {Ryde}, {Bayo Ar{\'a}n}, {Bensby}, {Casey},
  {Feltzing}, {Jofr{\'e}}, {Korn}, {Pancino}, {Damiani}, {Lanzafame}, {Lardo},
  {Monaco}, {Morbidelli}, {Smiljanic}, {Worley}, {Zaggia}, {Randich}, \&
  {Gilmore}}]{linelist_GES_v6}
{Heiter}, U., {Lind}, K., {Bergemann}, M., {et~al.} 2021, \aap, 645, A106

\bibitem[{{Jofr{\'e}} {et~al.}(2021){Jofr{\'e}}, {Petrucci}, {Maqueo Chew},
  {Ram{\'\i}rez}, {Saffe}, {Martioli}, {Buccino}, {Ma{\v{s}}ek}, {Garc{\'\i}a},
  {Canul}, \& {G{\'o}mez}}]{Jofre_2021}
{Jofr{\'e}}, E., {Petrucci}, R., {Maqueo Chew}, Y.~G., {et~al.} 2021, \aj, 162,
  291

\bibitem[{{Liu} {et~al.}(2020){Liu}, {Yong}, {Asplund}, {Wang}, {Spina},
  {Acu{\~n}a}, {Mel{\'e}ndez}, \& {Ram{\'\i}rez}}]{Liu_2020}
{Liu}, F., {Yong}, D., {Asplund}, M., {et~al.} 2020, \mnras, 495, 3961

\bibitem[{{Lodders}(2003)}]{Lodders_2003}
{Lodders}, K. 2003, \apj, 591, 1220

\bibitem[{{Lu} {et~al.}(2024){Lu}, {Minchev}, {Buck}, {Khoperskov},
  {Steinmetz}, {Libeskind}, {Cescutti}, {Freeman}, \& {Ratcliffe}}]{Lu_2024}
{Lu}, Y.~L., {Minchev}, I., {Buck}, T., {et~al.} 2024, \mnras, 535, 392

\bibitem[{Maia {et~al.}(2019)Maia, Mel{\'{e}}ndez, Lorenzo-Oliveira, Spina, \&
  Jofr{\'{e}}}]{Maia_2019}
Maia, M.~T., Mel{\'{e}}ndez, J., Lorenzo-Oliveira, D., Spina, L., \&
  Jofr{\'{e}}, P. 2019, \aap, 628, A126

\bibitem[{{Martos} {et~al.}(2023){Martos}, {Mel{\'e}ndez}, {Rathsam}, \&
  {Carvalho Silva}}]{Martos_2023}
{Martos}, G., {Mel{\'e}ndez}, J., {Rathsam}, A., \& {Carvalho Silva}, G. 2023,
  \mnras, 522, 3217

\bibitem[{Mel{\'{e}}ndez {et~al.}(2009)Mel{\'{e}}ndez, Asplund, Gustafsson, \&
  Yong}]{Melendez_2009}
Mel{\'{e}}ndez, J., Asplund, M., Gustafsson, B., \& Yong, D. 2009, \apj, 704,
  L66

\bibitem[{Mel{\'{e}}ndez {et~al.}(2014)Mel{\'{e}}ndez, Ram{\'{\i}}rez, Karakas,
  Yong, Monroe, Bedell, Bergemann, Asplund, Maia, Bean, do~Nascimento, Bazot,
  Alves-Brito, Freitas, \& Castro}]{Melendez_2014}
Mel{\'{e}}ndez, J., Ram{\'{\i}}rez, I., Karakas, A.~I., {et~al.} 2014, \apj,
  791, 14

\bibitem[{{Miquelarena} {et~al.}(2024){Miquelarena}, {Saffe}, {Flores},
  {Petrucci}, {Yana Galarza}, {Alacoria}, {Jaque Arancibia}, {Jofr{\'e}},
  {Montenegro Armijo}, \& {Gunella}}]{Miquelarena_2024}
{Miquelarena}, P., {Saffe}, C., {Flores}, M., {et~al.} 2024, \aap, 688, A73

\bibitem[{{Nieva} \& {Przybilla}(2012)}]{Nieva_Przybilla_2012}
{Nieva}, M.~F. \& {Przybilla}, N. 2012, \aap, 539, A143

\bibitem[{{Nissen}(2015)}]{Nissen_2015}
{Nissen}, P.~E. 2015, \aap, 579, A52

\bibitem[{{Nissen} {et~al.}(2020){Nissen}, {Christensen-Dalsgaard},
  {Mosumgaard}, {Silva Aguirre}, {Spitoni}, \& {Verma}}]{nissen_2020}
{Nissen}, P.~E., {Christensen-Dalsgaard}, J., {Mosumgaard}, J.~R., {et~al.}
  2020, \aap, 640, A81

\bibitem[{{Oh} {et~al.}(2018){Oh}, {Price-Whelan}, {Brewer}, {Hogg}, {Spergel},
  \& {Myles}}]{Oh_2018}
{Oh}, S., {Price-Whelan}, A.~M., {Brewer}, J.~M., {et~al.} 2018, \apj, 854, 138

\bibitem[{{Plez}(2012)}]{turbospectrum}
{Plez}, B. 2012, {Turbospectrum: Code for spectral synthesis}, Astrophysics
  Source Code Library, record ascl:1205.004

\bibitem[{{Plotnikova, A.} {et~al.}(2024){Plotnikova, A.}, {Spina, L.},
  {Ratcliffe, B.}, {Casali, G.}, \& {Carraro, G.}}]{Plotnikova_2024}
{Plotnikova, A.}, {Spina, L.}, {Ratcliffe, B.}, {Casali, G.}, \& {Carraro, G.}
  2024, \aap, 691, A298

\bibitem[{{Prantzos} {et~al.}(2023){Prantzos}, {Abia}, {Chen}, {de Laverny},
  {Recio-Blanco}, {Athanassoula}, {Roberti}, {Vescovi}, {Limongi}, {Chieffi},
  \& {Cristallo}}]{Prantzos_2023}
{Prantzos}, N., {Abia}, C., {Chen}, T., {et~al.} 2023, \mnras, 523, 2126

\bibitem[{Ram{\'{\i}}rez {et~al.}(2015)Ram{\'{\i}}rez, Khanal, Aleo, Sobotka,
  Liu, Casagrande, Mel{\'{e}}ndez, Yong, Lambert, \& Asplund}]{Ramirez_2015}
Ram{\'{\i}}rez, I., Khanal, S., Aleo, P., {et~al.} 2015, \apj, 808, 13

\bibitem[{{Ram{\'\i}rez} {et~al.}(2009){Ram{\'\i}rez}, {Mel{\'e}ndez}, \&
  {Asplund}}]{Ramirez_2009}
{Ram{\'\i}rez}, I., {Mel{\'e}ndez}, J., \& {Asplund}, M. 2009, \aap, 508, L17

\bibitem[{{Rampalli} {et~al.}(2024){Rampalli}, {Ness}, {Edwards}, {Newton}, \&
  {Bedell}}]{Rampalli_2024}
{Rampalli}, R., {Ness}, M.~K., {Edwards}, G.~H., {Newton}, E.~R., \& {Bedell},
  M. 2024, \apj, 965, 176

\bibitem[{{Rathsam} {et~al.}(2023){Rathsam}, {Mel{\'e}ndez}, \& {Carvalho
  Silva}}]{Rathsam_2023}
{Rathsam}, A., {Mel{\'e}ndez}, J., \& {Carvalho Silva}, G. 2023, \mnras, 525,
  4642

\bibitem[{{Saffe} {et~al.}(2017){Saffe}, {Jofr{\'e}}, {Martioli}, {Flores},
  {Petrucci}, \& {Jaque Arancibia}}]{Saffe_2017}
{Saffe}, C., {Jofr{\'e}}, E., {Martioli}, E., {et~al.} 2017, \aap, 604, L4

\bibitem[{{Shejeelammal} \& {Goswami}(2024)}]{Shejeela_galpy}
{Shejeelammal}, J. \& {Goswami}, A. 2024, \mnras, 527, 2323

\bibitem[{{Shejeelammal} {et~al.}(2024){Shejeelammal}, {Mel{\'e}ndez},
  {Rathsam}, \& {Martos}}]{Shejeela_2024}
{Shejeelammal}, J., {Mel{\'e}ndez}, J., {Rathsam}, A., \& {Martos}, G. 2024,
  \aap, 690, A107

\bibitem[{{Sousa} {et~al.}(2015){Sousa}, {Santos}, {Adibekyan}, {Delgado-Mena},
  \& {Israelian}}]{ARES}
{Sousa}, S.~G., {Santos}, N.~C., {Adibekyan}, V., {Delgado-Mena}, E., \&
  {Israelian}, G. 2015, \aap, 577, A67

\bibitem[{{Spina} {et~al.}(2018){Spina}, {Mel{\'e}ndez}, {Karakas}, {dos
  Santos}, {Bedell}, {Asplund}, {Ram{\'\i}rez}, {Yong}, {Alves-Brito}, {Bean},
  \& {Dreizler}}]{Spina_2018}
{Spina}, L., {Mel{\'e}ndez}, J., {Karakas}, A.~I., {et~al.} 2018, \mnras, 474,
  2580

\bibitem[{{Teske} {et~al.}(2016){Teske}, {Khanal}, \&
  {Ram{\'\i}rez}}]{Teske_2016}
{Teske}, J.~K., {Khanal}, S., \& {Ram{\'\i}rez}, I. 2016, \apj, 819, 19

\bibitem[{{Ting} {et~al.}(2019){Ting}, {Conroy}, {Rix}, \&
  {Cargile}}]{Ting_2019}
{Ting}, Y.-S., {Conroy}, C., {Rix}, H.-W., \& {Cargile}, P. 2019, \apj, 879, 69

\bibitem[{Tody(1986)}]{iraf}
Tody, D. 1986, in Instrumentation in Astronomy VI, ed. D.~L. Crawford, Vol.
  0627, International Society for Optics and Photonics (SPIE), 733 -- 748

\bibitem[{{Tsujimoto} \& {Baba}(2020)}]{Tsujimoto_Baba_2020}
{Tsujimoto}, T. \& {Baba}, J. 2020, \apj, 904, 137

\bibitem[{{Yana Galarza} {et~al.}(2021){Yana Galarza}, {L{\'o}pez-Valdivia},
  {Mel{\'e}ndez}, \& {Lorenzo-Oliveira}}]{Galarza_2021}
{Yana Galarza}, J., {L{\'o}pez-Valdivia}, R., {Mel{\'e}ndez}, J., \&
  {Lorenzo-Oliveira}, D. 2021, \apj, 922, 129

\end{thebibliography}
        
        \onecolumn
        
        \begin{appendix}

                \section{Atmospheric parameters and chemical abundances}
                
                
                \begingroup
                
                \setlength{\tabcolsep}{6pt}
                \renewcommand{\arraystretch}{.8}
                
                \tiny{\begin{longtable}{lllrlll}
                                \caption{Stellar parameters obtained automatically with NNs for the sample. Stars identified with  $^p$ are planet hosts. The complete table is available at CDS.} \label{tab:params_NN} \\
                                \toprule
                                Star & T$_{\rm eff}$ (K) & log g (dex) & [Fe/H] (dex) & v$_t$ (km/s) & vmac (km/s) & vsini (km/s) \\
                                \midrule
                                \endfirsthead
                                \caption[]{continued.} \\
                                \toprule
                                Star & T$_{\rm eff}$ (K) & log g (dex) & [Fe/H] (dex) & v$_t$ (km/s) & vmac (km/s) & vsini (km/s) \\
                                \midrule
                                \endhead
                                \midrule
                                \multicolumn{7}{r}{Continued on next page} \\
                                \midrule
                                \endfoot
                                \bottomrule
                                \endlastfoot
                                HIP 669 $^p$& 5908$\pm$12 & 4.51$\pm$0.02 & -0.133$\pm$0.009 & 1.08$\pm$0.02 & 3.44$\pm$0.03 & 1.96$\pm$0.07 \\
                                HIP 1954 & 5719$\pm$10 & 4.52$\pm$0.02 & -0.079$\pm$0.008 & 0.95$\pm$0.01 & 2.77$\pm$0.02 & 1.94$\pm$0.06 \\...\\
                \end{longtable}}
                
                \endgroup

                
                \begingroup

                \renewcommand{\arraystretch}{.8}

                \tiny{\begin{longtable}{lrrrrr}
                                \caption{Chemical abundances (dex) obtained automatically with NNs for the sample. Stars identified with  $^p$ are planet hosts. The complete table is available at CDS.} \label{tab:abundances} \\
                                \toprule
                                Star & \multicolumn{1}{c}{$[$Li/Fe$]$} & \multicolumn{1}{c}{$[$C/Fe$]$} & \multicolumn{1}{c}{$[$O/Fe$]$} & \multicolumn{1}{c}{$[$Na/Fe$]$} & \multicolumn{1}{c}{$[$Mg/Fe$]$} \\
                                \midrule
                                \endfirsthead
                                \caption[]{continued.} \\
                                \toprule
                                \endhead
                                \midrule
                                \multicolumn{6}{r}{Continued on next page} \\
                                \midrule
                                \endfoot
                                \bottomrule
                                \endlastfoot
                                HIP 669 $^p$ & 1.192$\pm$0.018 & -0.068$\pm$0.027 & 0.080$\pm$0.018 & -0.051$\pm$0.009 & -0.006$\pm$0.016 \\
                                HIP 1954 & 0.381$\pm$0.016 & -0.027$\pm$0.017 & 0.049$\pm$0.009 & -0.025$\pm$0.012 & 0.002$\pm$0.012 \\...\\
                                
                \end{longtable}}
                
                \endgroup

                
                \tiny{\begin{longtable}{lrrrc}
                                \caption[Solar abundances obtained automatically with Neural Networks using spectra of the Moon, Vesta and Ganymede, with Vesta as the reference solar spectrum.]{Solar abundances obtained automatically with NNs using spectra of the Moon, Vesta, and Ganymede, with Vesta as the reference solar spectrum. The last column shows the zero point error adopted for the element. The complete table is available at CDS.} \label{tab:solar_abundances} \\
                                \toprule
                                Element & 
                                \multicolumn{1}{c}{Moon} & \multicolumn{1}{c}{Vesta}  & \multicolumn{1}{c}{Ganymede} & 
                                \multicolumn{1}{c}{Error}\\
                                \midrule
                                \endfirsthead
                                \caption[]{continued.} \\
                                \toprule
                                Element & 
                                \multicolumn{1}{c}{Moon} & \multicolumn{1}{c}{Vesta}  & \multicolumn{1}{c}{Ganymede} & 
                                \multicolumn{1}{c}{Error}\\
                                \midrule
                                \endhead
                                \midrule
                                \multicolumn{5}{r}{Continued on next page} \\
                                \midrule
                                \endfoot
                                \bottomrule
                                \endlastfoot
                                $[$Li/Fe$]$ & 0.000$\pm$0.010 & 0.000$\pm$0.002 & 0.000$\pm$0.118 & 0.003 \\
                                $[$C/Fe$]$ & -0.014$\pm$0.013 & 0.002$\pm$0.001 & 0.021$\pm$0.013 & 0.018 \\ ...\\
                \end{longtable}}

                \begin{longtable}{ccc} 
                        \renewcommand{\arraystretch}{0.1} 
                        \includegraphics[width=0.45\linewidth]{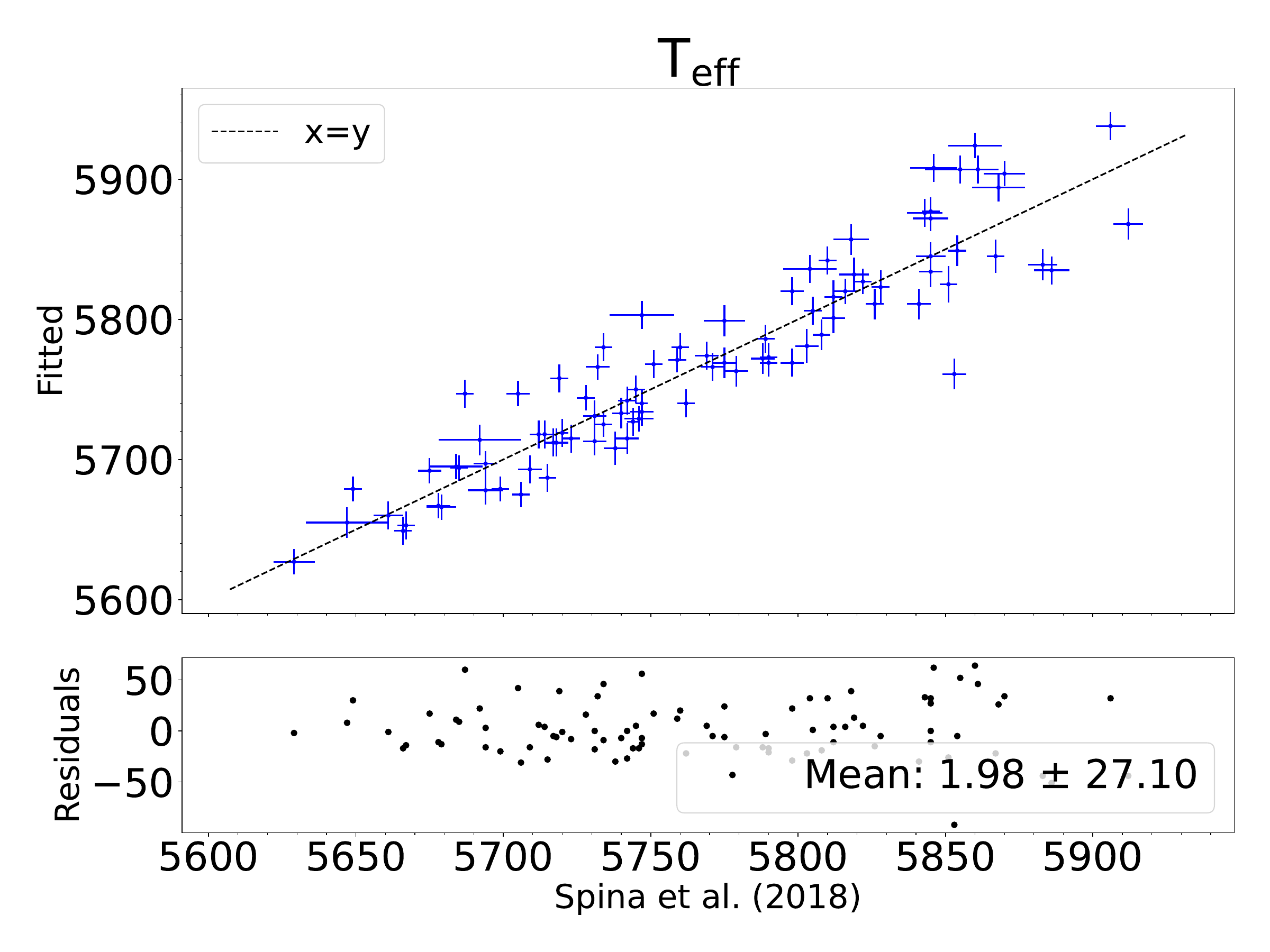} &  \includegraphics[width=0.45\linewidth]{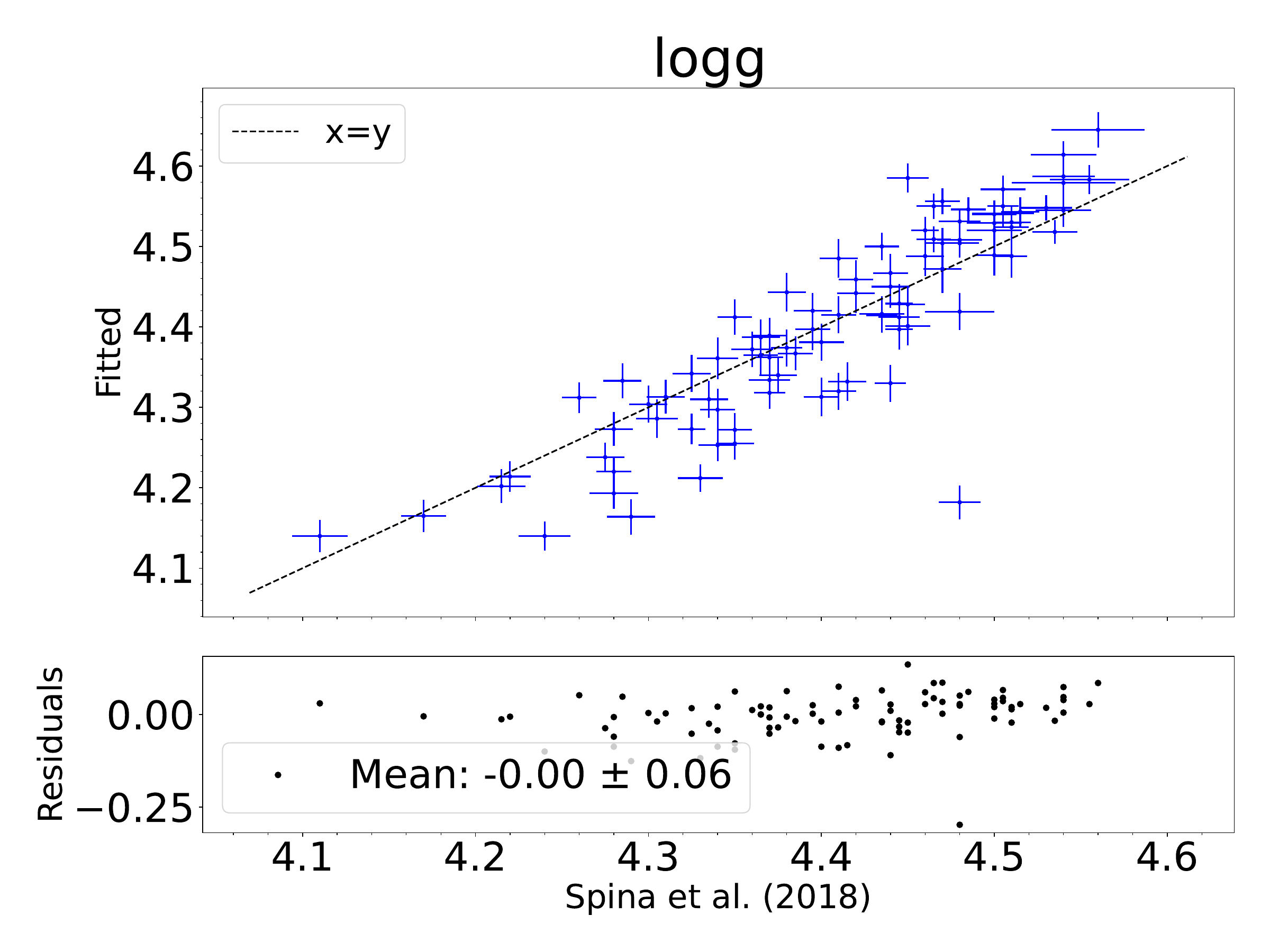}\\ 
                        \includegraphics[width=0.45\linewidth]{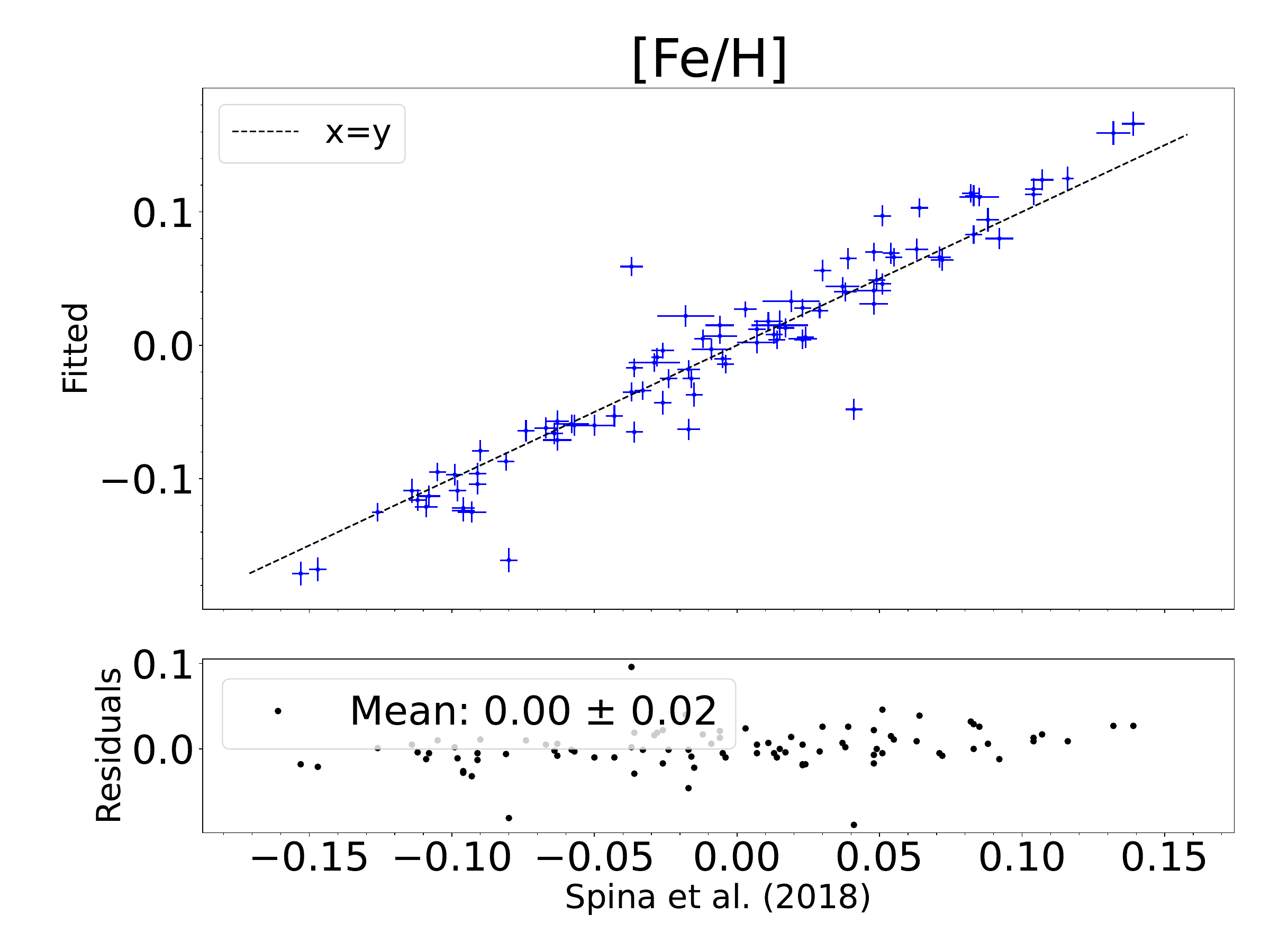} &  \includegraphics[width=0.45\linewidth]{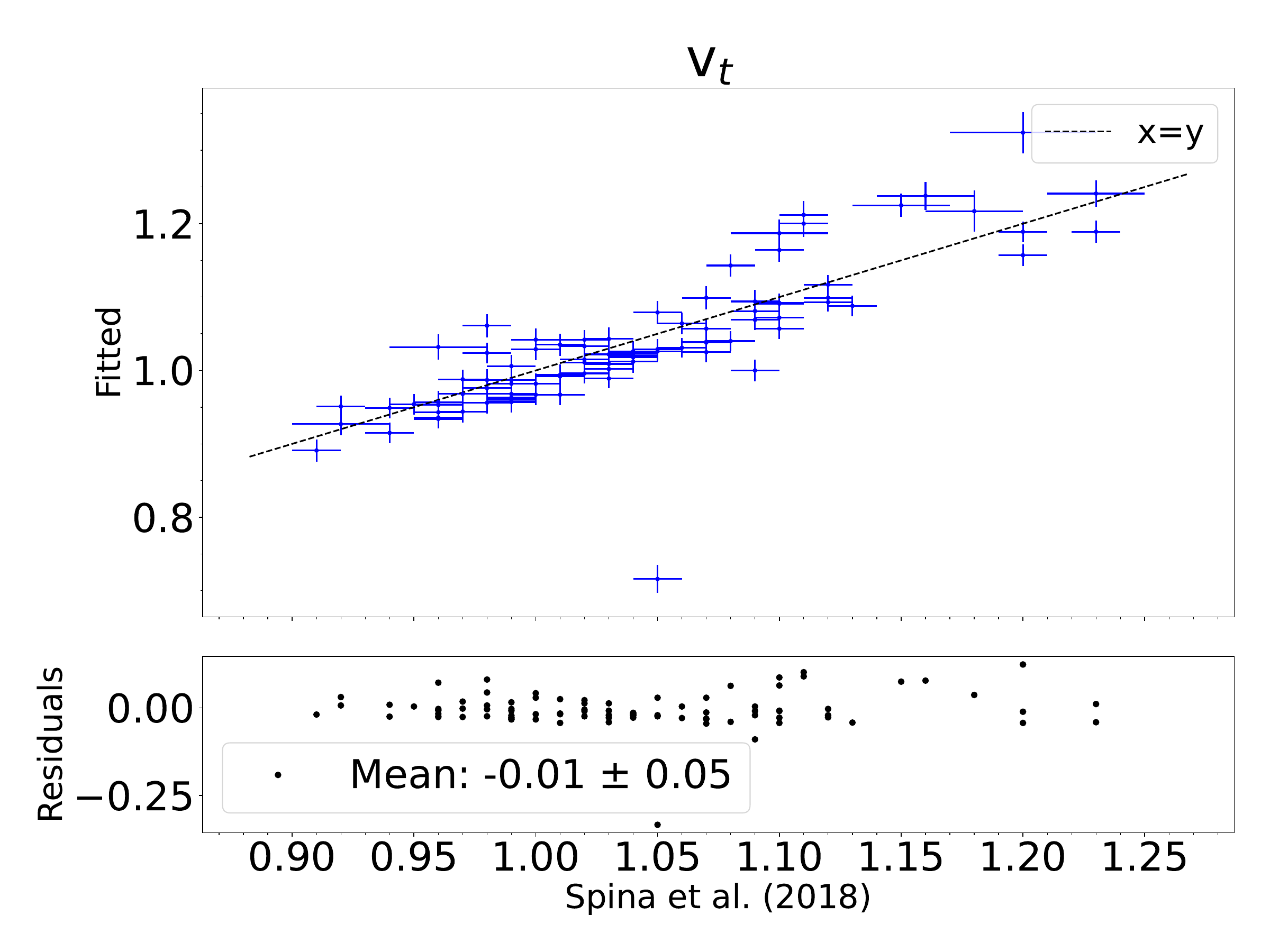}\\ 
                        \includegraphics[width=0.45\linewidth]{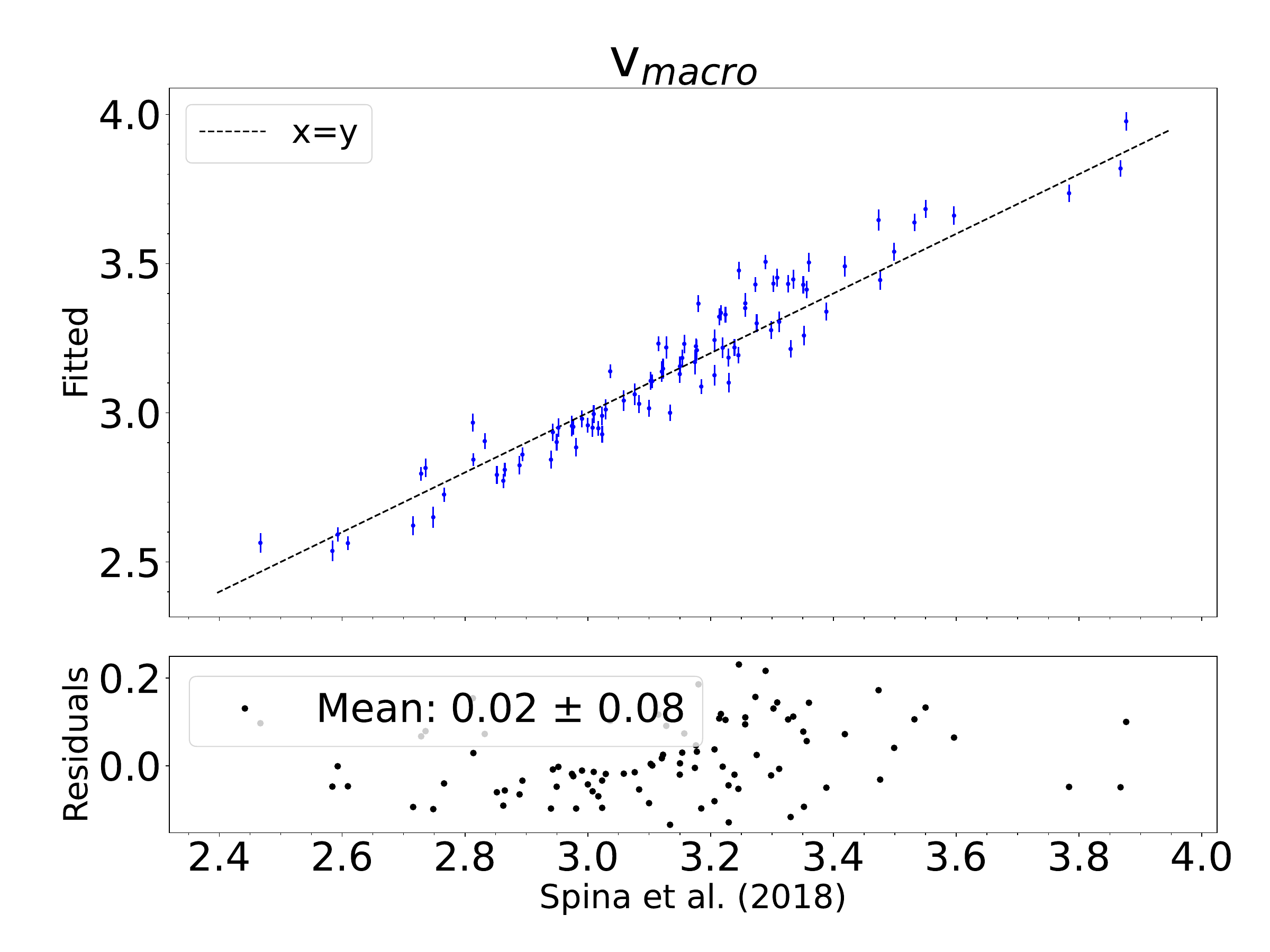} &  \includegraphics[width=0.45\linewidth]{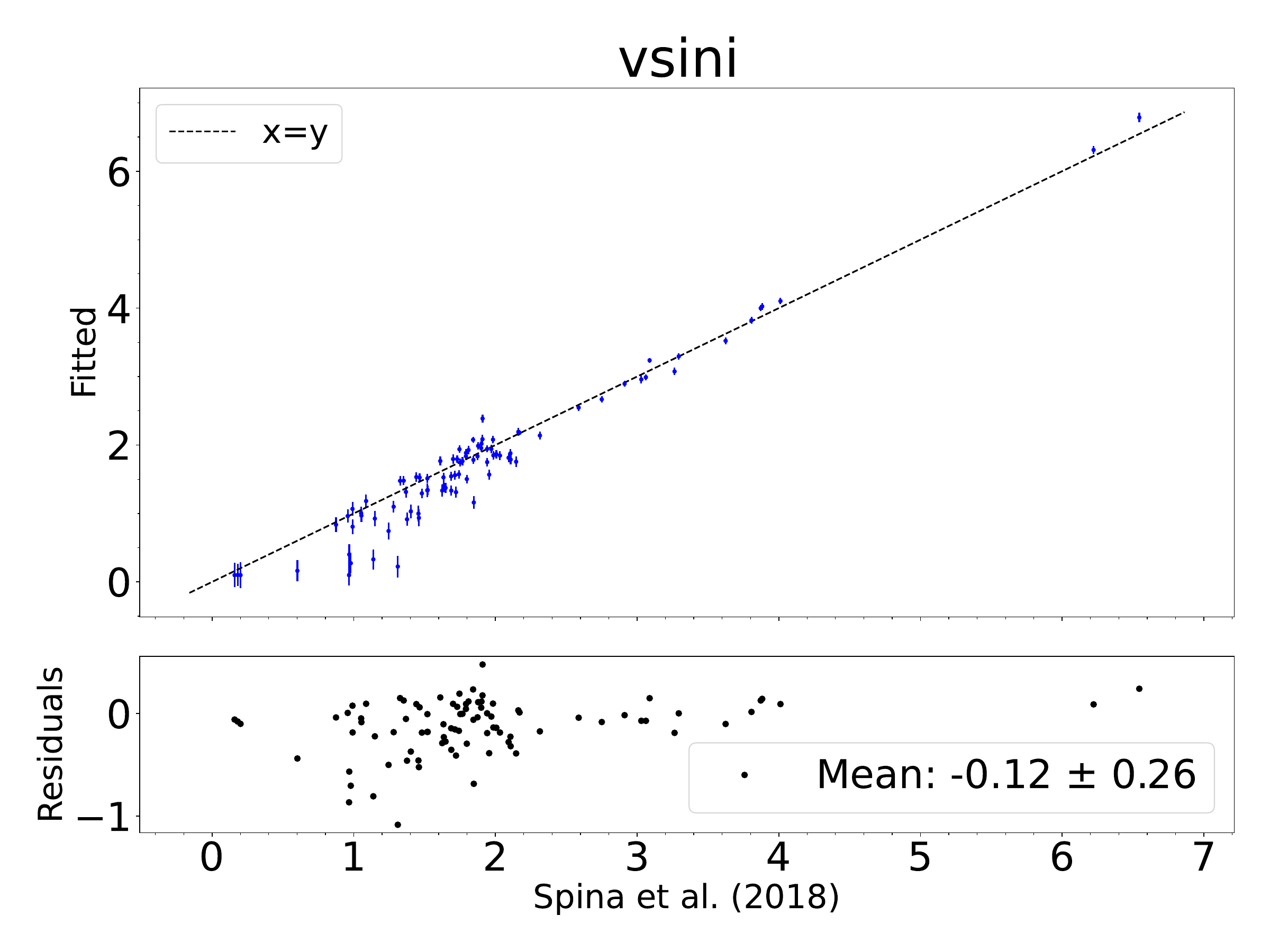}\\
                \end{longtable}
                \captionsetup{font=tiny}
                \captionof{figure}{Atmospheric parameters obtained automatically with the NN compared with the results of \citet{Spina_2018}. The bottom panel shows the residuals, as well as their average and standard deviation.}
                \label{fig:result_params}
                
                \begin{longtable}{ccc} 
                        \includegraphics[width=0.3\linewidth]{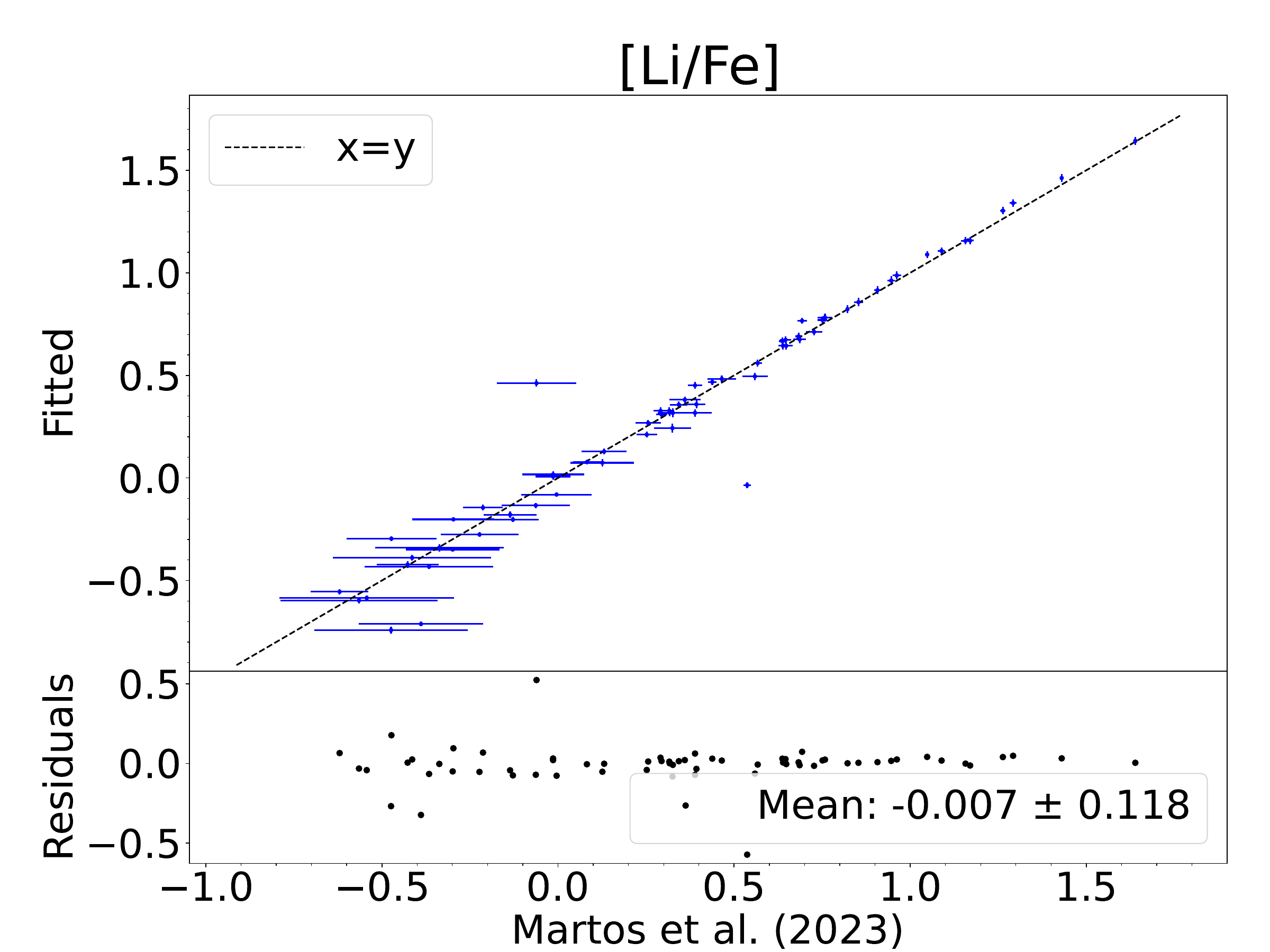} & 
                        \includegraphics[width=0.3\linewidth]{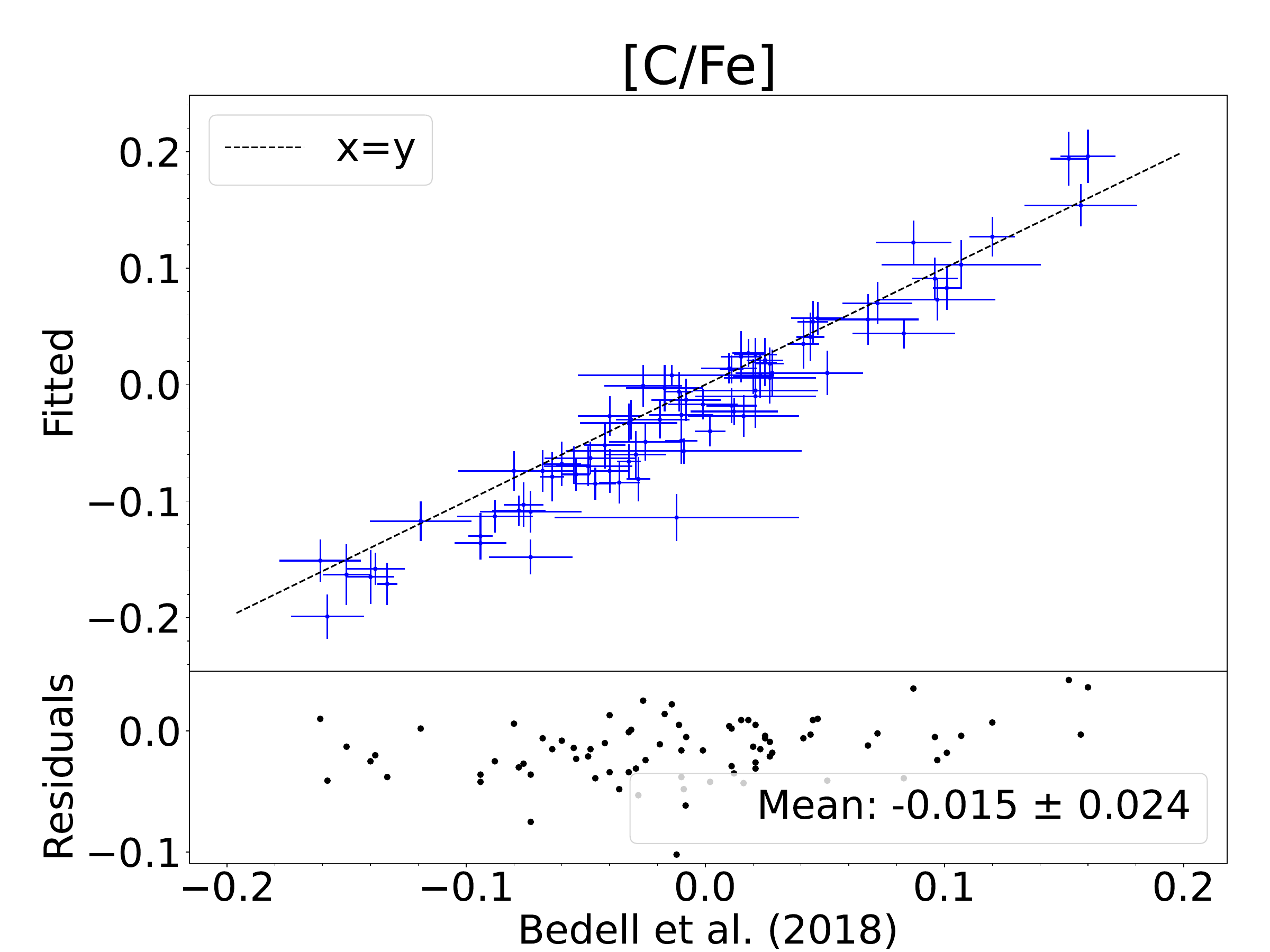} &  \includegraphics[width=0.3\linewidth]{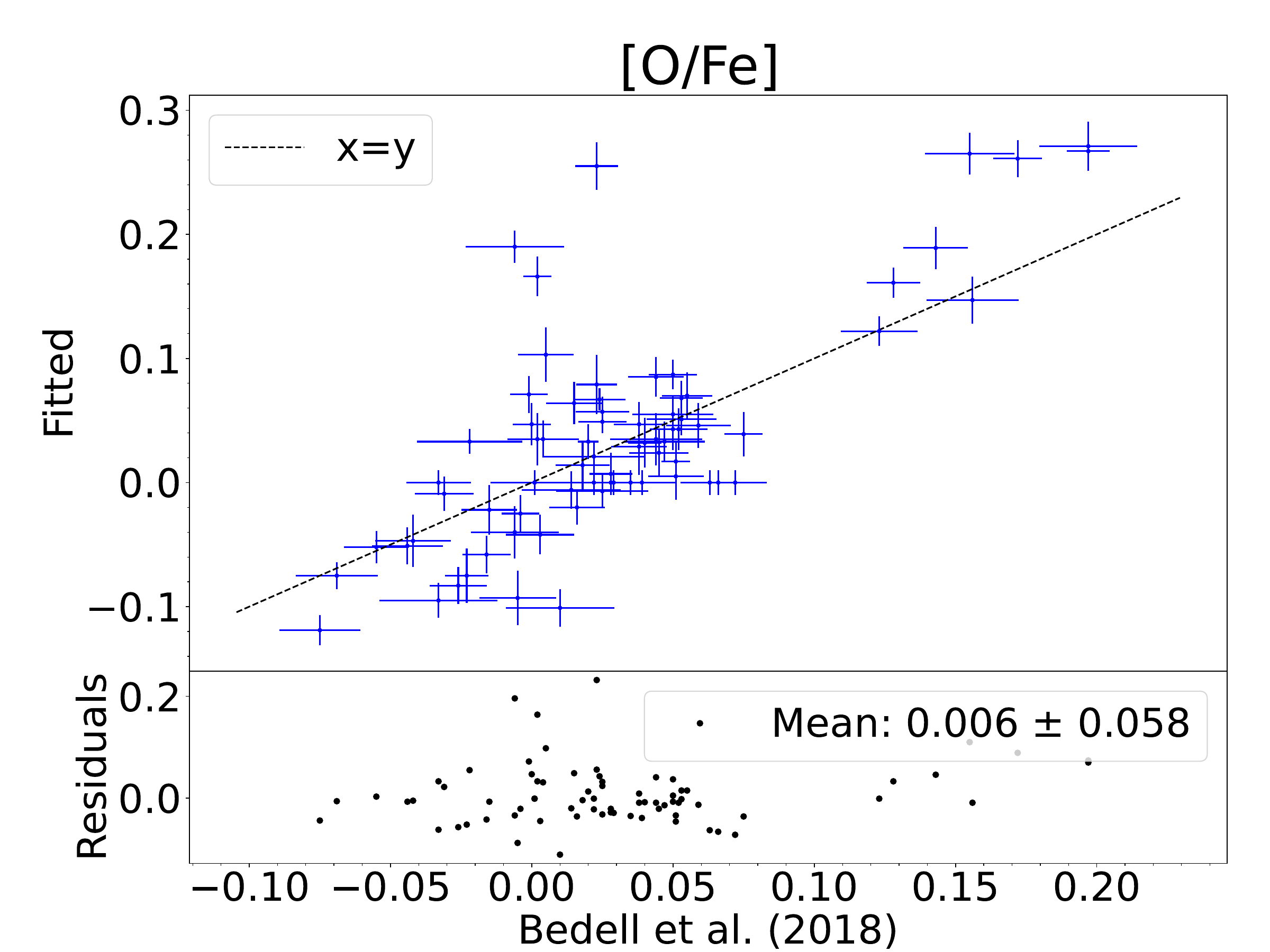}\\ 
                        \includegraphics[width=0.3\linewidth]{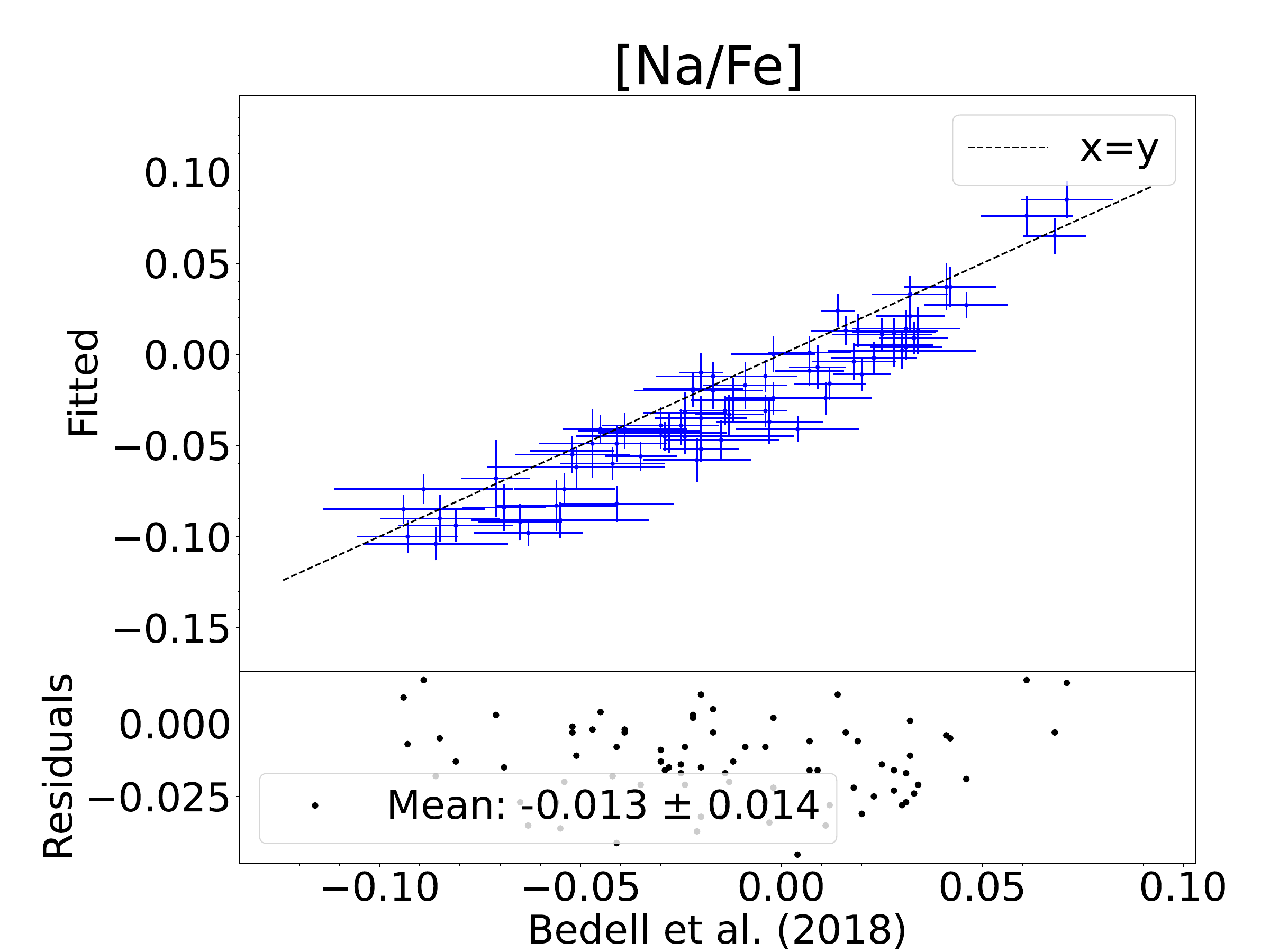} &  \includegraphics[width=0.3\linewidth]{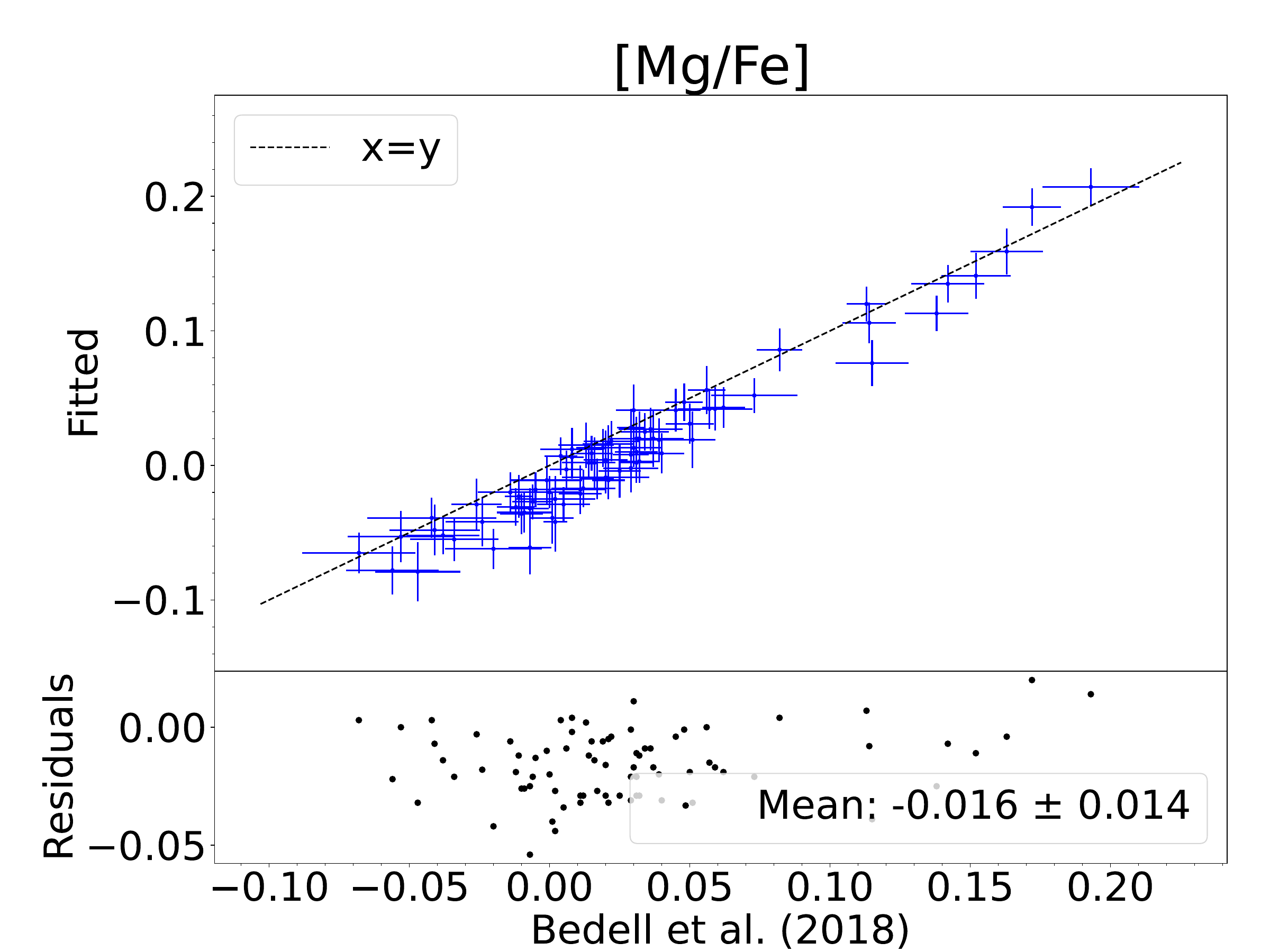}&
                        \includegraphics[width=0.3\linewidth]{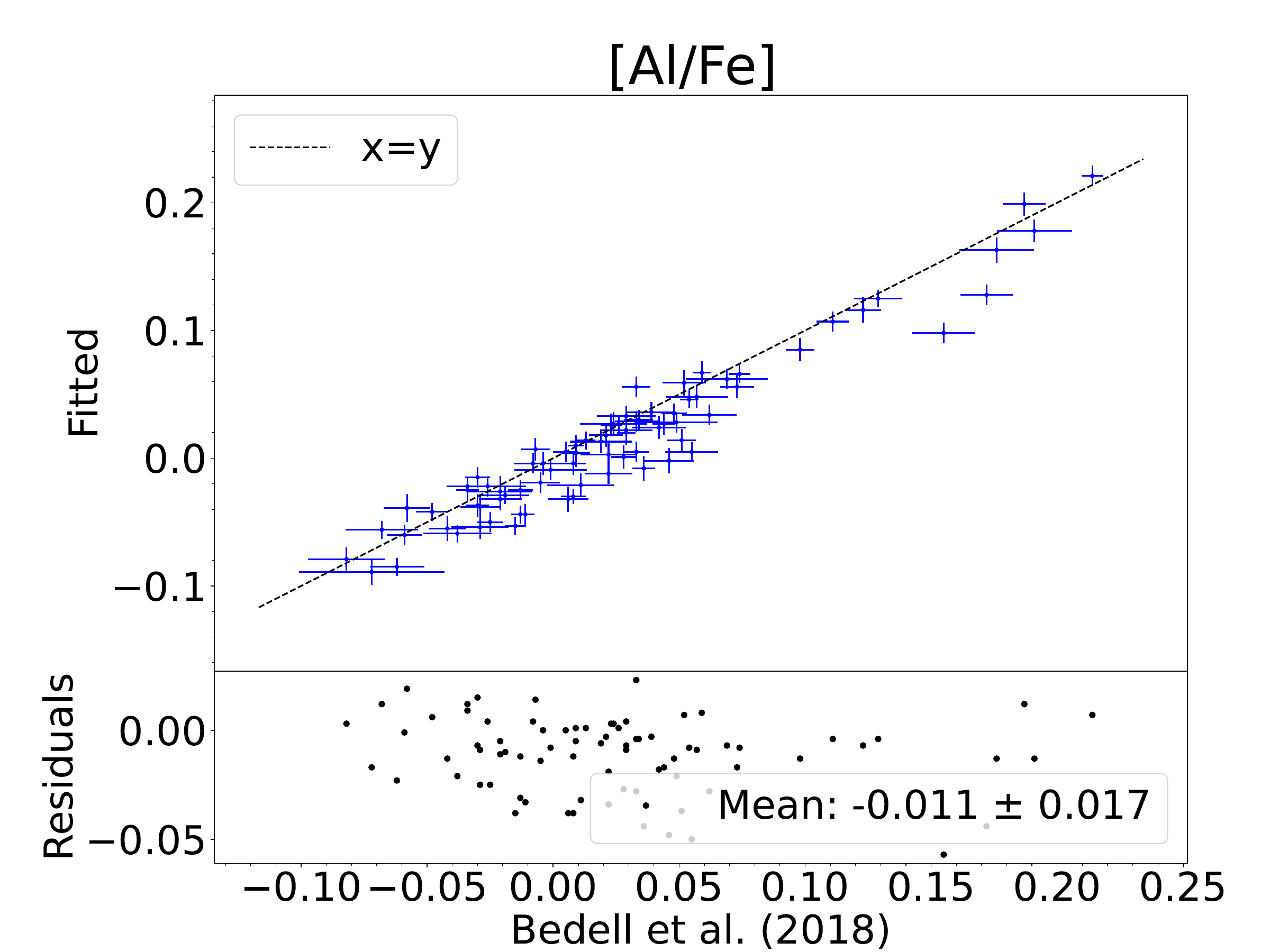} \\   \includegraphics[width=0.3\linewidth]{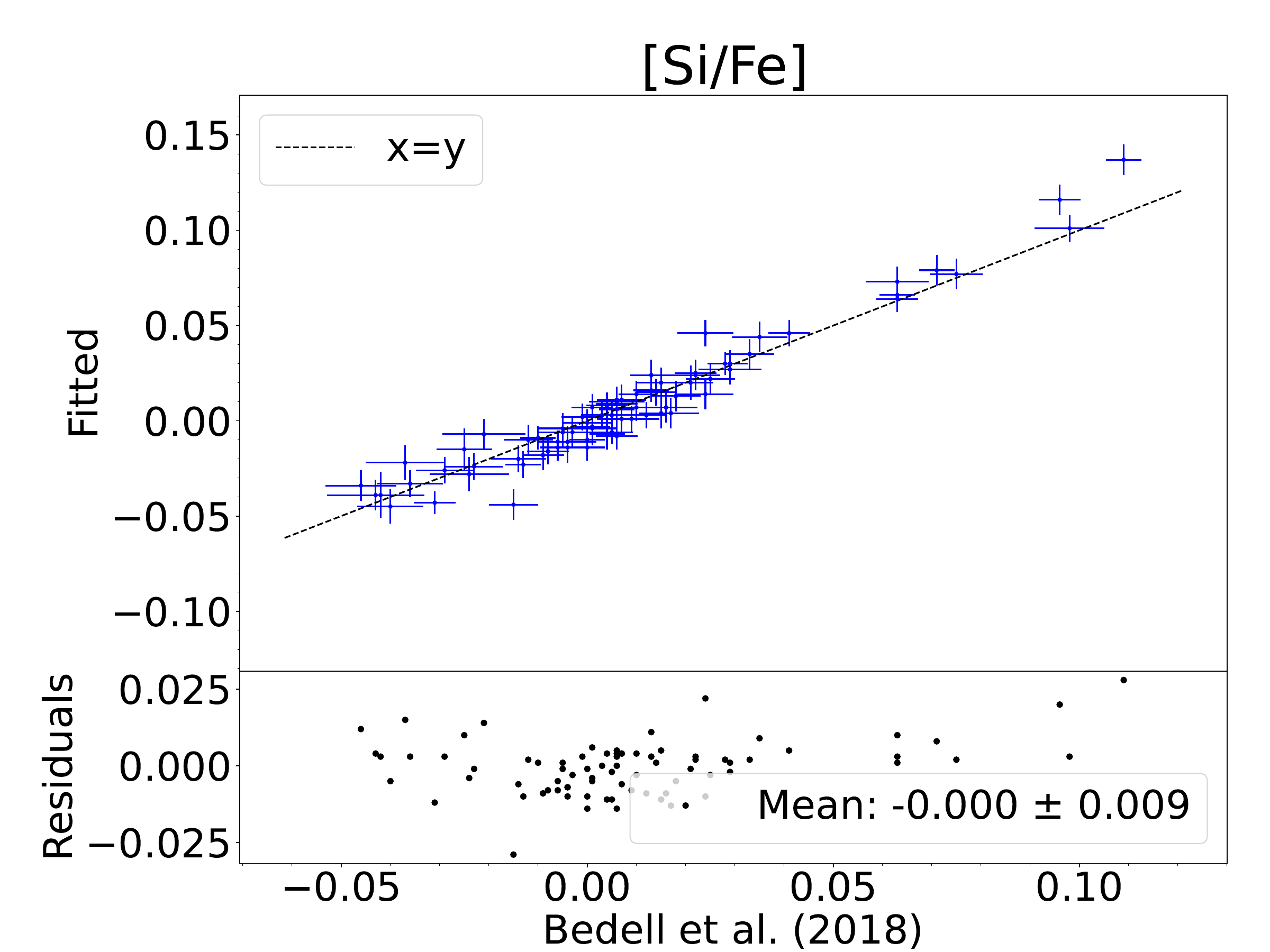}&
                        \includegraphics[width=0.3\linewidth]{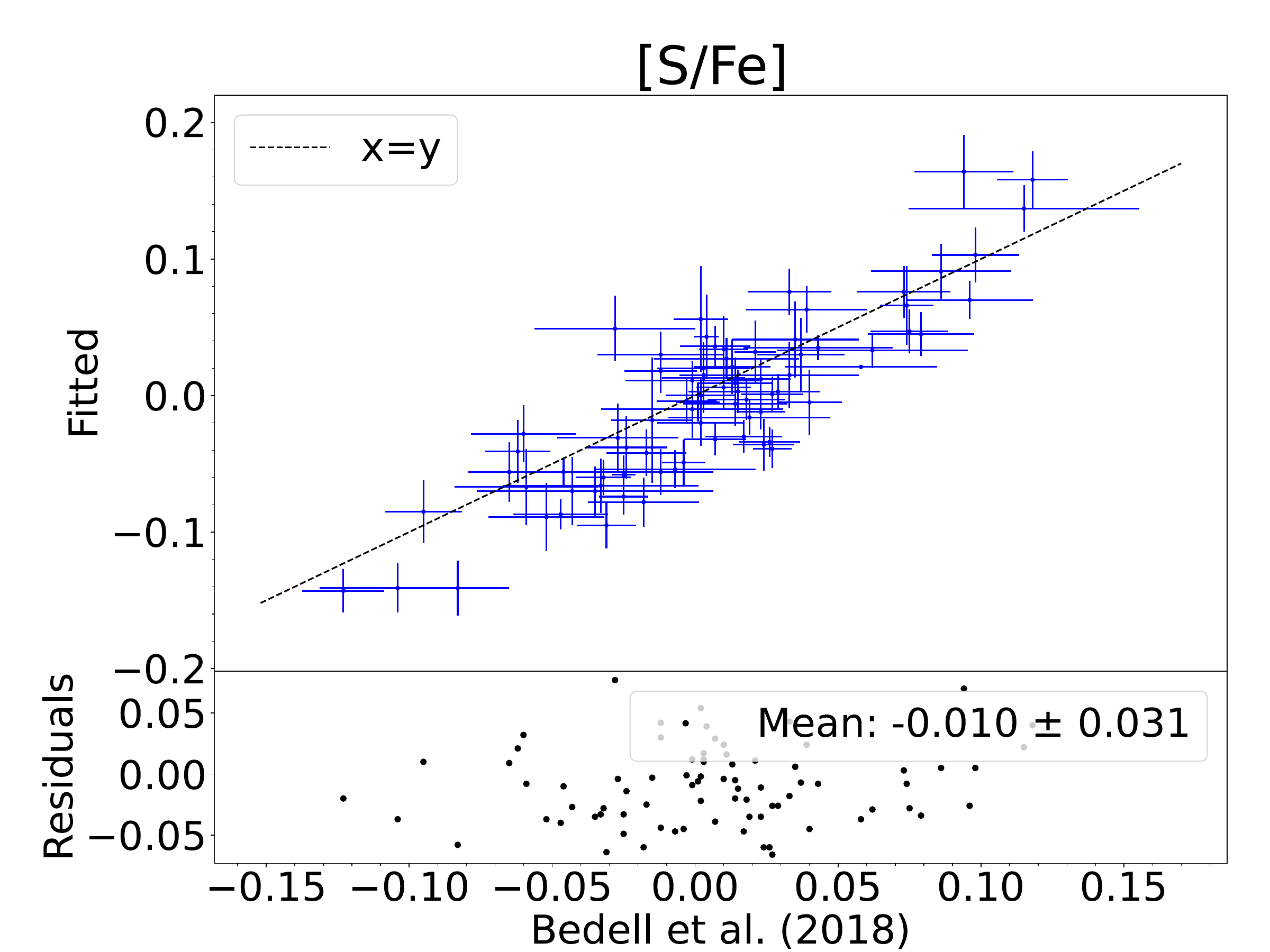} &  \includegraphics[width=0.3\linewidth]{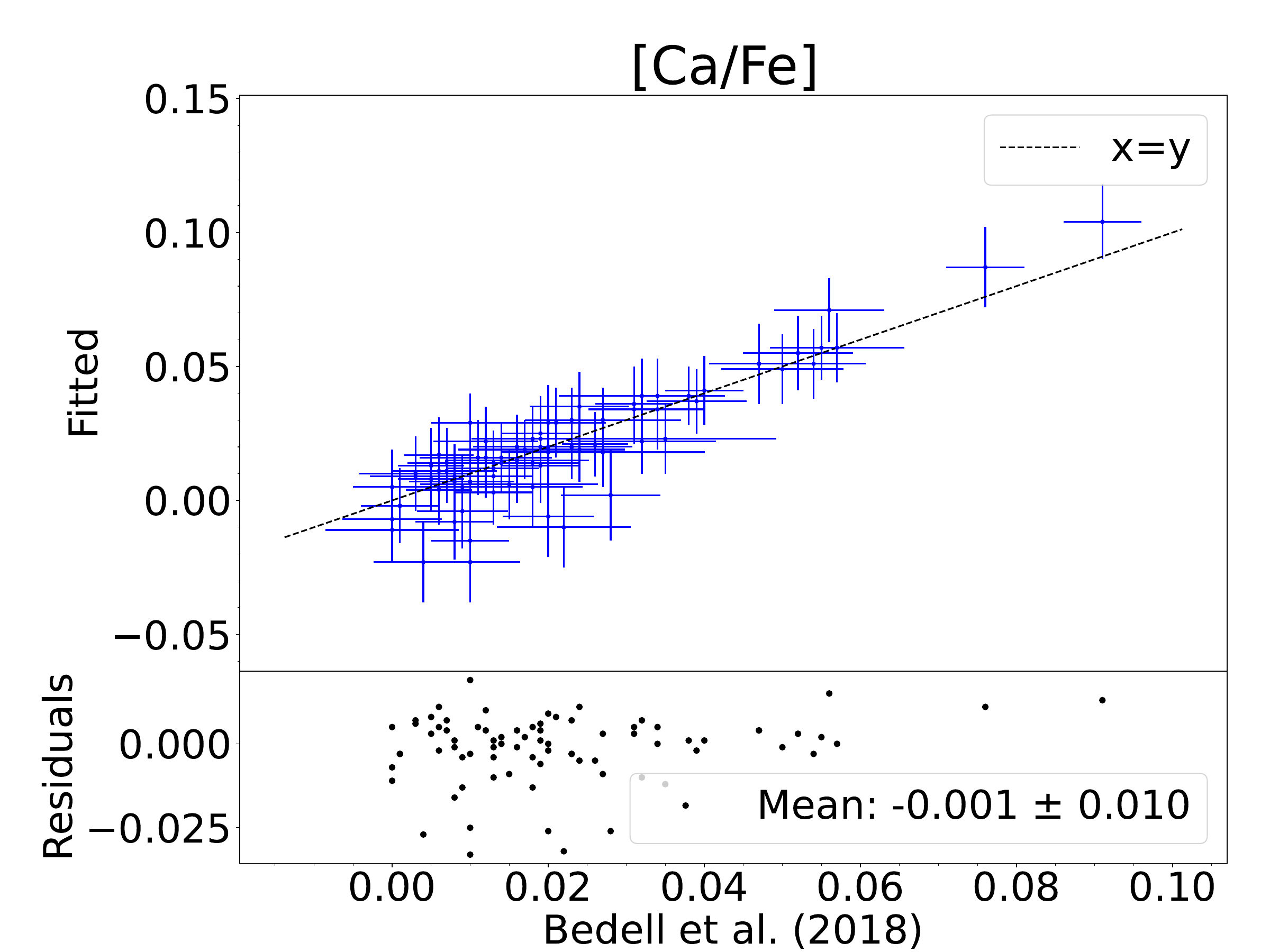}\\  \includegraphics[width=0.3\linewidth]{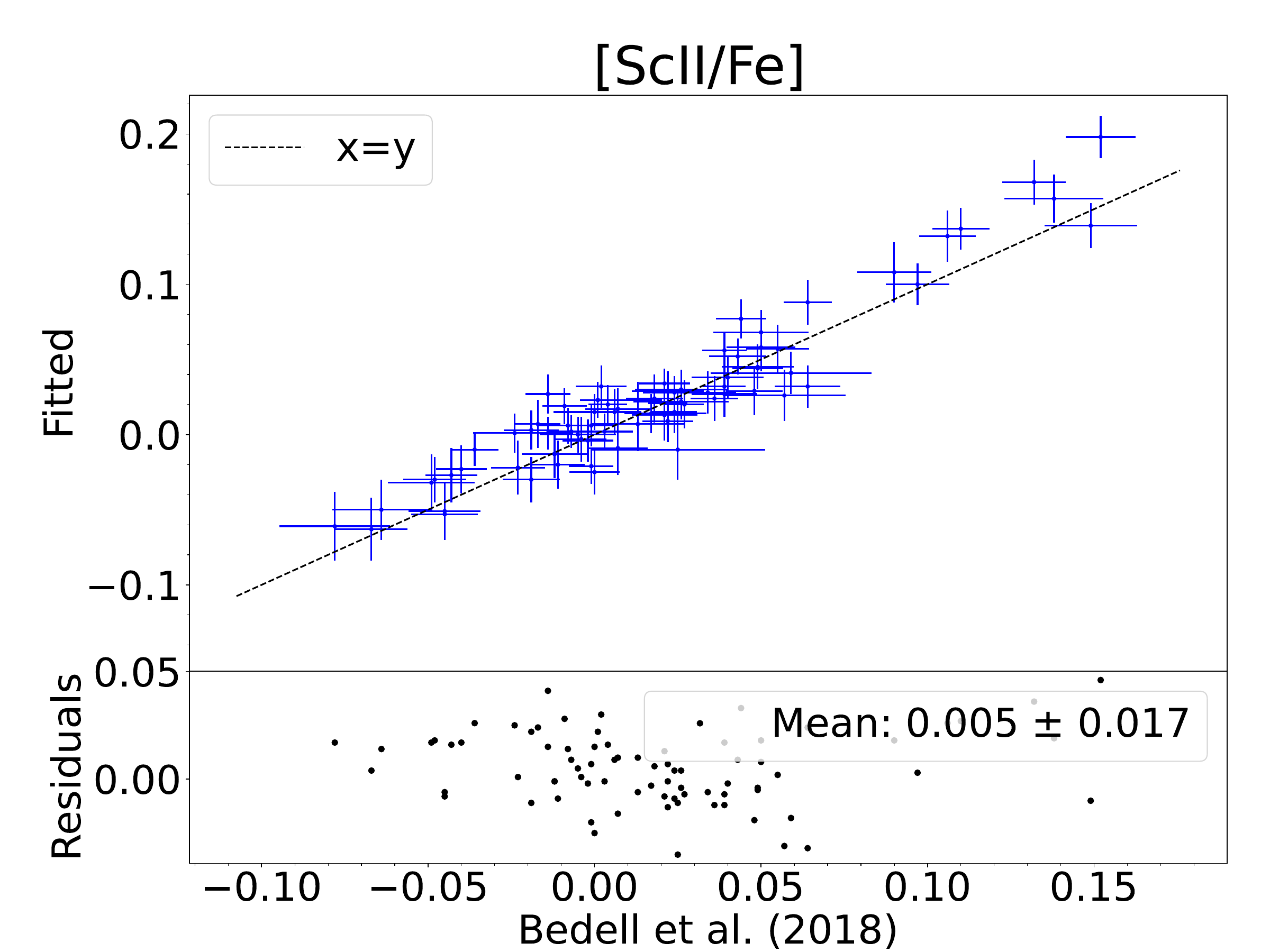}&
                        \includegraphics[width=0.3\linewidth]{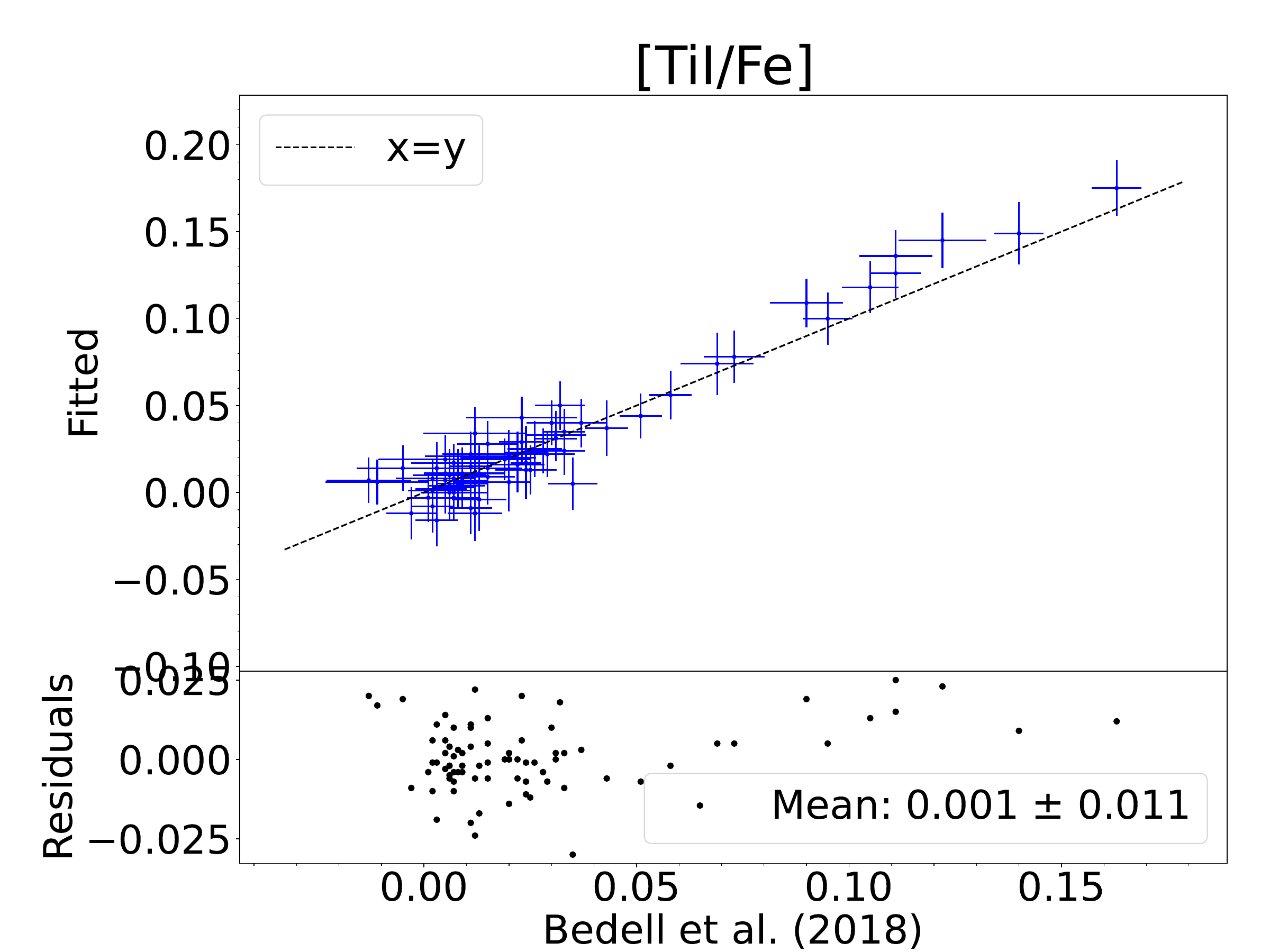} & 
                        \includegraphics[width=0.3\linewidth]{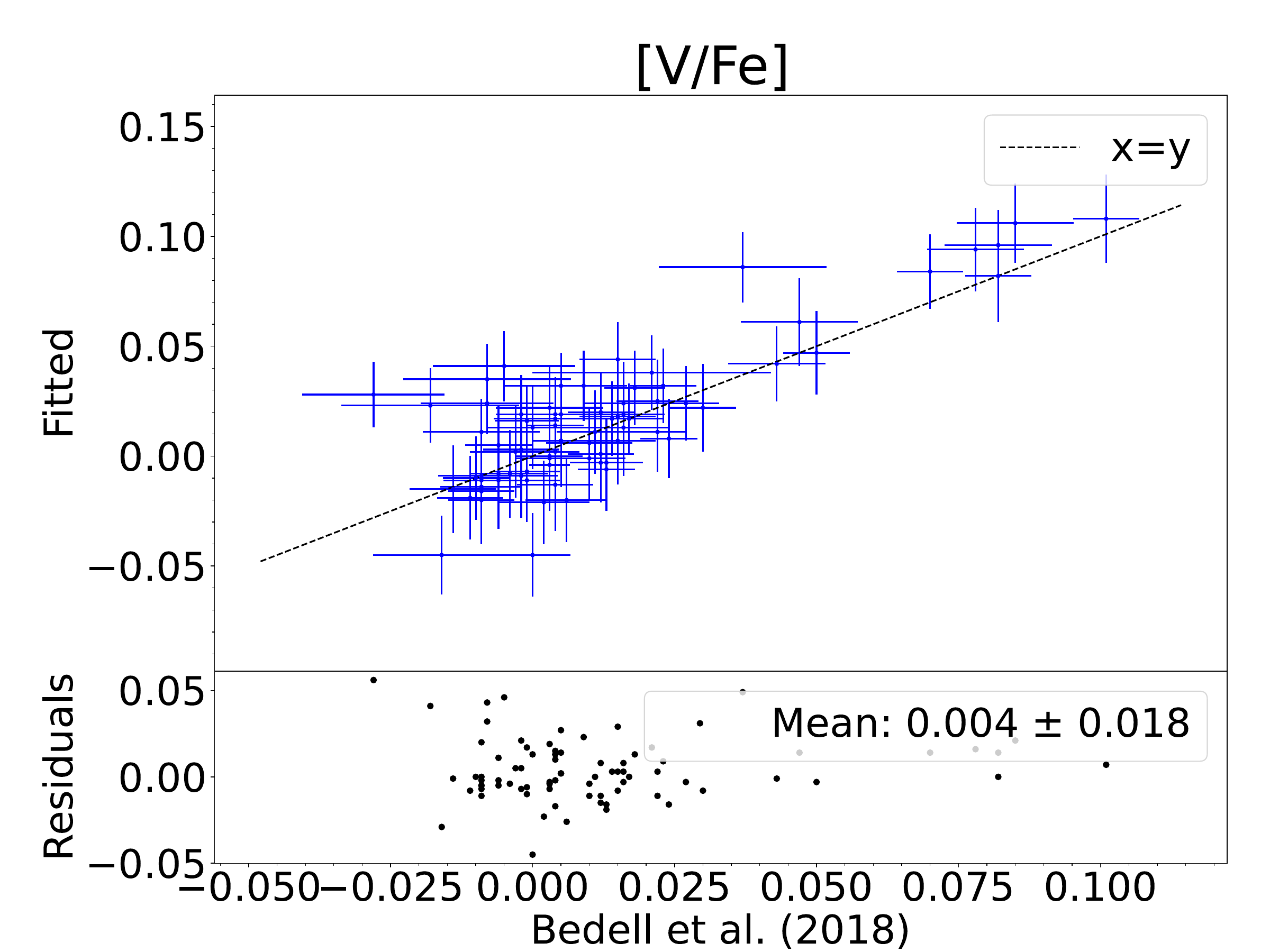} \\  \includegraphics[width=0.3\linewidth]{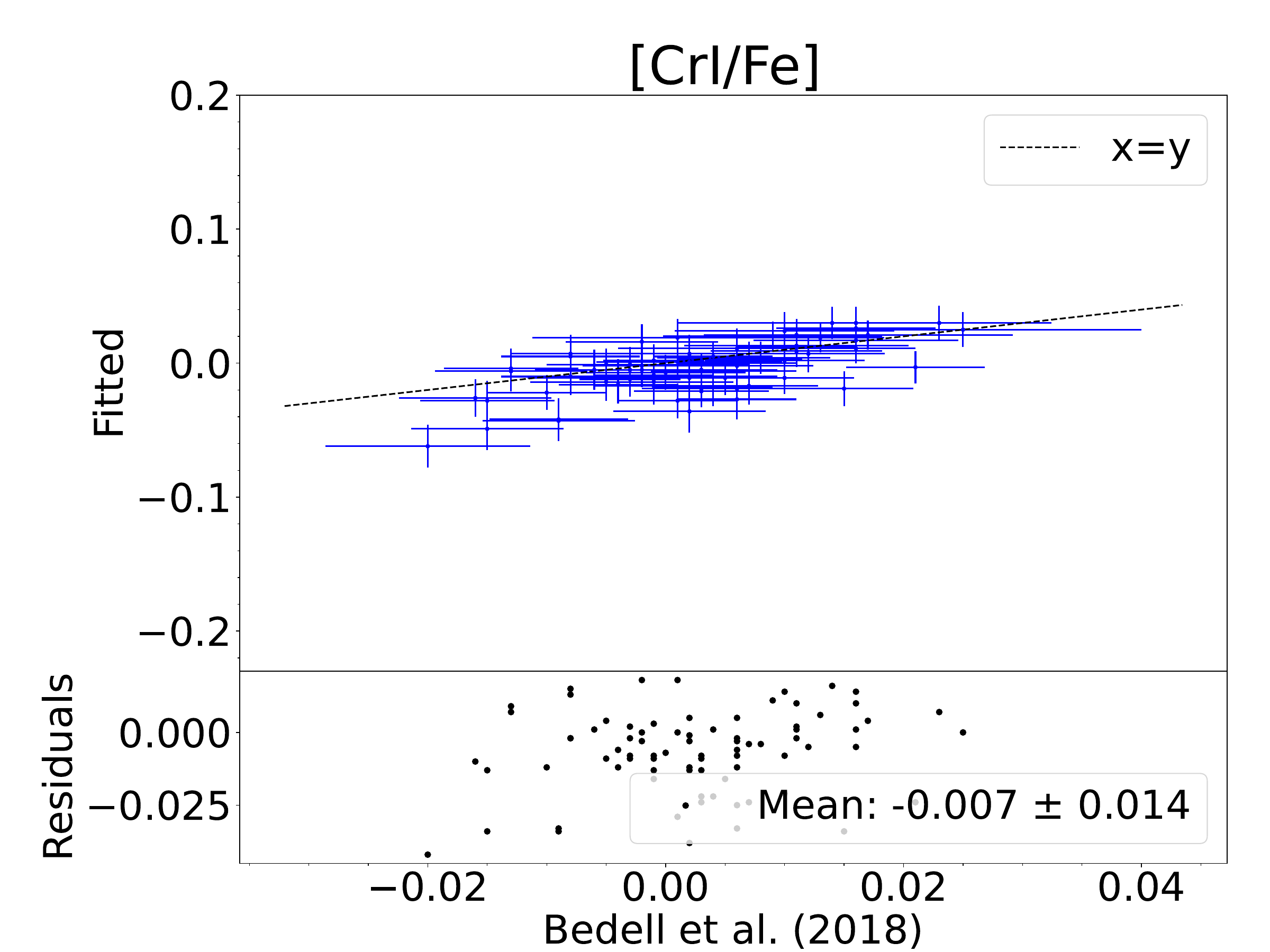}&
                        \includegraphics[width=0.3\linewidth]{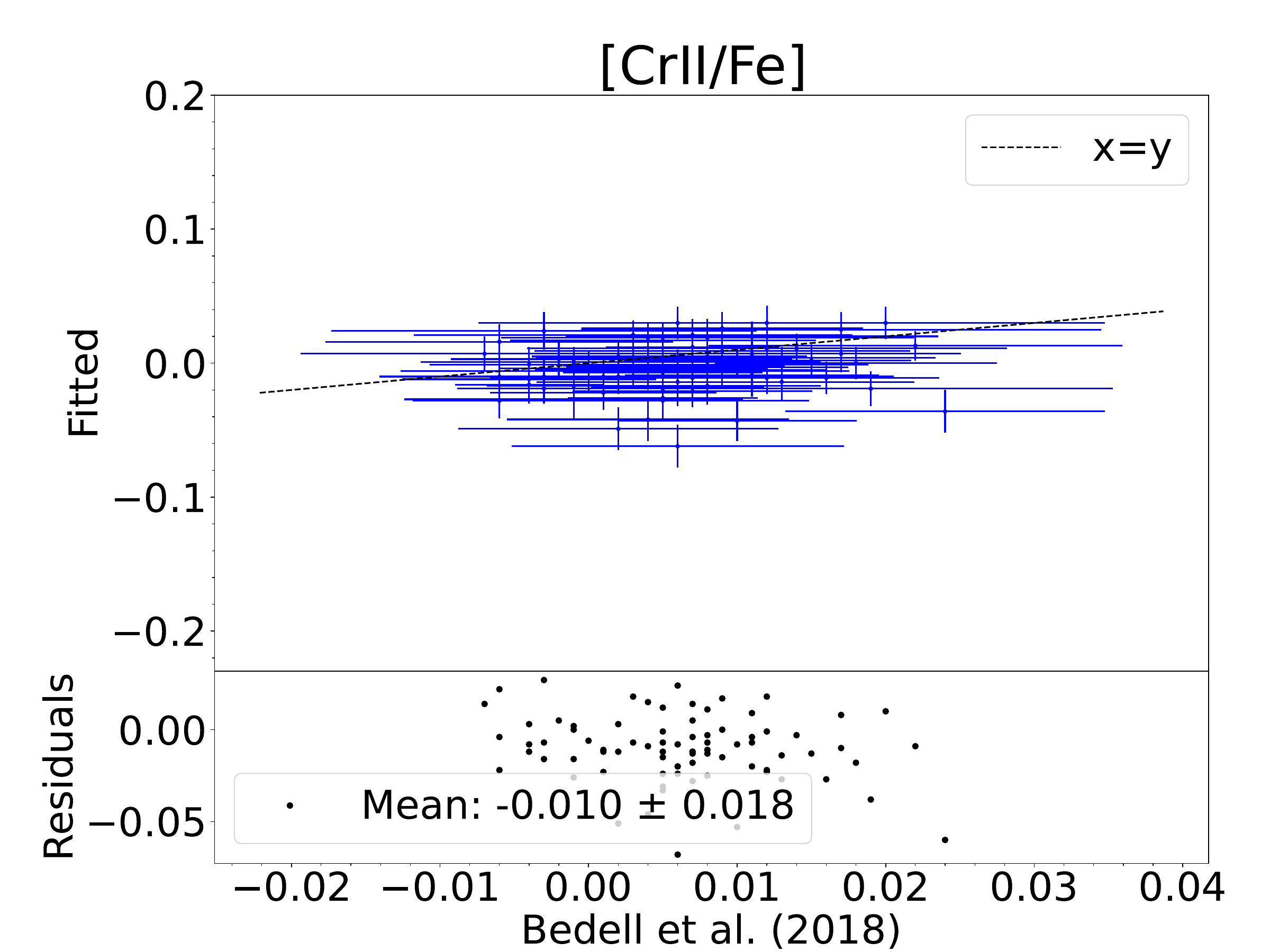} &  \includegraphics[width=0.3\linewidth]{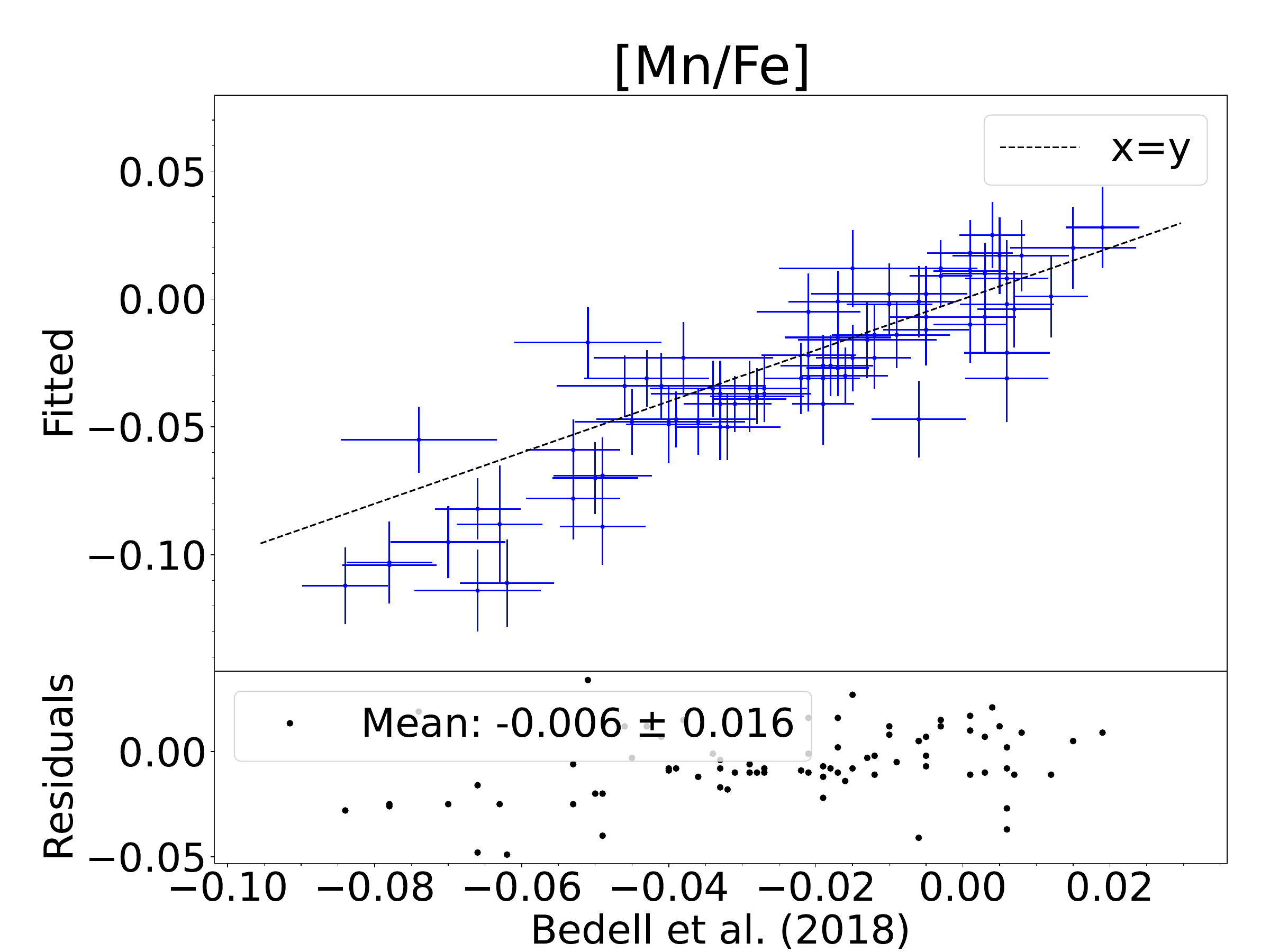}\\
                        \includegraphics[width=0.3\linewidth]{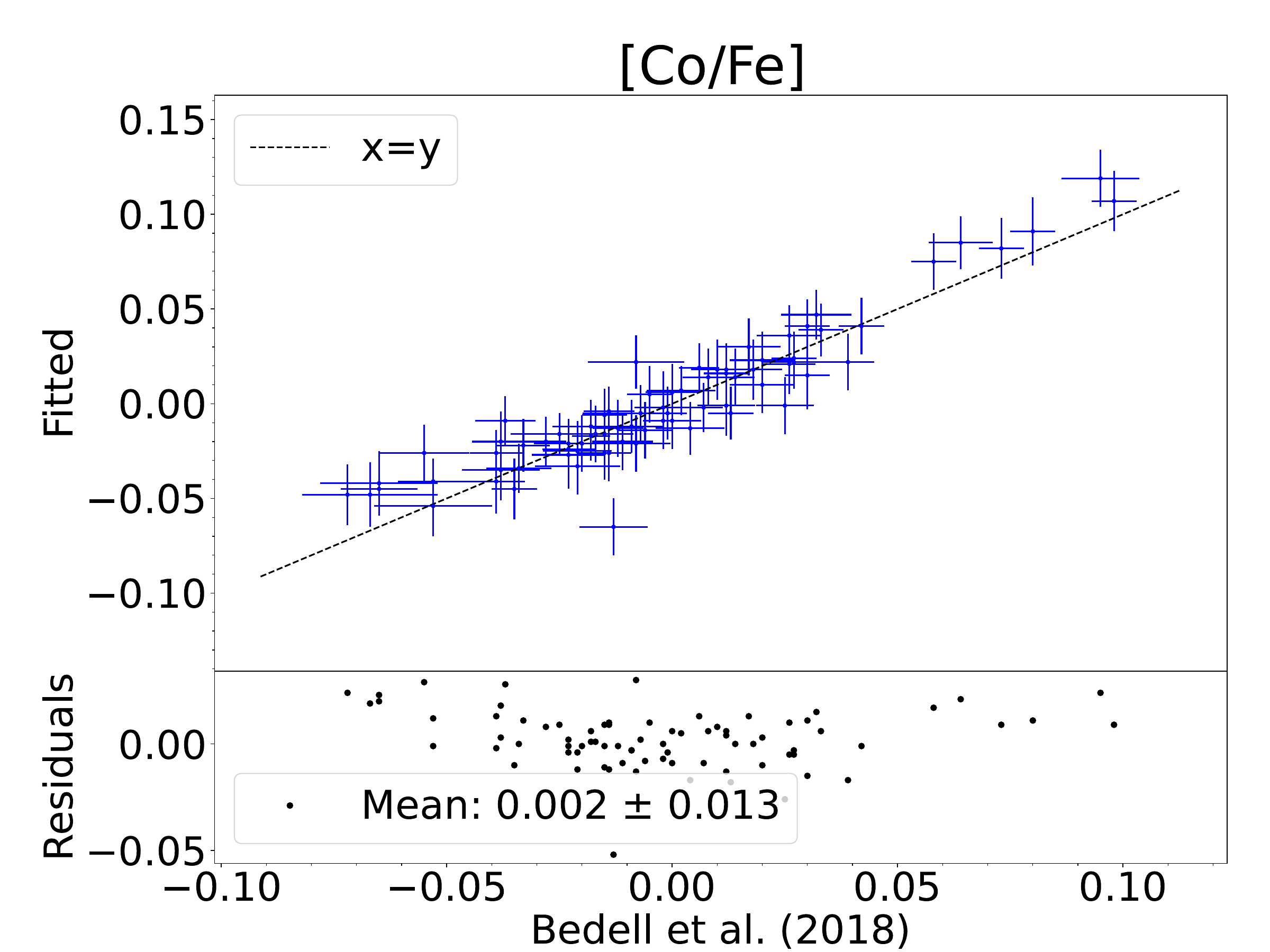} &  \includegraphics[width=0.3\linewidth]{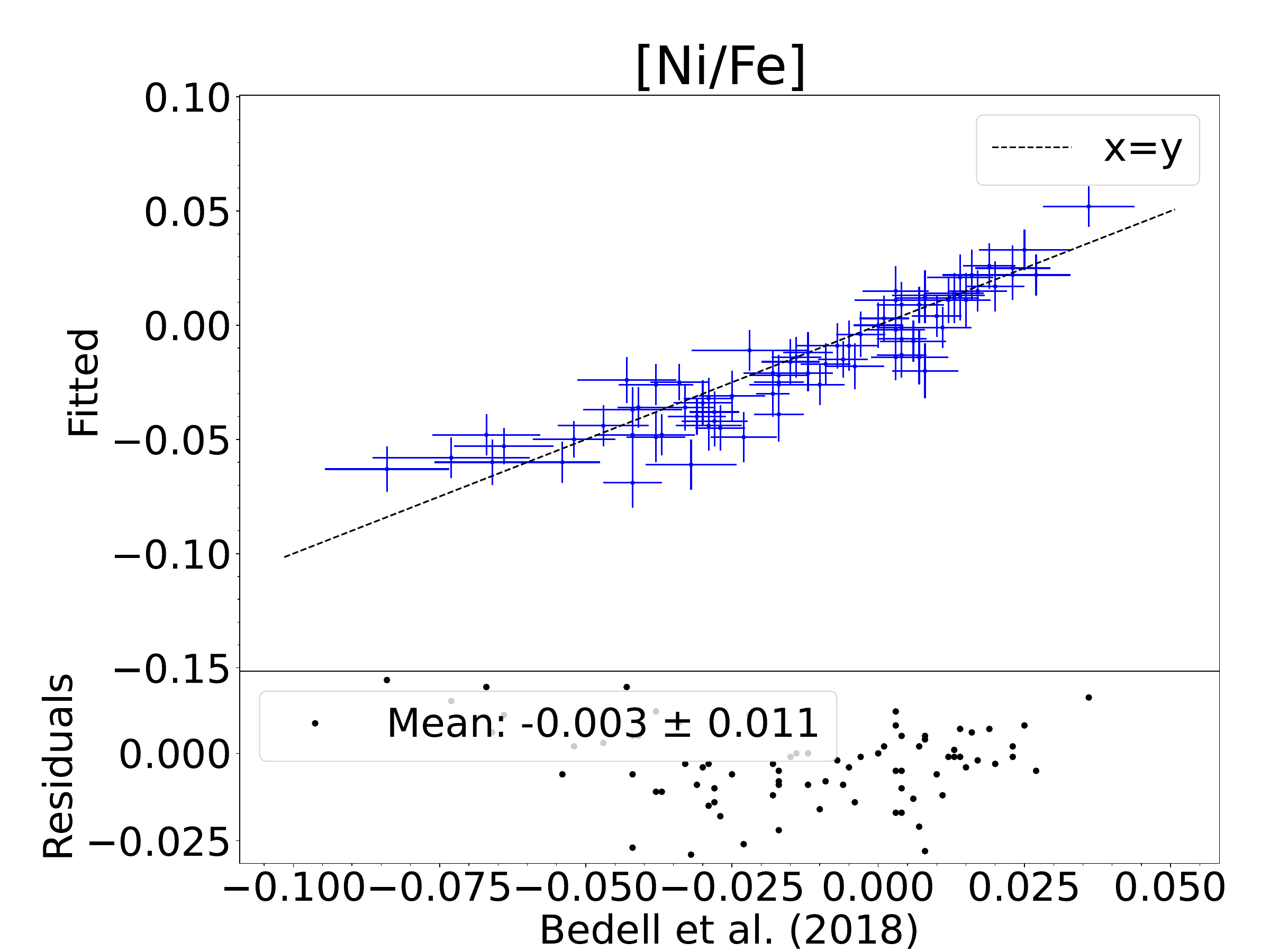}&
                        \includegraphics[width=0.3\linewidth]{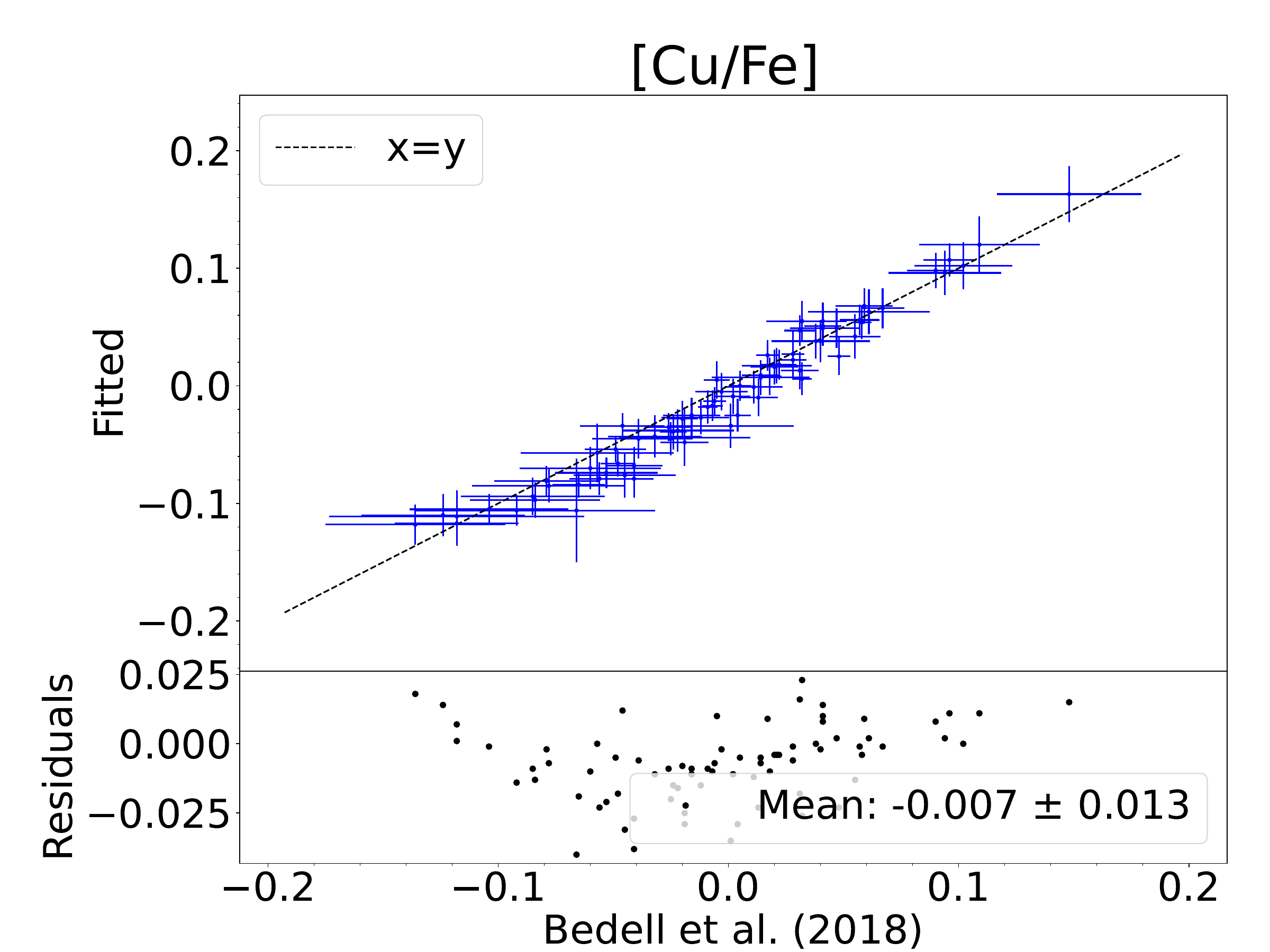} \\  \includegraphics[width=0.3\linewidth]{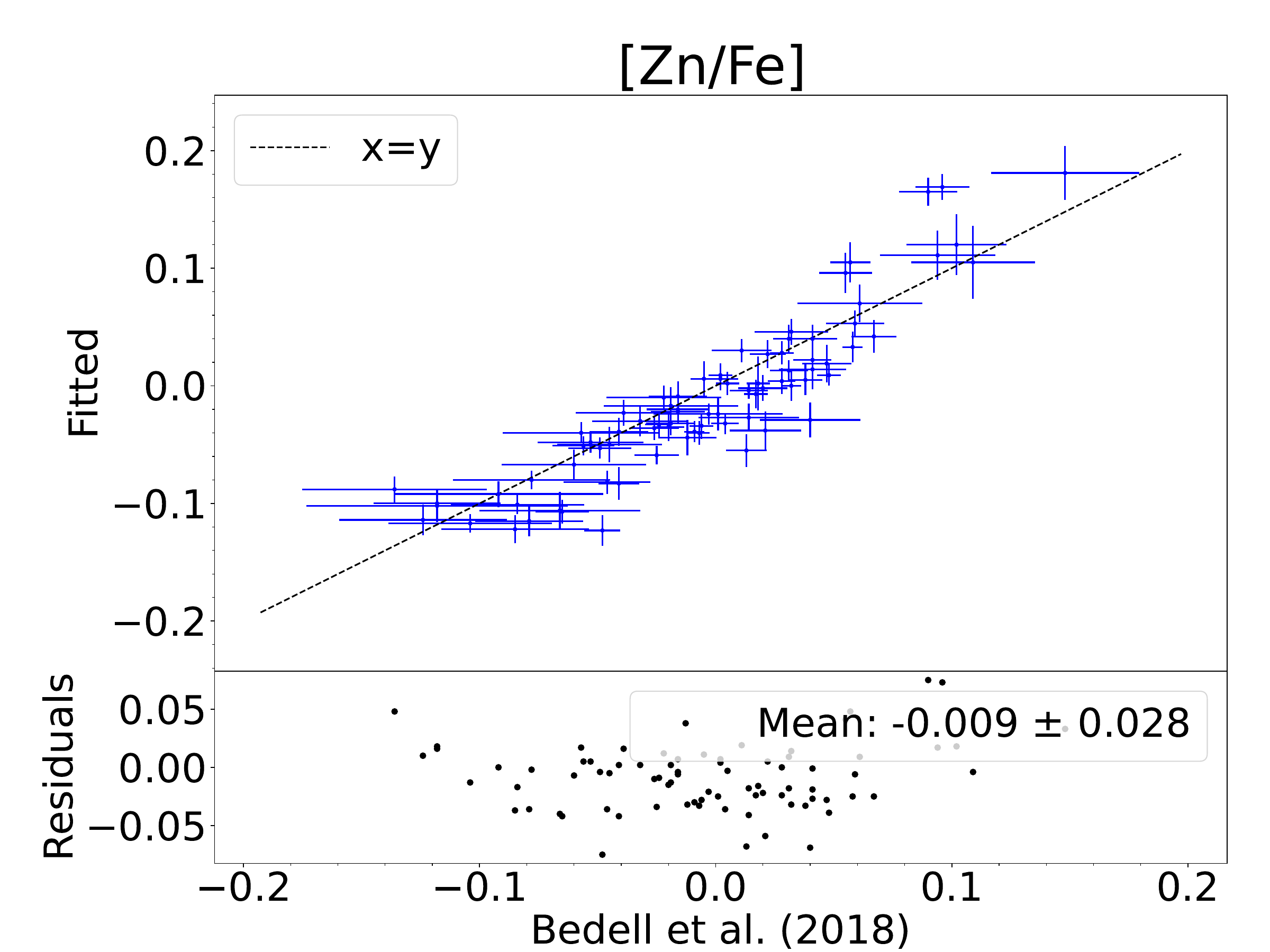}&
                        \includegraphics[width=0.3\linewidth]{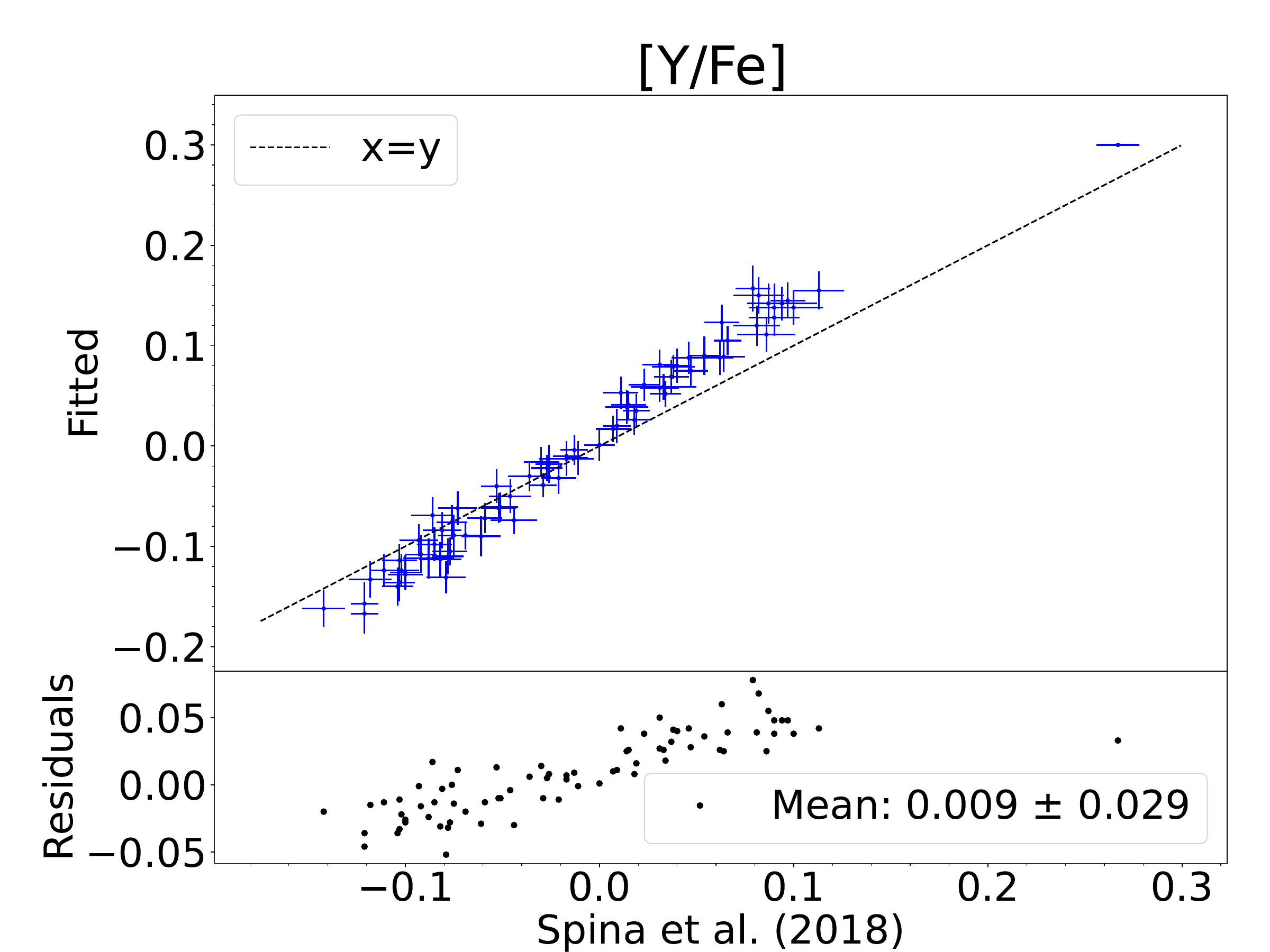} &  \includegraphics[width=0.3\linewidth]{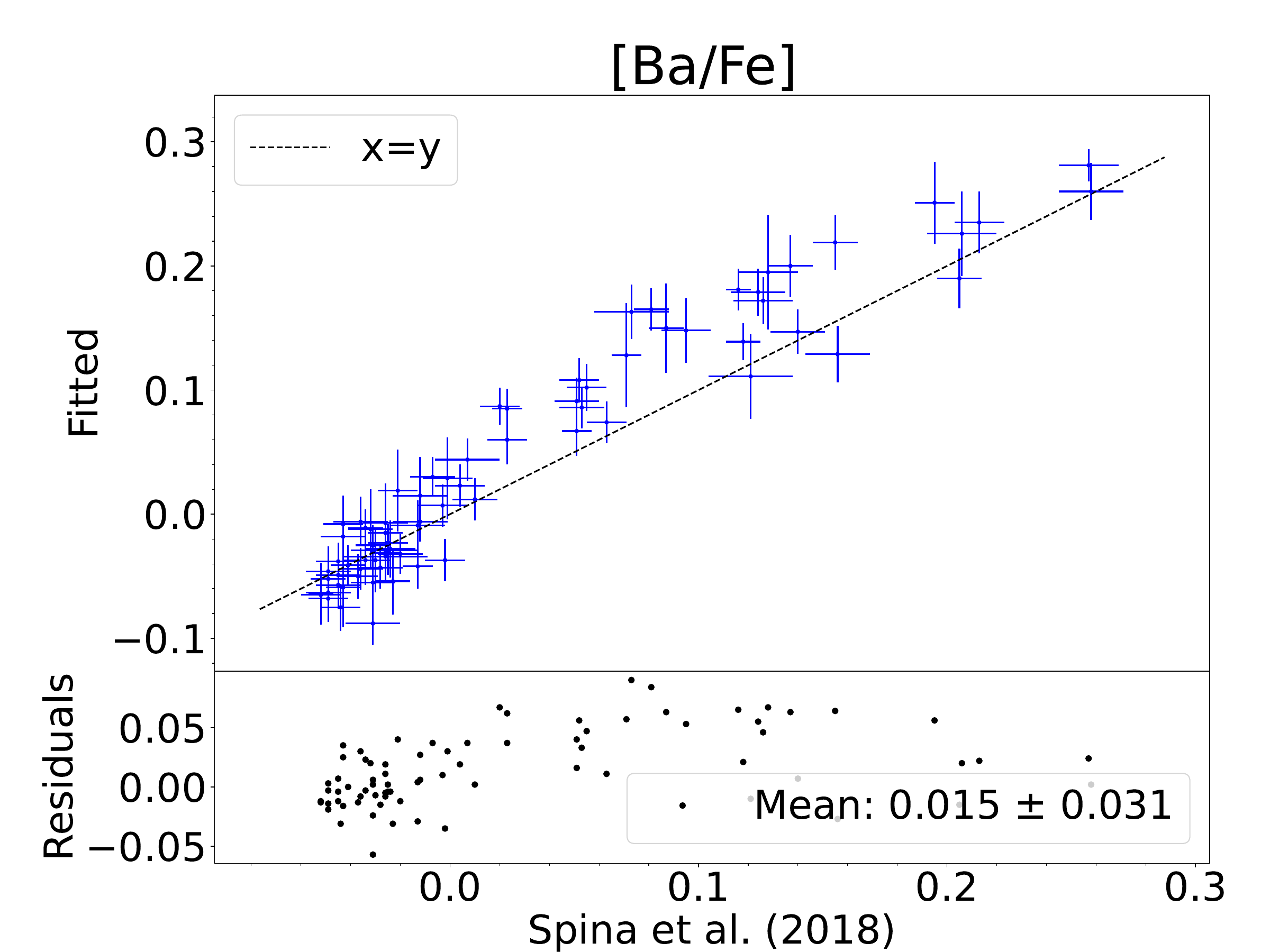} \\ 
                \end{longtable}
                \captionof{figure}{Comparison of the abundances obtained automatically with \citet{Martos_2023} for Li, \citet{Spina_2018} for Y and Ba and \citet{Bedell_2018} for the other elements. The bottom panel shows the residuals, as well as their average and standard deviation.}
                \label{fig:comp_abunds}
                
                \newpage

                \section{Fits with stellar age}
                
                
                \tiny{\begin{longtable}{lllllll}
                                \caption{Coefficients of the linear fit of the chemical abundances versus age. The complete table is available at CDS.} \label{tab:coefs_GCE_linear} \\
                                \toprule
                                Element & A & $\sigma_A$ & B & $\sigma_B$ & $\chi^2_{red}$ & $\sigma_{resid}$ \\
                                \midrule
                                \endfirsthead
                                \caption[]{Coefficients of the linear fit of the chemical abundances versus age.} \\
                                \toprule
                                Element & A & $\sigma_A$ & B & $\sigma_B$ & $\chi^2_{red}$ & $\sigma_{resid}$\\
                                \midrule
                                \endhead
                                \midrule
                                \multicolumn{6}{r}{Continued on next page} \\
                                \midrule
                                \endfoot
                                \bottomrule
                                \endlastfoot
                                $[$Li/Fe$]$ & -0.3174 & 0.0282 & 1.8651 & 0.1464 & 20.4 & 0.500\\
                                $[$C/Fe$]$ & 0.0246 & 0.0019 & -0.1663 & 0.0101 & 4.3 & 0.042\\...\\
                                
                \end{longtable}}

                
                \begin{longtable}{ccc} 
                        \renewcommand{\arraystretch}{0.1} 
                        \includegraphics[width=0.33\linewidth]{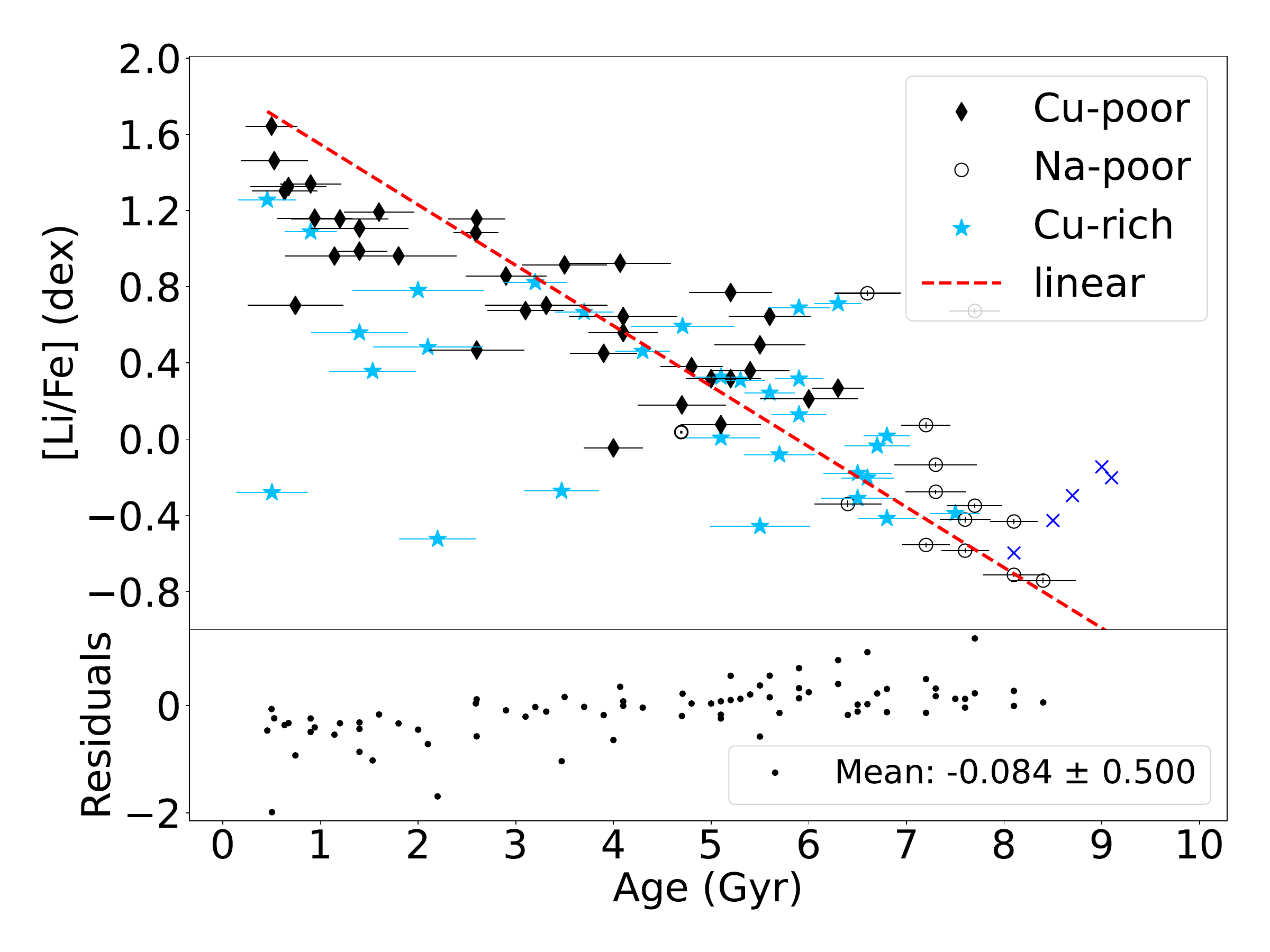} & \hspace{-0.6cm} \includegraphics[width=0.33\linewidth]{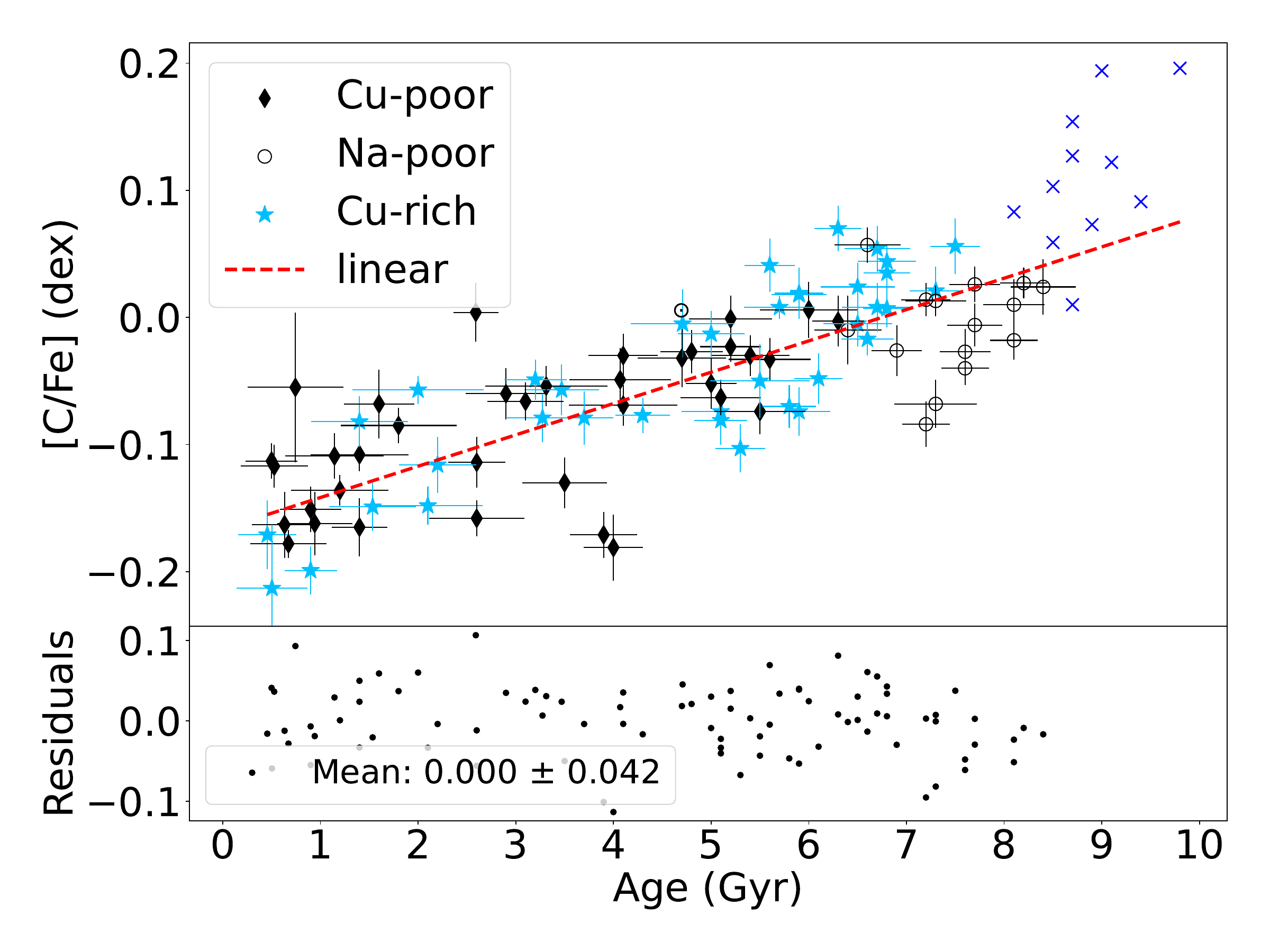}& \hspace{-0.6cm}
                        \includegraphics[width=0.33\linewidth]{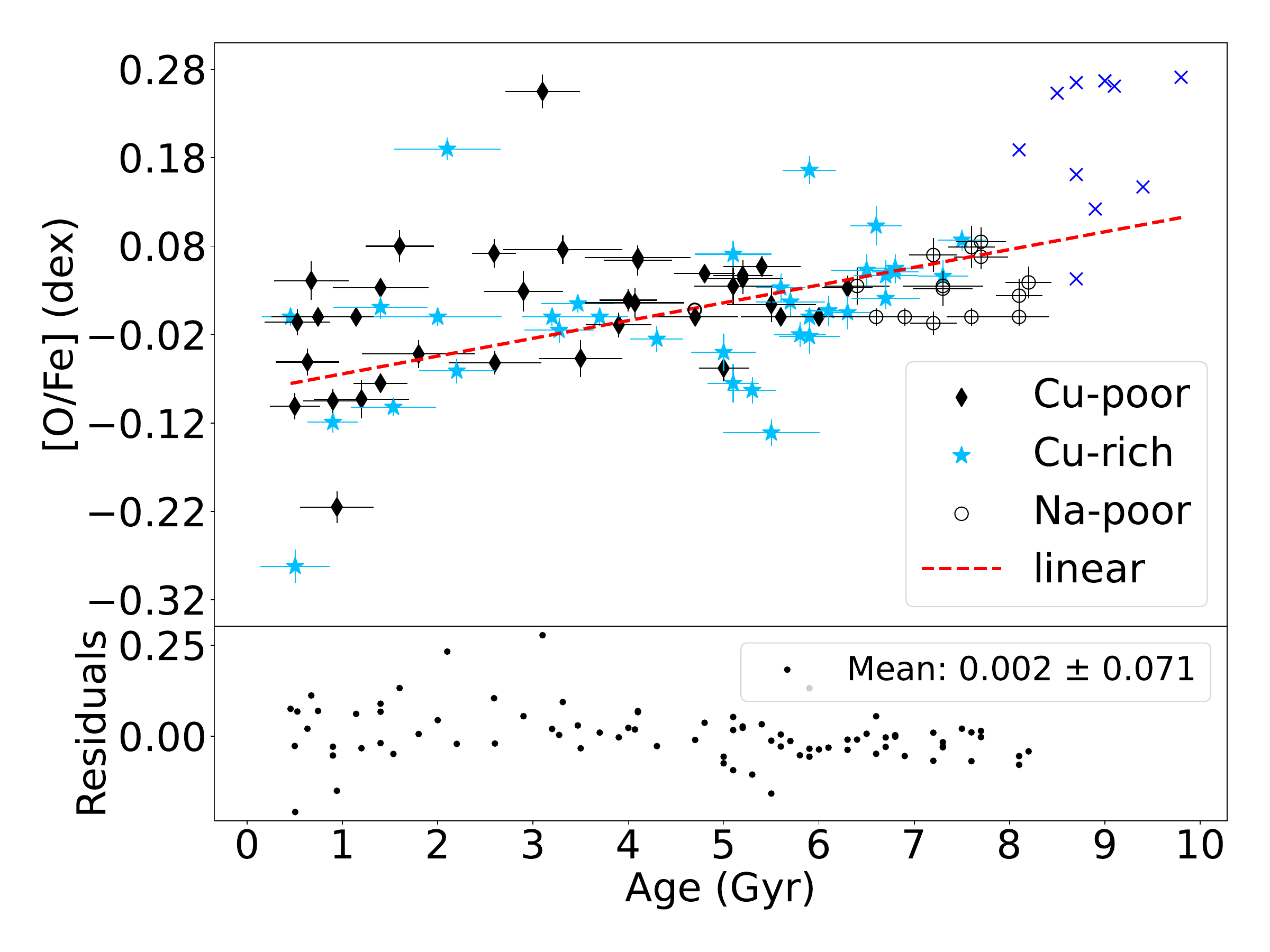} \\   
                        \includegraphics[width=0.33\linewidth]{fig/Na_Fe_age_linear.pdf}& \hspace{-0.6cm}
                        \includegraphics[width=0.33\linewidth]{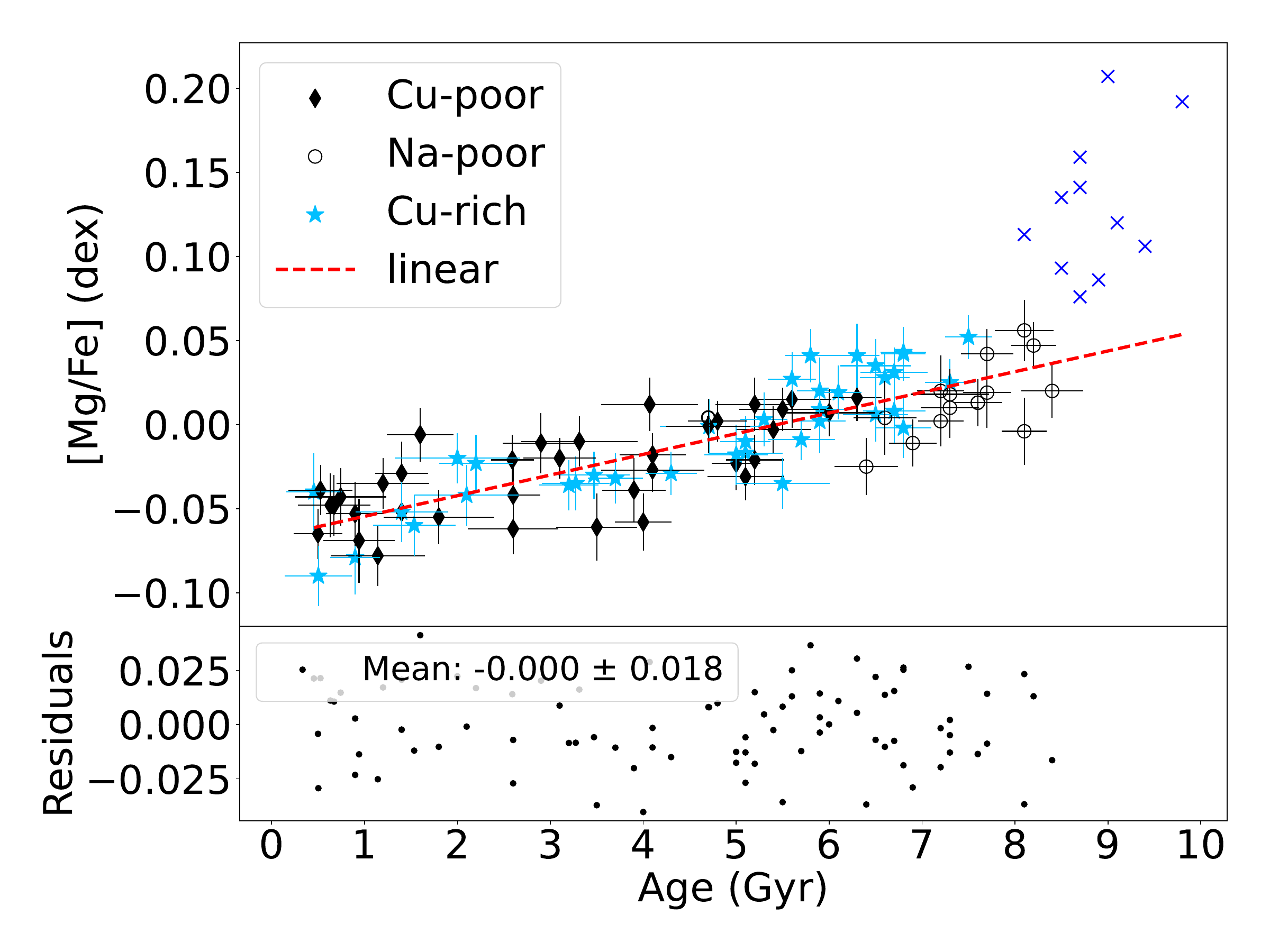} & \hspace{-0.6cm}  \includegraphics[width=0.33\linewidth]{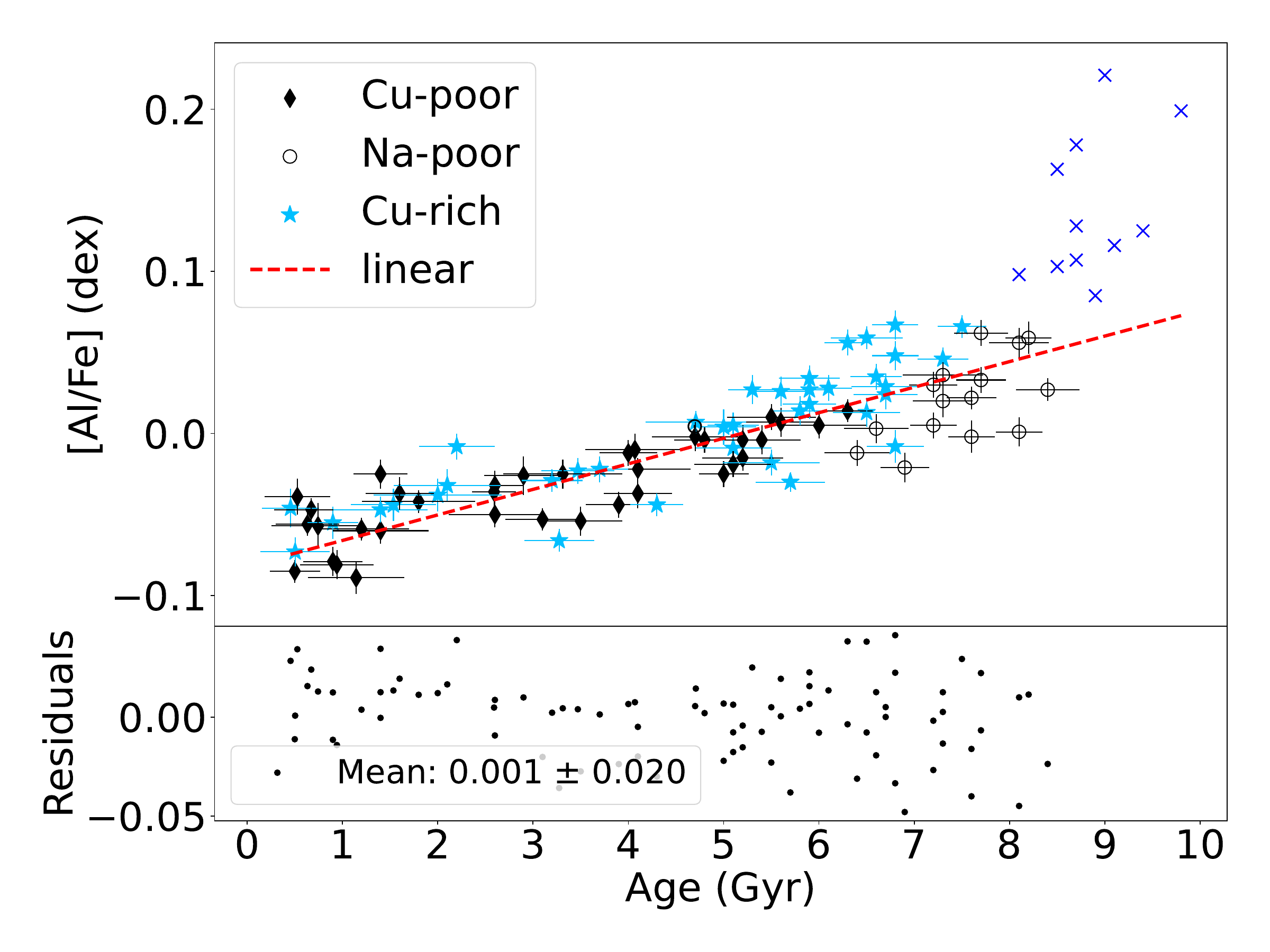}\\
                        \includegraphics[width=0.33\linewidth]{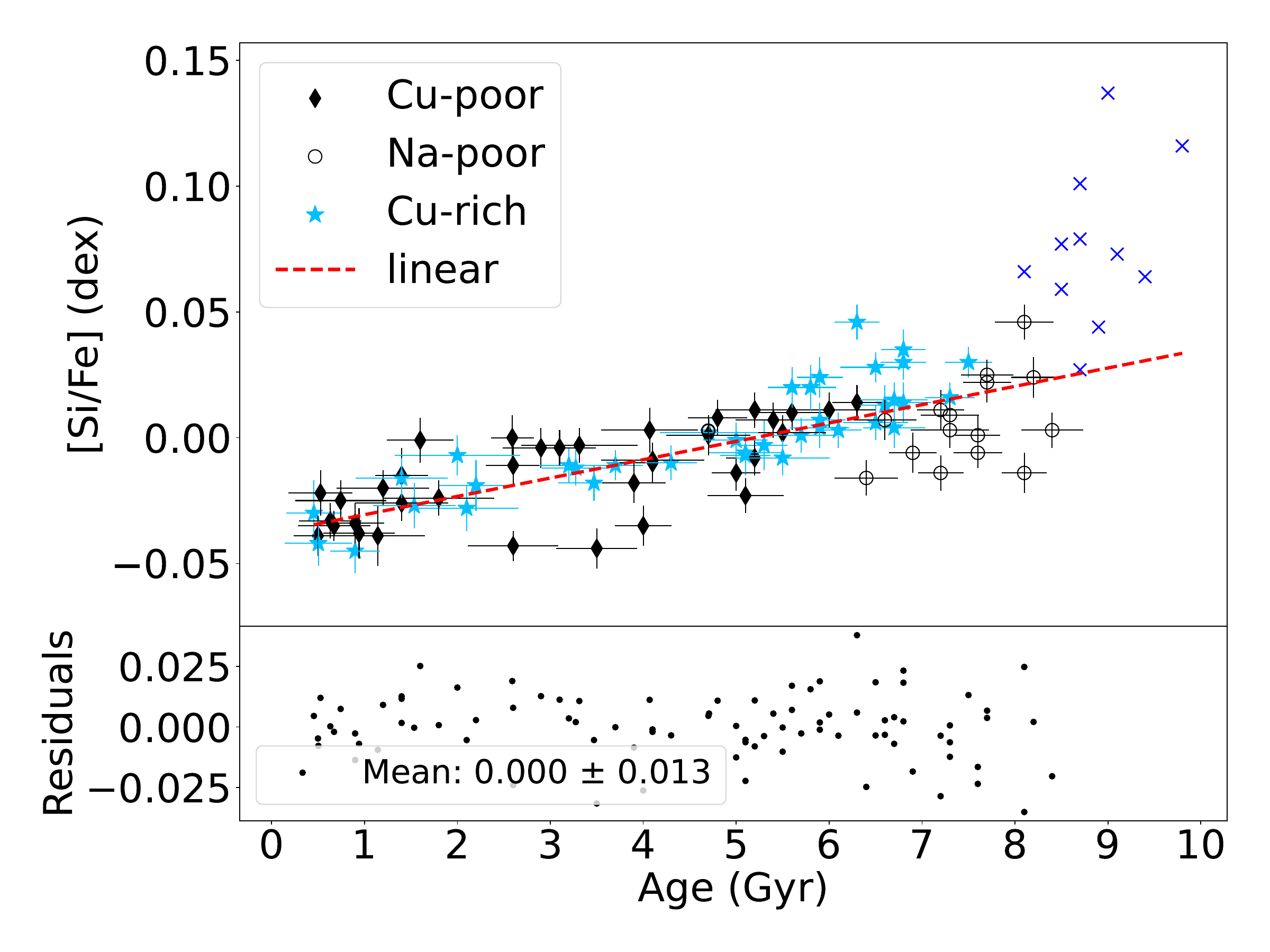} & \hspace{-0.6cm}  \includegraphics[width=0.33\linewidth]{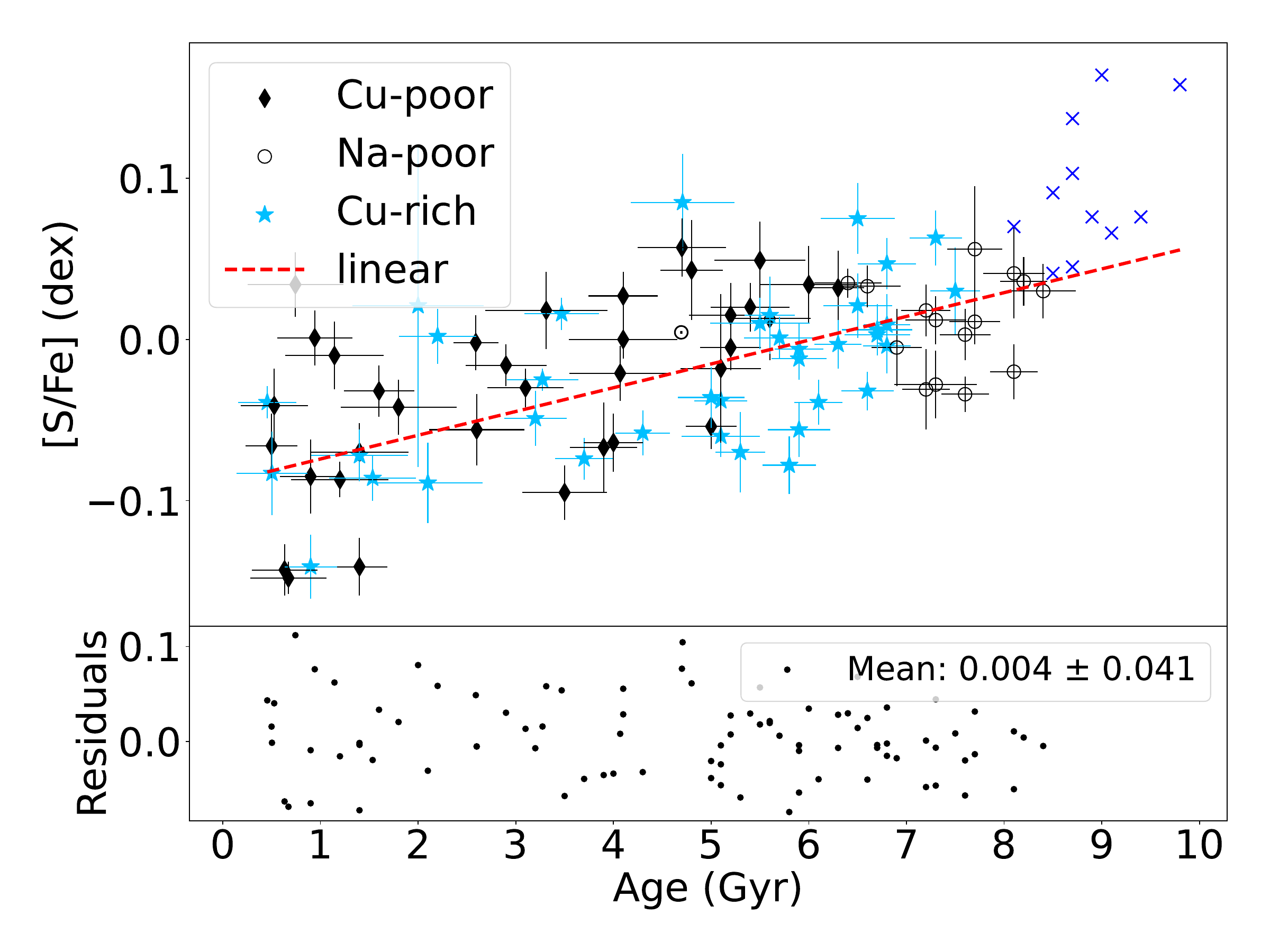}& \hspace{-0.6cm}
                        \includegraphics[width=0.33\linewidth]{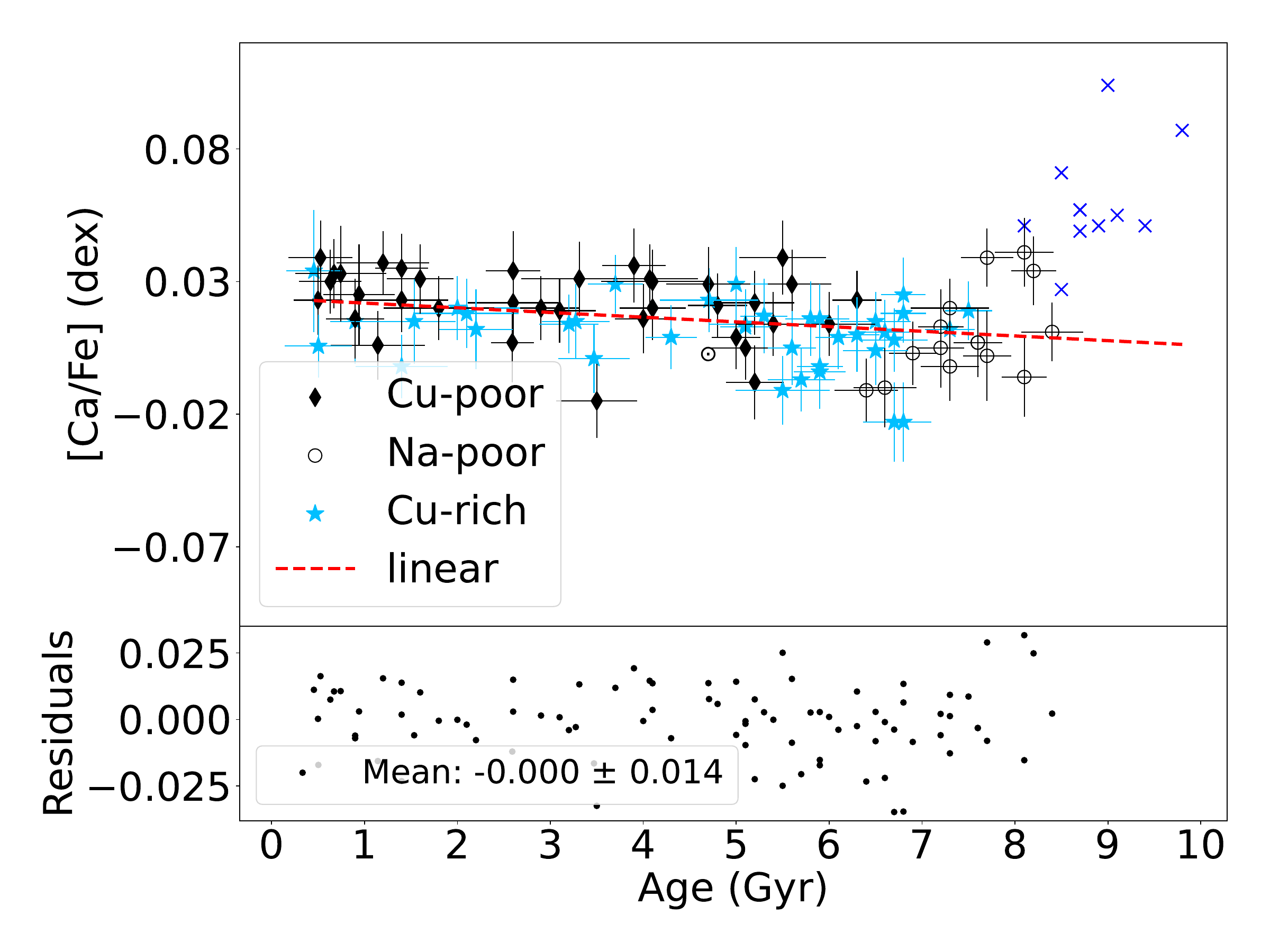} \\  \includegraphics[width=0.33\linewidth]{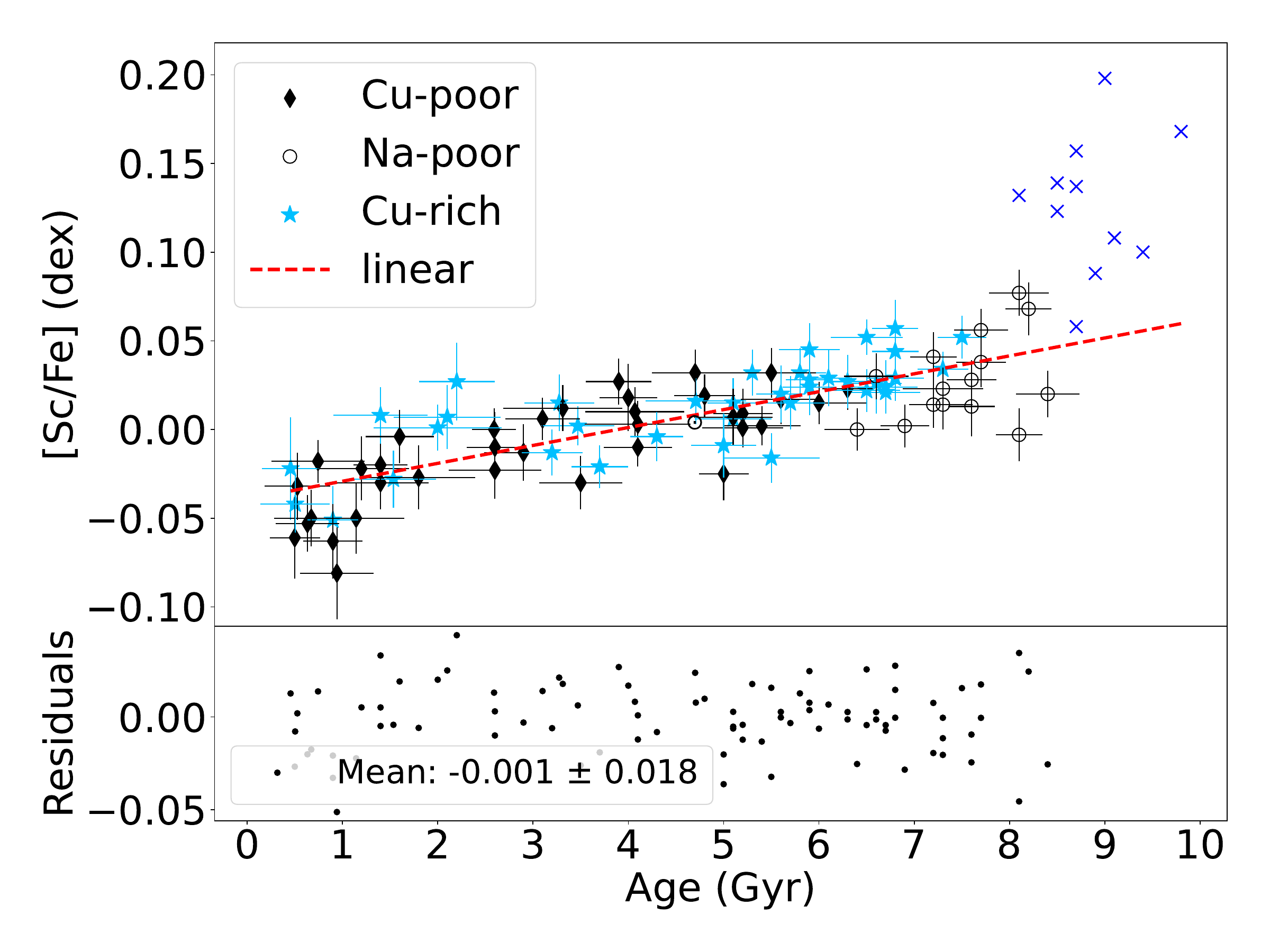}& \hspace{-0.6cm}
                        \includegraphics[width=0.33\linewidth]{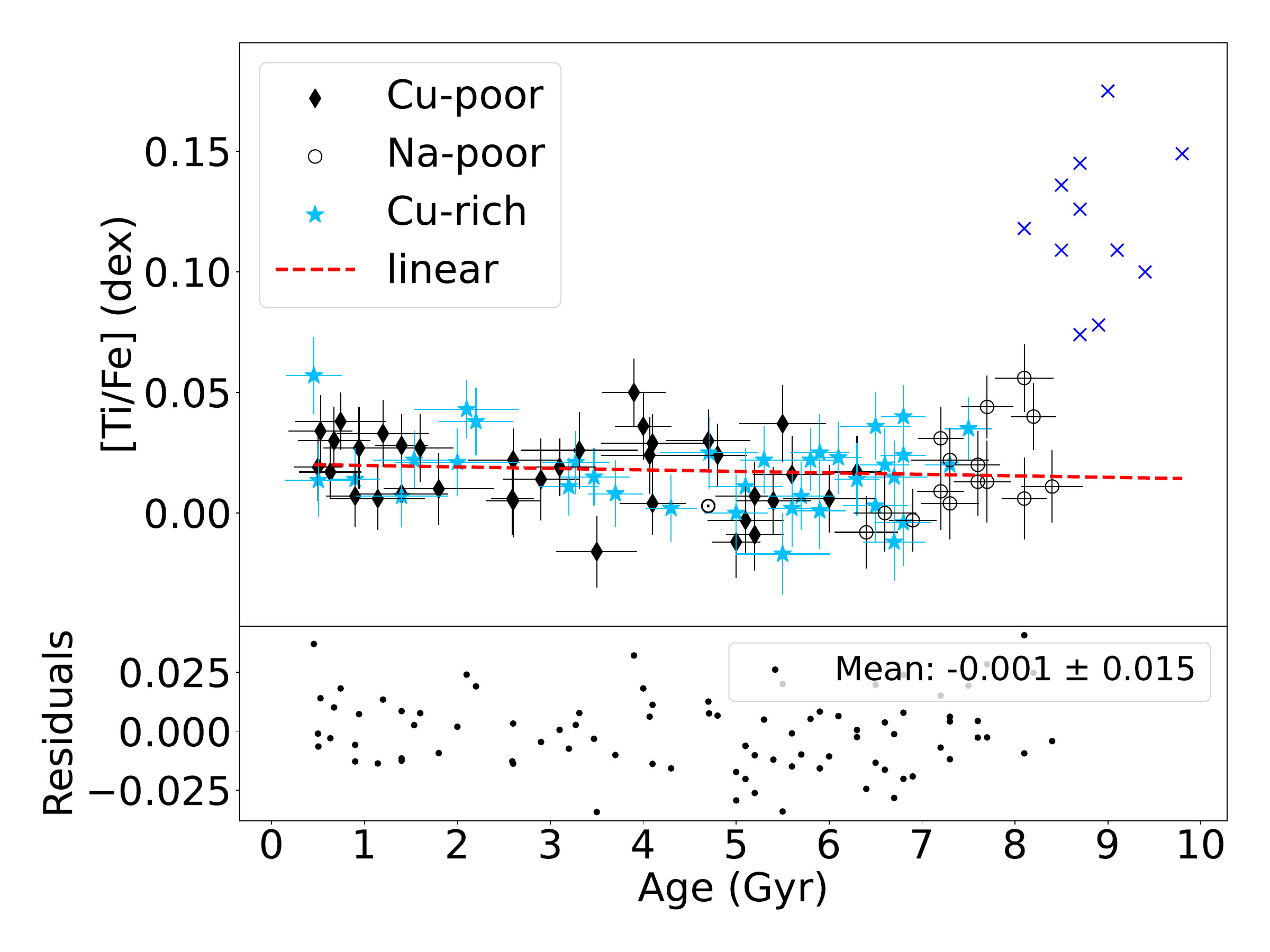} & \hspace{-0.6cm}  \includegraphics[width=0.33\linewidth]{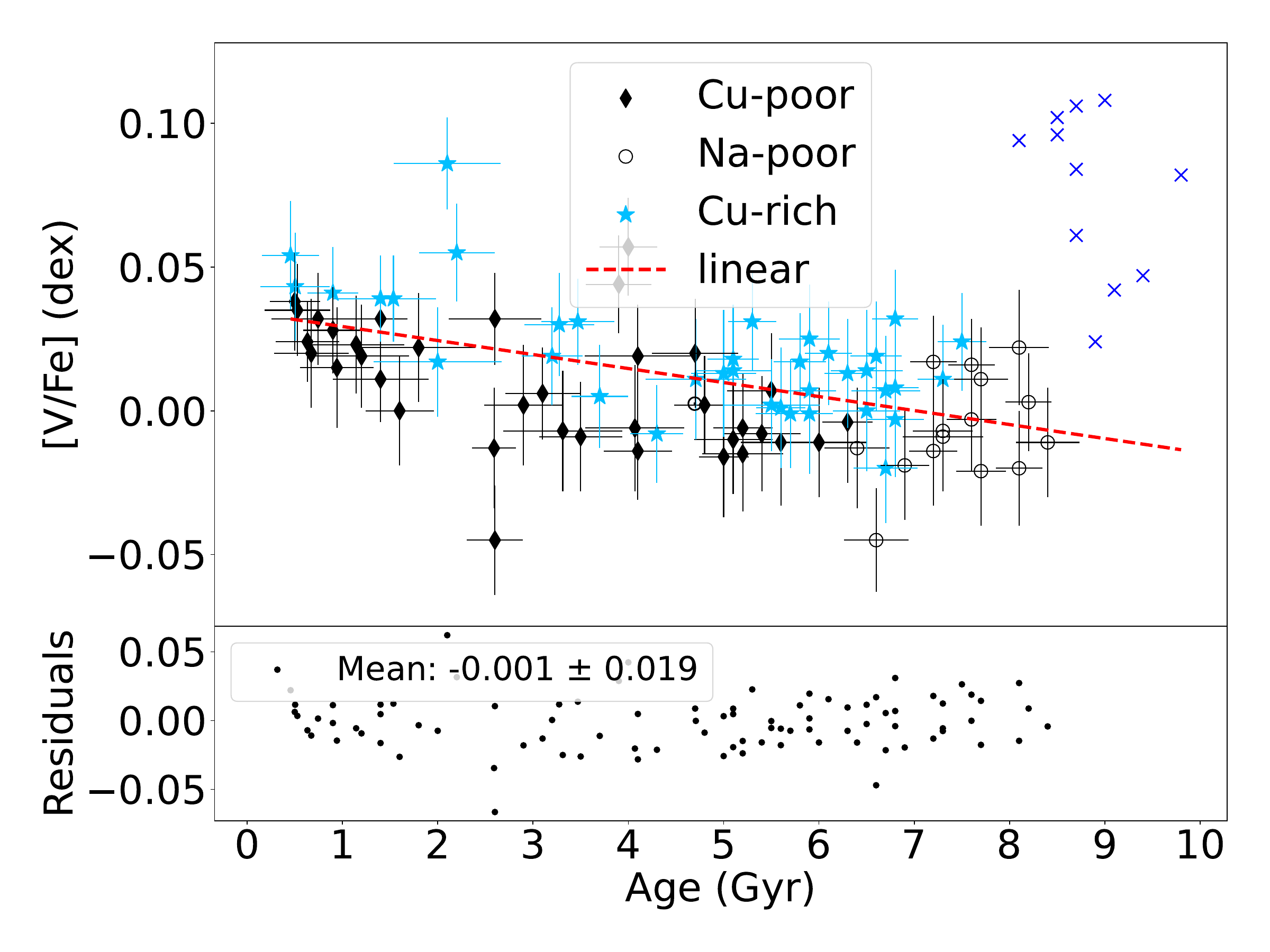}\\
                        \includegraphics[width=0.33\linewidth]{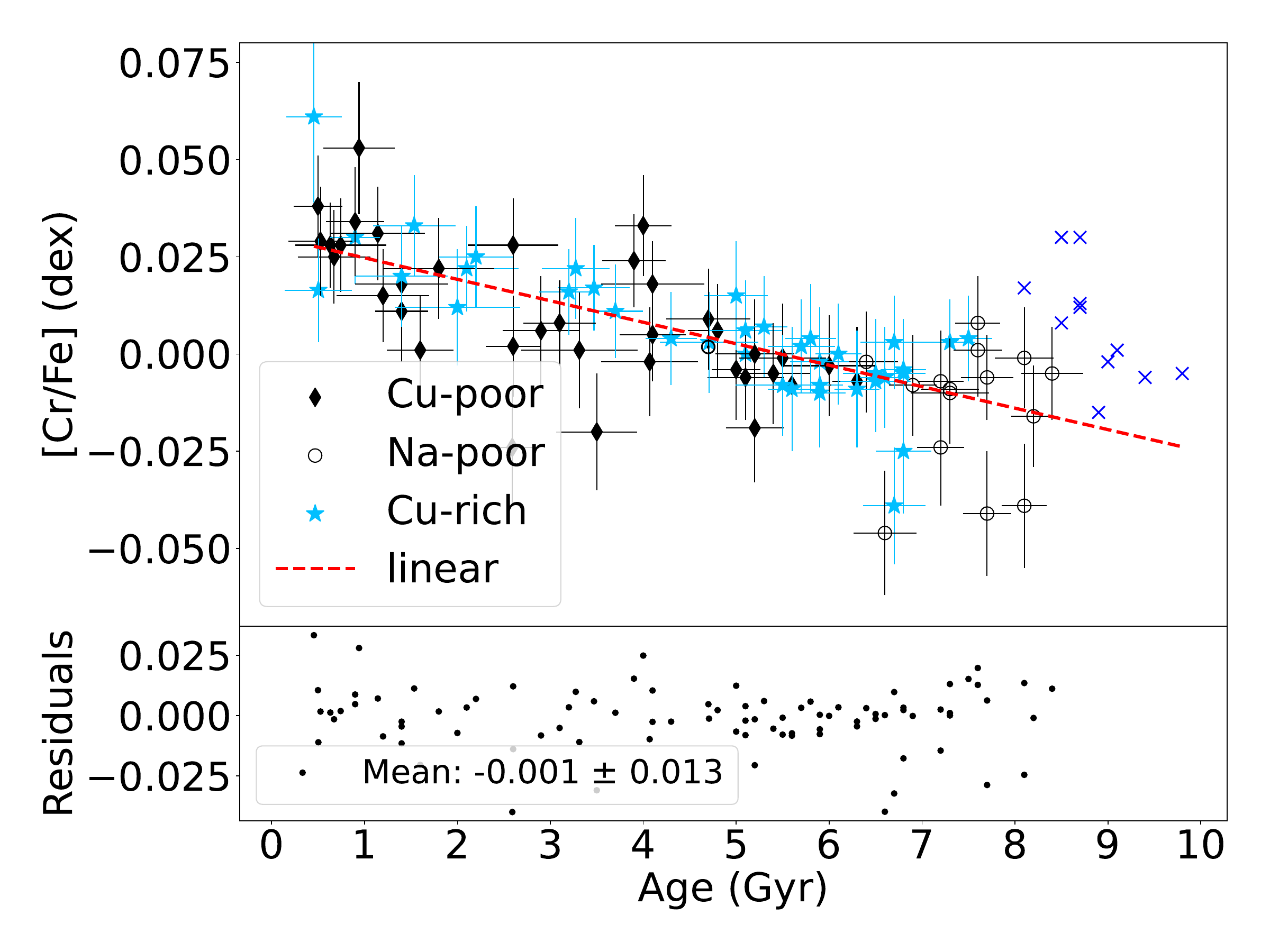} & \hspace{-0.6cm}  \includegraphics[width=0.33\linewidth]{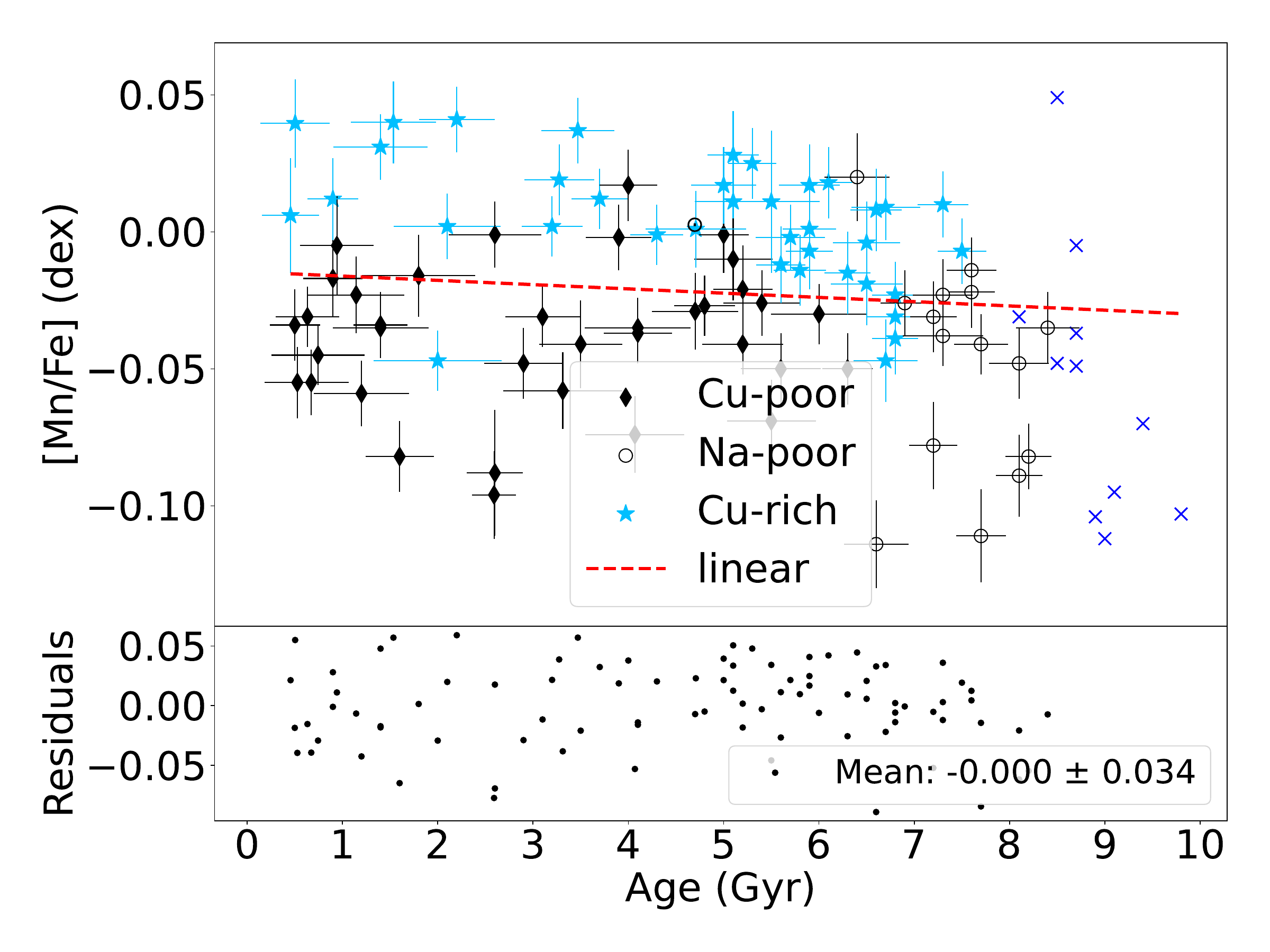}& \hspace{-0.6cm}
                        \includegraphics[width=0.33\linewidth]{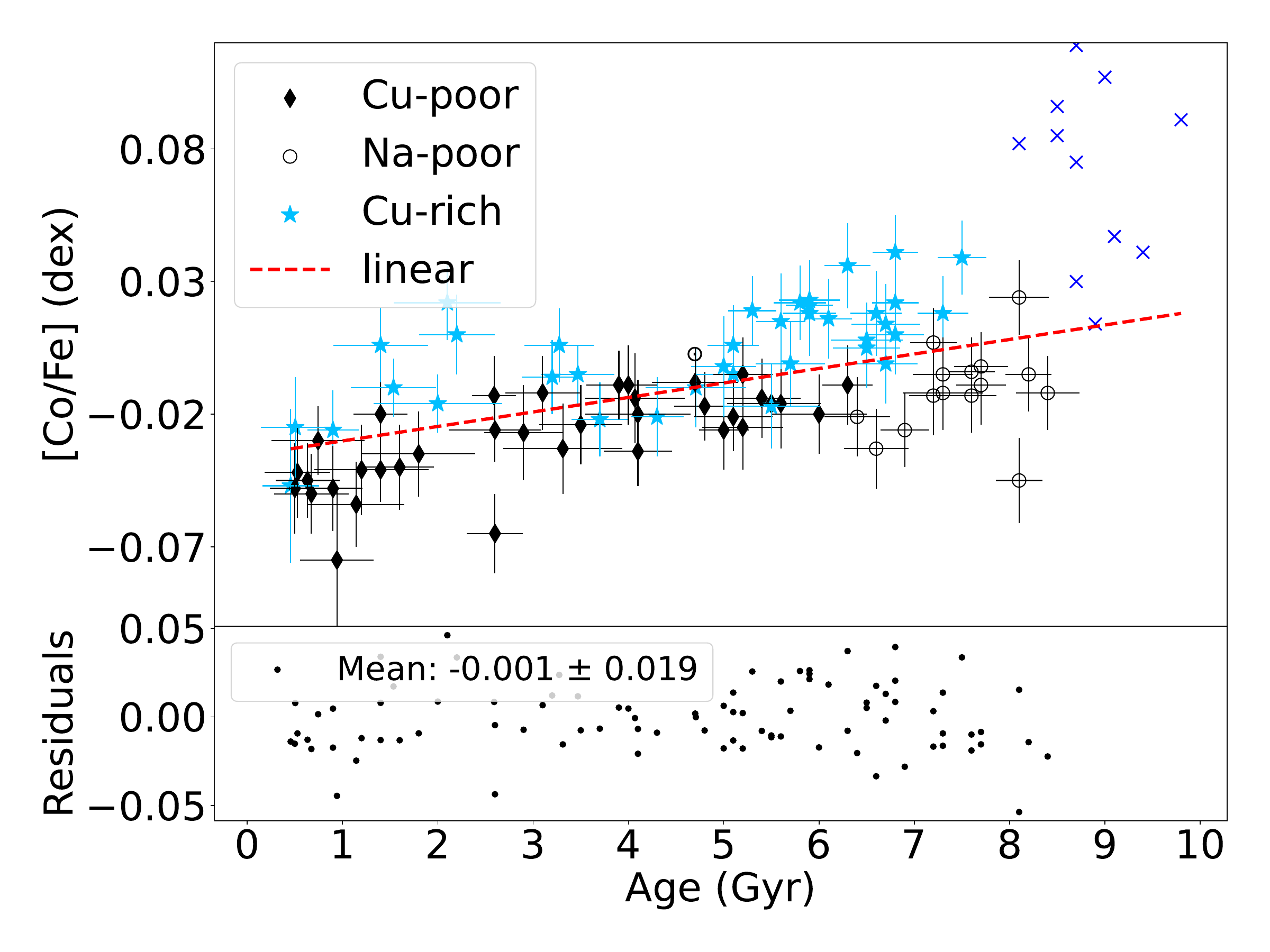} \\  \includegraphics[width=0.33\linewidth]{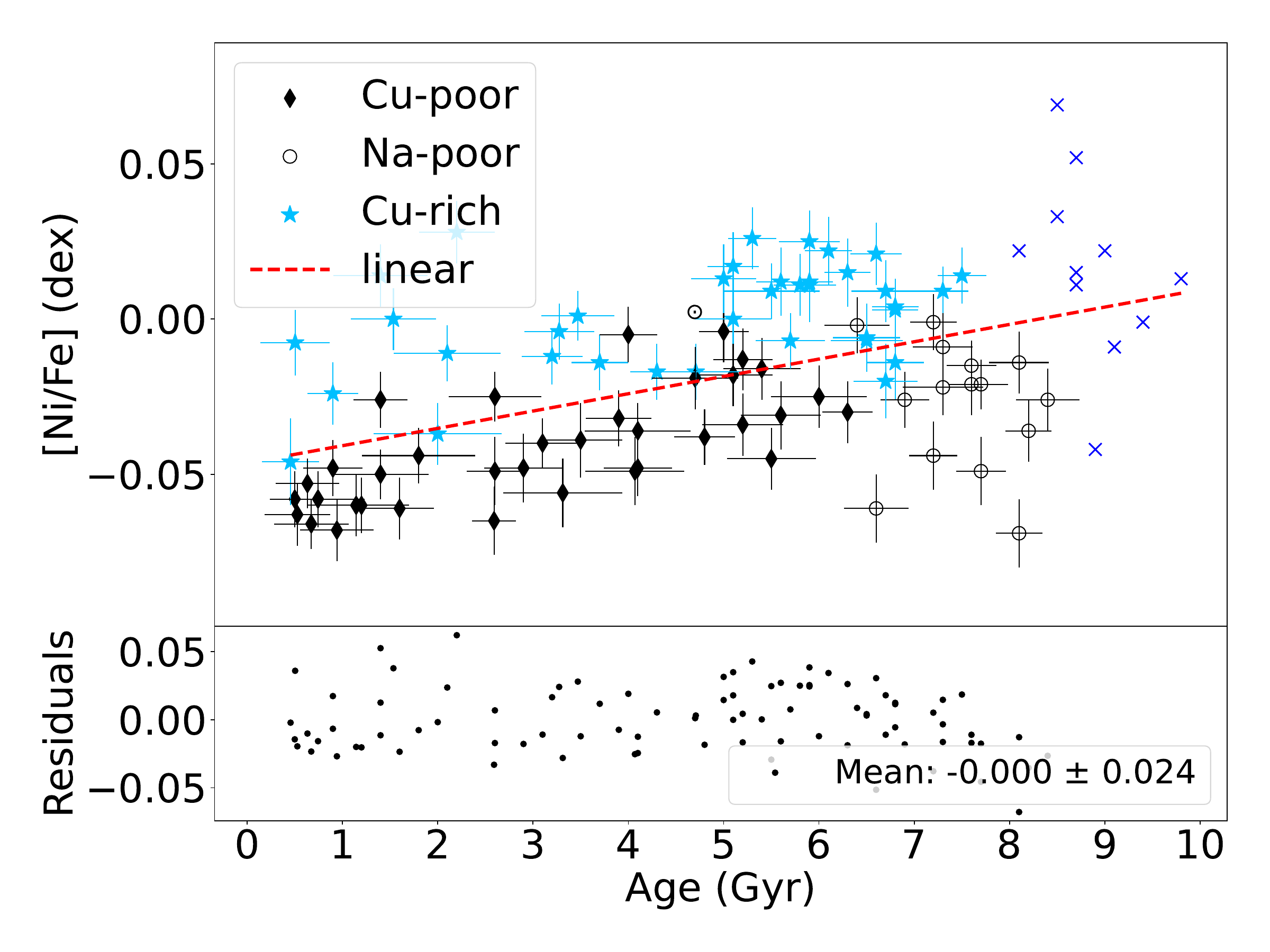}& \hspace{-0.6cm}
                        \includegraphics[width=0.33\linewidth]{fig/Cu_Fe_age_linear.pdf} & \hspace{-0.6cm}  \includegraphics[width=0.33\linewidth]{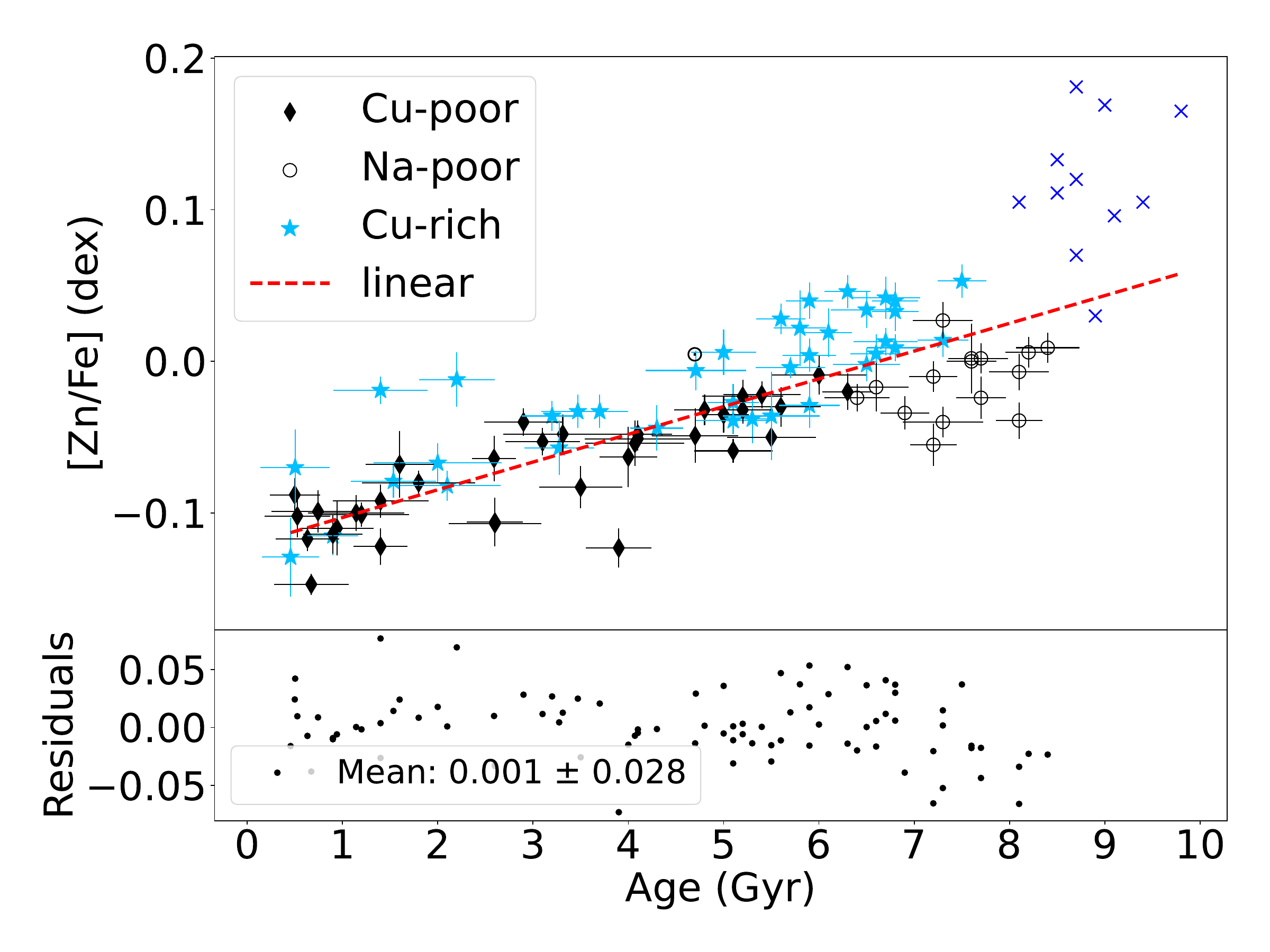}\\
                        \includegraphics[width=0.33\linewidth]{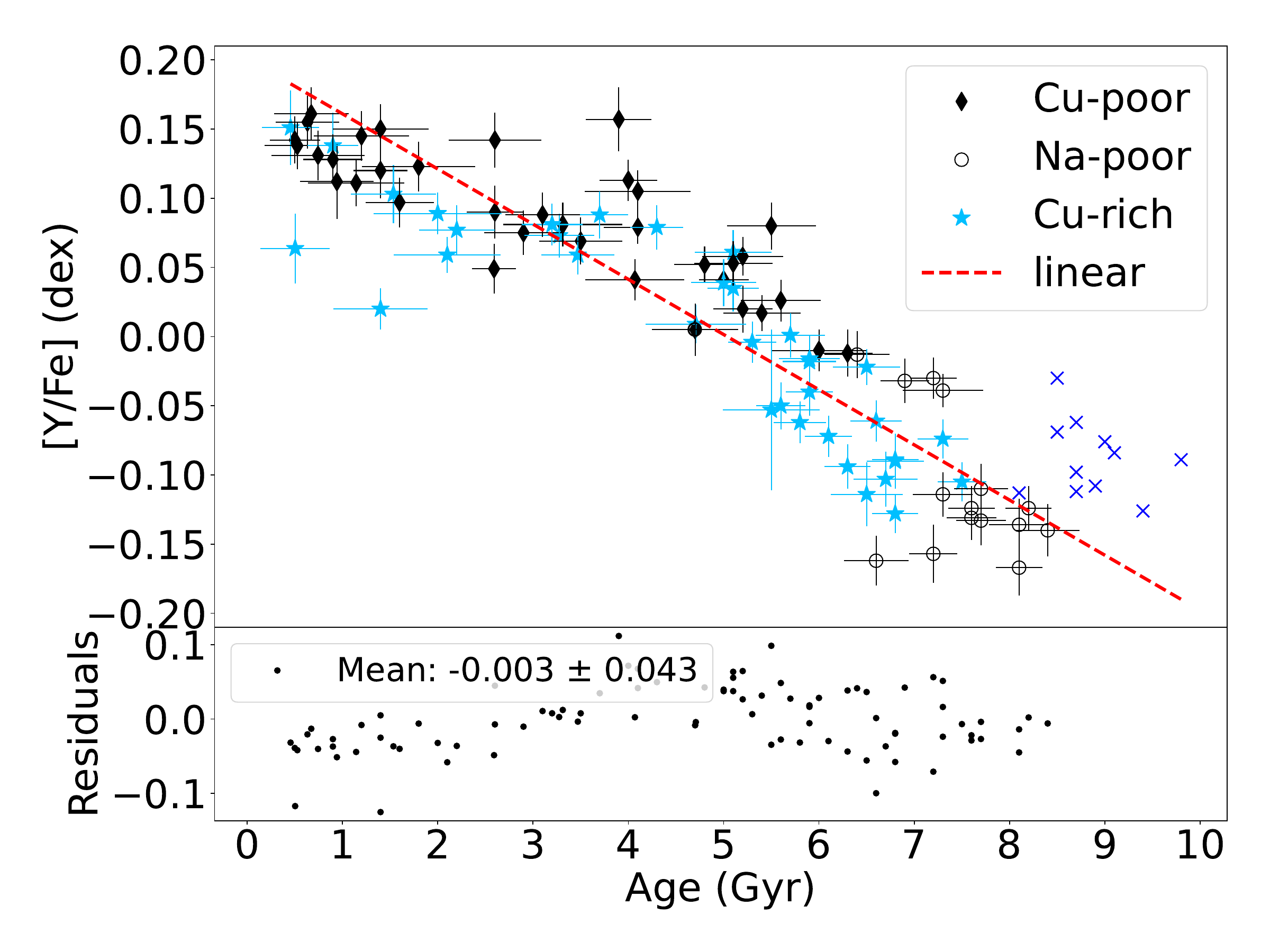} & \hspace{-0.6cm}  \includegraphics[width=0.33\linewidth]{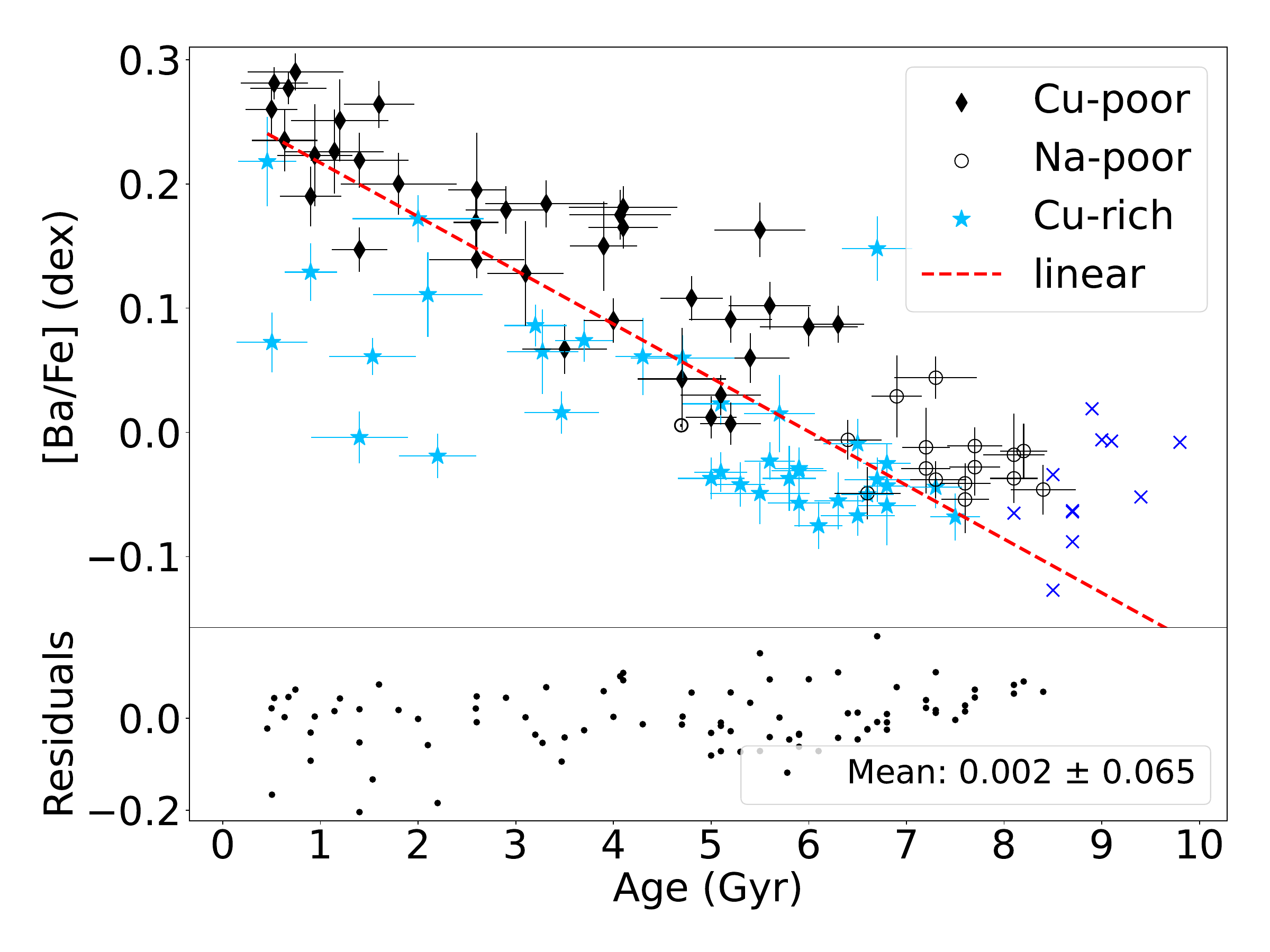}& \hspace{-0.6cm} \\
                \end{longtable}
                \captionsetup{font=tiny}
                \captionof{figure}{Linear fit of the abundance ratios [X/Fe] versus age for all 20 elements. The stars represented by blue crosses are from the thick disk and were not considered in the linear fit (red dashed line). The black empty circles are stars from the distinct population identified in the [Na/Fe] plot, and the blue filled circles and black diamonds are stars from the Cu-rich and Cu-poor populations, respectively. The bottom panel shows the residuals, as well as their average and standard deviation.}
                \label{fig:linear_fit_GCE}

        \end{appendix}

\end{document}